\documentclass[12pt]{article}
\usepackage{epsfig}
\usepackage{amssymb,amsmath}
\usepackage{setspace}
\pdfoutput=1
\usepackage{subfigure}
\usepackage{amssymb,amsmath}
\usepackage{graphicx}
\usepackage{color}
\usepackage{cancel}
\usepackage[colorlinks=true
,urlcolor=blue
,citecolor=blue
,linkcolor=blue
,pagecolor=blue
,linktocpage=true
,pdfproducer=medialab
]{hyperref}
\def\bi{\begin{itemize}}
\def\ei{\end{itemize}}
\newcommand{\Z}{{\mathbb Z}}

\def\tst{\tilde t}
\def\ttau{\tilde \tau}

\def\alt{\lesssim}
\def\agt{\gtrsim}

\newcommand\prd[3]{{\it Phys.\ Rev.\ }{\bf D #1} (#2) #3}
\newcommand\prl[3]{{\it Phys.\ Rev.\ Lett.\ }{\bf #1} (#2) #3}
\newcommand\npb[3]{{\it Nucl.\ Phys.\ }{\bf B #1} (#2) #3}
\newcommand\plb[3]{{\it Phys.\ Lett.\ }{\bf B #1} (#2) #3}

\newcommand\app[3]{{\it Astropart.\ Phys.\ }{\bf #1} (#2) #3}
\newcommand\jhep[3]{{\it J. High Energy Phys.\ }{\bf #1} (#2) #3}

\def\tst{\tilde t}
\def\ttau{\tilde \tau}

\usepackage{amsmath}

\newcommand{\bea}{\begin{eqnarray}}
\newcommand{\eea}{\end{eqnarray}}

\newcommand{\beq}{\begin{equation}}
\newcommand{\eeq}{\end{equation}}

 \makeatletter
\def\alt{\mathrel{\mathpalette\gl@align<}}
\def\agt{\mathrel{\mathpalette\gl@align>}}
\def\gl@align#1#2{\lower.6ex\vbox{\baselineskip\z@skip\lineskip\z@
\ialign{$\m@th#1\hfil##\hfil$\crcr#2\crcr\sim\crcr}}} \makeatother

\usepackage{longtable}

\begin{document}
%
\vspace*{1.0cm}

\begin{center}
\baselineskip 20pt {\Large\bf
A Realistic Intersecting D6-Brane Model after the First LHC Run
}
\vspace{1cm}
\end{center}

\begin{center}
{\large
Tianjun Li$^{a,b,}$\footnote{E-mail: tli@itp.ac.cn},
D.V. Nanopoulos$^{c,d}$\footnote{E-mail: dimitri@physics.tamu.edu},
Shabbar Raza$^{a,}$\footnote{E-mail: shabbar@itp.ac.cn}, 
Xiao-Chuan Wang$^{a,}$\footnote{E-mail: xcwang@itp.ac.cn}
} \vspace{.5cm}

{\baselineskip 20pt \it $^a$
State Key Laboratory of Theoretical Physics and Kavli Institute for Theoretical Physics China (KITPC),
Institute of Theoretical Physics, Chinese Academy of Sciences, Beijing 100190, P. R. China \\
}
{\it $^b$
School of Physical Electronics, University of Electronic Science and Technology of China,\\
Chengdu 610054, P. R. China \\
}

{\it $^c$
George P. and Cynthia W. Mitchell Institute for Fundamental Physics, Texas A$\&$M University, College Station, Texas 77843, USA\\
}

{\it $^d$
Astroparticle Physics Group, Houston Advanced Research Center (HARC), Mitchell Campus, Woodlands, Texas 77381, USA
and Academy of Athens, Division of Natural Sciences, 28 Panepistimiou Avenue, Athens 10679, Greece
}

\vspace{.5cm}

\vspace{1.5cm} {\bf Abstract}
\end{center}

With the Higgs boson mass around 125 GeV and the LHC supersymmetry search constraints,
we revisit a  three-family Pati-Salam model from intersecting D6-branes in
Type IIA string theory on the $\mathbf{T^6/(\Z_2\times \Z_2)}$
orientifold which has a realistic phenomenology. We systematically scan the
parameter space for $\mu<0$ and $\mu>0$, and find
that the gravitino mass is generically heavier than about 2 TeV for 
both cases due to the Higgs mass low bound 123 GeV. In particular, we identify 
a region of parameter space with the electroweak fine-tuning as small as 
$\Delta_{EW} \sim$ 24-32 (3-4$\%$). In the viable parameter space which 
is consistent with all the current constraints,
the mass ranges for gluino, the first two-generation
squarks and sleptons are respectively $[3, ~18]$ TeV, $[3, ~16]$ TeV, and $[2, ~7]$ TeV.
For the third-generation sfermions, the light stop satisfying $5\sigma$ WMAP bounds 
via neutralino-stop coannihilation has mass from 0.5 to 1.2 TeV, and 
the light stau can be as light as 800 GeV. We also show various coannihilation and resonance 
scenarios through which the observed dark matter relic density is achieved. Interestingly,
the certain portions of parameter space has excellent $t$-$b$-$\tau$ and $b$-$\tau$ Yukawa 
coupling unification. Three regions of parameter space are highlighted as well where 
the dominant component of the lightest neutralino is a bino, wino or higgsino. 
We discuss various scenarios in which such solutions may avoid recent astrophysical bounds 
in case if they satisfy or above observed relic density bounds. 
Prospects of finding higgsino-like neutralino in direct and indirect searches are also
studied. And we display six tables of benchmark points depicting various interesting features 
of our model. Note that the lightest neutralino can be heavy up to 2.8 TeV,
and there exists a natural region of parameter space from 
low-energy fine-tuning definition with heavy gluino
and first two-generation squarks/sleptons, we point out that
the 33 TeV and 100 TeV proton-proton colliders are indeed needed to probe our D-brane model.


\thispagestyle{empty}

\newpage

\addtocounter{page}{-1}

\baselineskip 18pt

\section{Introduction}

String theory is one of the most promising candidates for quantum gravity.
Thus, the string phenomenology goal is 
 to construct the Standard Model (SM) or Supersymmetric SMs (SSMs)  
from string theory with moduli stabilization and without chiral
exotics, and try to make unique predictions which can probed at
the Large Hadron Collider (LHC) and other future experiments.
It is well-known that four kinds of string models have 
been studied extensively: (1) The heterotic $E_8\times E_8$ string model building. 
The SSMs can be constructed via the orbifold 
compactifications~\cite{Buchmuller:2005jr, Lebedev:2006kn, Kim:2006hw} 
and the Calabi-Yau manifold 
compactifications~\cite{Braun:2005ux, Bouchard:2005ag}.
(2)  The free fermionic string model building. The realistic models 
with clean particle spectra such as the standard-like models,
Pati-Salam models, and flipped $SU(5)$ models have been constructed at 
the Kac-Moody level one~\cite{AEHN, Faraggi:1989ka,
Antoniadis:1990hb, LNY, Cleaver:2001ab}.
(3) The D-brane model building. Two kinds of such
models have been studied: (i) Intersecting D-brane models~\cite{Berkooz:1996km,
Ibanez:2001nd, Blumenhagen:2001te, CSU, Cvetic:2002pj, Cvetic:2004ui, Cvetic:2004nk, 
Cvetic:2005bn, Chen:2005ab, Chen:2005mj, Blumenhagen:2005mu}; (ii) 
Orientifolds of Gepner 
models~\cite{Dijkstra:2004ym, Dijkstra:2004cc}.  
(4) The F-theory model building for the $SU(5)$, flipped $SU(5)$, 
and $SU(3)_C\times SU(2)_L \times SU(2)_R \times U(1)_{B-L}$ 
models~\cite{Vafa:1996xn, Donagi:2008ca,
Beasley:2008dc, Beasley:2008kw, Donagi:2008kj, Font:2008id, Jiang:2009zza, 
Jiang:2009za, Li:2009cy}. 

For the intersecting D-brane model building, the realistic SM fermion Yukawa couplings can 
be realized only in the Pati-Salam models. The three-family Pati-Salam models
have been constructed systematically in
Type IIA string theory on the $\mathbf{T^6/(\Z_2\times \Z_2)}$
orientifold with intersecting D6-branes~\cite{Cvetic:2004ui}, and 
two of us (TL and DVN) with Chen and Mayes
found that one model has a realistic phenomenology: the tree-level 
gauge coupling unification is realized naturally at the string scale,  
the Pati-Salam gauge symmetry can be broken to the SM close to the string
scale, the small number of extra chiral exotic states  may be decoupled
via the Higgs mechanism and strong dynamics, the SM fermion masses and mixings
can be explained, the low-energy supersymmetric particle spectra might potentially
be tested at the LHC, and the observed dark matter relic density may be 
generated for the lightest neutralino 
as the lightest supersymmetric particle (LSP), etc~\cite{Chen:2007px, Chen:2007zu}. 
As far as we know, this is indeed one of the best globally 
consistent string models.

On the other hand, for the first run of the LHC, the big success is obviously the discovery of 
a SM-like Higgs boson with mass $m_h$ around 125 GeV 
in July 2012~\cite{Aad:2012tfa, Chatrchyan:2012ufa}, which is 
a little bit too large for the Minimal SSM (MSSM). Such large Higgs boson mass
in the MSSM requires the multi-TeV top squarks with small mixing or TeV-scale 
top squarks with large mixing. In addition, 
the LHC supersymmetry (SUSY) searches have given strong constraints on 
the pre-LHC viable parameter space. For instance, the gluino mass $m_{\tilde g}$ should
be heavier than about 1.7 TeV if the first two-generation squark mass $m_{\tilde q}$ 
is around the gluino mass $m_{\tilde q} \sim m_{\tilde g}$, and heavier than about 1.3 TeV
for $m_{\tilde q} \gg m_{\tilde g}$~\cite{Chatrchyan:2013wxa, Aad:2014wea}.

Therefore, we should update the phenomenological study of this intersecting D-brane model.
 For this purpose, we have systematically scan the viable parameter space by
considering $\mu<0$ and $\mu>0$ scenarios where $\mu$ is the bilinear Higgs mass term. 
We show that there indeed exists such viable parameter space 
 which satisfies the collider and astrophysical bounds including the Higgs boson mass
in the range $[123, 127]$ GeV. In particular, the absolute value of $\mu$ 
can be as small as 300 GeV in a region of parameter space, where the electroweak fine-tuning (EWFT) is small around $\Delta_{EW} \sim$ 24-32 (3-4$\%$).
 We identify another region of parameter space with $|\mu|\lesssim$ 500 GeV and $\Delta_{EW}\lesssim$ 300, where gluino masses are from 3 to 7 TeV, 
 and the first two-generation squarks and sleptons are in the mass ranges of 
 $[4,~7]$ TeV and $[2,~4]$ TeV, respectively. Because such parameter space is 
 natural from the low-energy fine-tuning definition while the gluino and first 
 two-generation squarks/sleptons are out of the reach of 14 TeV LHC,
 this will provide a strong motivation for the
33 TeV and 100 TeV proton-proton colliders.
There is some visible preference to achieve the viable parameter space
 consistent with constraints for $\mu <0$ case, but this is just an artifact of lack of statistics 
for $\mu>0$. Moreover,  in order to have the Higgs boson mass from 123 GeV to 127 GeV, 
and satisfy the LHC low bounds on sparticles and the B-physics bounds, we require 
gravitino mass $\gtrsim$ 2 TeV for both cases of $\mu <0$ and $\mu>0$. We also present graphs in 
neutralino-sparticle planes showing various coannihilation scenarios such as neutralino-stau, 
neutralino-stop, neutralino-gluino, and $A$-resonance solutions. The solutions, which are consistent with the 
observed relic density, have gluino masses from 3 to 18 TeV. 
We also note that in our present data consistent with all bounds, 
the first two generation squarks are in the mass range $[3,~16]$ TeV and the
 first two generation sleptons can be heavier than 2 TeV but less than 6 TeV.  On the other hand for third
family squarks, the NLSP light stop satisfying $5\sigma$ WMAP bounds is in the mass of 0.5-1.2 TeV, in case of third
family slepton, the light stau can be as light as 800 GeV. We have checked status of $t$-$b$-$\tau$ and
$b$-$\tau$ Yukawa unification (YU) scenarios with both signs of $\mu$ in our data. 
For $\mu<0$ we find solutions with $10\%$ or better YU with typical heavy spectra. 
The best YU we have achieved in our data set is about $5\%$ consistent with all
the constraints including the observed dark matter relic density bound. 
On the other hand, we do not have better than $12\%$ YU $t$-$b$-$\tau$  for $\mu>0$ case. 
Since we did not perform any dedicate searches to study YU in this project 
otherwise we may have solutions with much better YU. Relaxing the $t$-$b$-$\tau$ YU constraint 
to $b$-$\tau$ YU, we have plenty of solutions with $100\%$ YU. For the points
with $\Omega h^2 \gtrsim$ 1 where the lightest neutralino is almost a pure bino, 
we introduce a lighter state axino $\tilde a$ as the LSP. Thus, the
lightest neutralino is the Next to the LSP (NLSP) and can decay to axino
via  $\tilde \chi_{1}^{0}\rightarrow \gamma \tilde a$.
We calculate the lifetime of the NLSP neutralino for various choices of 
the axion decay constant $f_a$ 
in our data. For $f_a > 10^{14}$ GeV, the lifetime of the NLSP bino is more than 1 second 
and may be ruled out by Big Bang Nucleosynthesis (BBN) constraints. We also note that in our data, 
there are solutions where the lightest neutralino can be a 
bino, wino, or higgsino type. The lightest neutralino masses are
more than 1 TeV for both cases ($\mu<0$ and $\mu>0$) in the wino-type solutions, while 
they are less than 1 TeV in the bino-type solutions and 
 in the mass range of 150-600 GeV in the higgsino-type solutions. 
Recent studies showed that the scenario with pure wino as dark matter is 
under siege~\cite{Fan:2013faa, Cohen:2013ama}. 
In our model, the relic density of the wino dominant lightest neutralino can be smaller 
than the correct relic density, and then the above constraint can be escaped. 
Otherwise,  to solve this problem,
we suggest that the wino dominant neutralino is the NLSP and may decay to $\tilde a \gamma$ 
and hence fulfil the relic density bounds, or we may invoke 
R-parity violation. Similarly, the higgsino-type solutions suffer underabundance of 
relic density problem. In such a case we assume that the higgsino-type neutralino 
makes up only a fraction of the dark matter relic density and the remaining
abundance is comprised of axions. 
We also display graphs for direct and indirect searches for dark matter for our 
higgsino-like solutions and show that these solutions will be observed 
or ruled out by the XENON1T experiment. Finally, we present 
six tables of benchmark points, three for each sign of $\mu$. 
These points depict various interesting scenarios of our model, 
namely points with minimum EWFT, various coannihilation 
and resonance solutions, bino-type, wino-type and higgsino-type solutions.
Furthermore, because the lightest neutralino can be heavier than 1 TeV and up to about 2.8 TeV,
how to search for such scenario at the 14 TeV LHC is still a challenging question. In short,
we do need the 33 TeV and 100 TeV proton-proton colliders 
to probe such D-brane model.

This paper is organized as follows. In Section \ref{sec:scan} we outline details 
of the supersymmetry breaking (SSB) 
parameters, the range of values employed in our scan, 
the scanning procedure and the relevant experimental constraints 
that we have employed.
In Section \ref{ft} we briefly describe our definition of EWFT and High scale (GUT) 
fine-tuning. We discuss results
of our scans in Section \ref{results}. 
A summary and conclusions are given in Section \ref{conclusions}.

\section{Phenomenological constraints and scanning procedure}
\label{sec:scan}

In our realistic intersecting D-brane model, if we do not consider
CP violation, the supersymmetry breaking (SSB) soft terms from the non-zero F-terms $F^{u^i}$ and $F^s$
can be parametrized by $\Theta_1$, $\Theta_2$, $\Theta_3$, $\Theta_4\equiv \Theta_s$,
and gravitino mass $m_{3/2}$ where $\sum_{i=1}^4\Theta^2_i=1$~\cite{Chen:2007zu}.
Thus, we can reparametrize $\Theta_i$ with $i=1,~2,~3$ in terms of $\gamma_1$, $\gamma_2$ and $\Theta_4$ as follows 
\begin{align}
\alpha & \equiv 2 \pi \gamma_{1} ~,~\nonumber\\
\beta &\equiv 2 \pi \gamma_{2} ~,~\nonumber  \\
\Theta_1 &= \cos(\beta)\cos(\alpha)\sqrt{1-\Theta_{4}^{2}} ~,~\nonumber\\
\Theta_2 &= \cos(\beta)\sin(\alpha)\sqrt{1-\Theta_{4}^{2}} ~,~\nonumber \\
\Theta_3 &= \sin(\beta)\sqrt{1-\Theta_{4}^{2}} ~.~
\end{align}
Thus, the supersymmetry breaking soft terms are~\cite{Chen:2007zu} 
\begin{align}
M_1 &=(0.519\Theta_1+0.346\Theta_{2}+0.866\Theta_{3})m_{3/2} ~,~\nonumber \\
M_2 &=(0.866\Theta_2 - 0.866\Theta_{4} )m_{3/2} ~,~\nonumber \\
M_3 &=(0.866\Theta_2 + 0.866\Theta_{3})m_{3/2} ~,~\nonumber \\
A_0 &=(-1.111\Theta_{1}-0.621\Theta_{2}+0.245\Theta_{3}-0.245\Theta_{4})m_{3/2} ~,~\nonumber \\
m_L &=\sqrt{(1.0+0.899\Theta_{1}^{2}-0.518\Theta_{2}^{2}-0.849\Theta_{3}^{2}-1.418\Theta_{4}^{2}-0.557\Theta_{1}\Theta_2-0.557\Theta_{3}\Theta_{4})}m_{3/2} ~,~\nonumber \\
m_R &=\sqrt{(1.0-1.418\Theta_{1}^{2}-0.849\Theta_{2}^{2}-0.518\Theta_{3}^{2}+0.899\Theta_{4}^{2}-0.557\Theta_{1}\Theta_{2}-0.557\Theta_{3}\Theta_{4})}m_{3/2} ~,~\nonumber\\
m_{H_u}&=m_{H_d}=\sqrt{(1.0-(1.5 \Theta_{3}^2)-(1.5 \Theta_{4}^2))}m_{3/2}~, 
\label{ssb}
\end{align}
where $M_{1,2,3}$ are the gauginos masses respectively for
$U(1)_Y$, $SU(2)_L$ and $SU(3)_c$ gauge groups, $A_0$ is the trilinear scalar coupling, $m_L$ and $m_R$
are the soft mass terms respectively for the left-handed and right-handed
squarks and sleptons, and $m_{H_{u,d}}$ are the SSB Higgs soft mass terms.

We employ the ISAJET~7.84 package~\cite{ISAJET}
 to perform random scans over the parameter space
 given below.
In this package, the weak scale values of gauge and third
 generation Yukawa couplings are evolved to
 $M_{\rm GUT}$ via the MSSM renormalization group equations (RGEs)
 in the $\overline{DR}$ regularization scheme.
We do not strictly enforce the unification condition
 $g_3=g_1=g_2$ at $M_{\rm GUT}$, since a few percent deviation
 from unification can be assigned to unknown GUT-scale threshold
 corrections~\cite{Hisano:1992jj}.
With the boundary conditions given at $M_{\rm GUT}$,
 all the SSB parameters, along with the gauge and Yukawa couplings,
 are evolved back to the weak scale $M_{\rm Z}$.

In evaluating Yukawa couplings the SUSY threshold
 corrections~\cite{Pierce:1996zz} are taken into account
 at the common scale $M_{\rm SUSY}= \sqrt{m_{\tst_L}m_{\tst_R}}$.
The entire parameter set is iteratively run between
 $M_{\rm Z}$ and $M_{\rm GUT}$ using the full 2-loop RGEs
 until a stable solution is obtained.
To better account for leading-log corrections, one-loop step-beta
 functions are adopted for gauge and Yukawa couplings, and
 the SSB parameters $m_i$ are extracted from RGEs at appropriate scales
 $m_i=m_i(m_i)$.
The RGE-improved 1-loop effective potential is minimized
 at an optimized scale $M_{\rm SUSY}$, which effectively
 accounts for the leading 2-loop corrections.
Full 1-loop radiative corrections are incorporated
 for all sparticle masses.

The requirement of radiative electroweak symmetry breaking
 (REWSB)~\cite{Ibanez:1982fr} puts an important theoretical
 constraint on the parameter space.
Another important constraint comes from limits on the cosmological
 abundance of stable charged particle~\cite{Beringer:1900zz}.
This excludes regions in the parameter space where charged
 SUSY particles, such as $\ttau_1$ or $\tst_1$,
 become the LSP.
We accept only those solutions for which one of the neutralinos
 is the LSP and saturates the dark matter relic abundance bound
 observed by WMAP9.

We have performed Markov-chain Monte Carlo (MCMC) scans
 for the following parameter range
\begin{align}
0\leq & \gamma_1  \leq 1  ~,~\nonumber \\
0\leq & \gamma_2  \leq 1 ~,~\nonumber \\
0\leq &  \Theta_4  \leq 1 ~,~\nonumber \\
1\leq & m_{3/2}  \leq 10 \, \rm{TeV} ~,~\nonumber \\
2\leq & \tan\beta  \leq 60~,
 \label{input_param_range}
\end{align}
where  $\tan\beta$ is the ratio of the vacuum expectation values (VEVs) of two Higgs fields.
We use $m_t = 173.3\, {\rm GeV}$~\cite{:2009ec}, and $m_b^{\overline{DR}}(M_{\rm Z})=2.83$ GeV
 which is hard-coded into ISAJET. 
We have done our scans with  both $\mu<0$ and $\mu > 0$,
and find that our results are not too sensitive to one
 or two sigma variation in the value of $m_t$~\cite{bartol2}.

In scanning the parameter space, we employ the Metropolis-Hastings
 algorithm as described in \cite{Belanger:2009ti}.
The collected data points  all satisfy the requirement of REWSB,
 with the neutralino in each case being the LSP.
After collecting the data, we require the following bounds (inspired by the LEP2 experiment) on particle masses:
\begin{eqnarray} 
m_{\tilde t_1},m_{\tilde b_1} \gtrsim 100 \; {\rm GeV} ~,~\\
m_{\tilde \tau_1} \gtrsim 105 \; {\rm GeV}  ~,~\\
m_{\tilde \chi_{1}^{\pm}} \gtrsim 103 \; {\rm GeV}~.~
\end{eqnarray}

We also use IsaTools package~\cite{bsg, bmm} and Ref.~\cite{mamoudi}
 to implement the following B-physics constraints:
\begin{eqnarray}
0.8\times 10^{-9} \leq{\rm BR}(B_s \rightarrow \mu^+ \mu^-) 
  \leq 6.2 \times10^{-9} \;(2\sigma)~~&\cite{Aaij:2012nna}~,~& 
\\ 
2.99 \times 10^{-4} \leq 
  {\rm BR}(b \rightarrow s \gamma) 
  \leq 3.87 \times 10^{-4} \; (2\sigma)~~&\cite{Amhis:2012bh}~,~&  
\\
0.15 \leq \frac{
 {\rm BR}(B_u\rightarrow\tau \nu_{\tau})_{\rm MSSM}}
 {{\rm BR}(B_u\rightarrow \tau \nu_{\tau})_{\rm SM}}
        \leq 2.41 \; (3\sigma)~~&\cite{Asner:2010qj}~.~&  
\end {eqnarray}
In addition to above constraints, we impose the following bounds from the LHC and WMAP9
\begin{eqnarray}
m_h  = 123-127~{\rm GeV}~~&\cite{ATLAS, CMS}~,~&  \\ 
m_{\tilde{g}} \gtrsim  1.7 \, {\rm TeV}\ ({\rm for}\ m_{\tilde{g}}\sim m_{\tilde{q}}) &
\cite{Chatrchyan:2013wxa, Aad:2014wea} ~,~\\
m_{\tilde{g}}\gtrsim 1.3 \, {\rm TeV}\ ({\rm for}\ m_{\tilde{g}}\ll m_{\tilde{q}}) &
\cite{Chatrchyan:2013wxa, Aad:2014wea}~,~\\
 0.0913 \leq \Omega_{\rm CDM}h^2 (\rm WMAP9) \leq 0.1363   \; (5\sigma)~~&\cite{WMAP9}~.~&
\end{eqnarray}
As far as the muon anomalous magnetic moment $a_{\mu}$ is concerned, we require that the benchmark
points are at least as consistent with the data as the Standard Model.
\section{Fine-Tuning \label{ft}}
We use the latest (7.84) version of  ISAJET \cite{ISAJET} to calculate the  fine-tuning (FT) conditions at the 
electroweak scale (EW) $M_{EW}$ and at the high scale ($M_{HS}$). Brief description of these parameters is given 
in this section.

The Z boson mass $M_Z$, after including the one-loop effective potential contributions to the tree level MSSM Higgs potential,
is given by the following relation:
\begin{equation}
\frac{M_Z^2}{2} =
\frac{(m_{H_d}^2+\Sigma_d^d)-(m_{H_u}^2+\Sigma_u^u)\tan^2\beta}{\tan^2\beta
-1} -\mu^2 \; ,
\label{eq:mssmmu}
\end{equation}
where $\Sigma_u^u$ and  $\Sigma_d^d$ are the contributions coming from the one-loop effective potential defined in \cite{Baer:2012mv}) and $\tan\beta \equiv {\langle H_u \rangle}/{\langle H_d\rangle}$. 
All parameters  in Eq. (\ref{eq:mssmmu}) are defined at the $M_{EW}$.

\subsection{Electroweak Scale Fine-Tuning}
\label{esft}

We follow \cite{Baer:2012mv} in order to measure the EW scale fine-tuning condition, the following definitions are used:
\begin{equation}
 C_{H_d}\equiv |m_{H_d}^2/(\tan^2\beta -1)|,\,\, C_{H_u}\equiv
|-m_{H_u}^2\tan^2\beta /(\tan^2\beta -1)|, \, \, C_\mu\equiv |-\mu^2 |,
\label{cc1}
\end{equation}
 with
each $C_{\Sigma_{u,d}^{u,d} (k)}$  less than some characteristic value of order $M_Z^2$.
Here, $k$ labels the SM and SUSY particles that contribute to the one-loop Higgs potential.
For the fine-tuning condition we have
\begin{equation}
 \Delta_{\rm EW}\equiv {\rm max}(C_k )/(M_Z^2/2)~.
\label{eq:ewft}
\end{equation}
It is important to note that $\Delta_{EW}$ depends only on the weak scale parameters of the theory, therefore fixed
by the particle spectrum. Hence, it is independent of how SUSY particle masses arise. Lower values of $\Delta_{EW}$
correspond to less fine tuning, for example, $\Delta_{EW}=10$ implies $\Delta_{EW}^{-1}=10\%$ fine tuning.
Moreover, this condition of EW scale fine-tuning is different from the fine-tuning definition
in Refs.~\cite{Ellis:1986yg,Barbieri:1987fn} beyond the tree level 
(for more details see \cite{Baer:2012up}).

\subsection{High Scale Fine-Tuning}
\label{hsft}
From Eq.~(\ref{eq:mssmmu}) it is evident that $\Delta_{EW}$ does not give any informations about the possible high scale 
origin of the parameters in the equation. In order to address fully the fine-tuning condition we need to write down
weak-scale parameter $m_{H_{u,d}}^2$ in Eq.~(\ref{eq:mssmmu}) and with their explicit dependence on the (HS) as:
 \begin{equation}
 m_{H_{u,d}}^2=
m_{H_{u,d}}^2(M_{HS}) +\delta m_{H_{u,d}}^2, \ \,\,\,
\mu^2=\mu^2(M_{HS})+\delta\mu^2.
\end{equation}
Here
$m_{H_{u,d}}^2(M_{HS})$ and $\mu^2(M_{HS})$ are the corresponding
parameters renormalized at the high scale, and
$\delta m_{H_{u,d}}^2$, and $\delta\mu^2$ measure how the given parameter is changed due to 
Renormalization Group Equation (RGE) evolution.
 Eq.~(\ref{eq:mssmmu}) can be re-expressed in the form
\begin{eqnarray}
\frac{m_Z^2}{2} &=& \frac{(m_{H_d}^2(M_{HS})+ \delta m_{H_d}^2 +
\Sigma_d^d)-
(m_{H_u}^2(M_{HS})+\delta m_{H_u}^2+\Sigma_u^u)\tan^2\beta}{\tan^2\beta -1}
\nonumber \\
&-& (\mu^2(M_{HS})+\delta\mu^2)\;.
\label{eq:FT}
\end{eqnarray}
As we did before, we follow Ref.~\cite{Baer:2012mv} and introduce the following parameters
\begin{eqnarray}
&B_{H_d}\equiv|m_{H_d}^2(M_{HS})/(\tan^2\beta -1)|,
B_{\delta H_d}\equiv |\delta m_{H_d}^2/(\tan^2\beta -1)|, \nonumber \\
&B_{H_u}\equiv|-m_{H_u}^2(M_{HS})\tan^2\beta /(\tan^2\beta -1)|, B_{\mu}\equiv|\mu^2(M_{HS})|, \nonumber \\
&B_{\delta H_u}\equiv|-\delta m_{H_u}^2\tan^2\beta /(\tan^2\beta -1)|,
  B_{\delta \mu}\equiv |\delta \mu^2|,
  \label{bb1}
\end{eqnarray}
and  the
high scale fine-tuning measure $\Delta_{\rm HS}$ is defined to be
\begin{equation}
\Delta_{\rm HS}\equiv {\rm max}(B_i )/(M_Z^2/2).
\label{eq:hsft}
\end{equation}

In short, $\Delta_{EW}$ includes information about the minimal amount of fine-tuning present in the 
low scale model for a given 
SUSY spectrum, while $\Delta_{HS}$ better represents the fine-tuning that is present in high scale model.

\section{Numerical Results}
\label{results}
\begin{figure}[htp!]
\centering
\subfiguretopcaptrue

\subfigure{
\includegraphics[totalheight=5.5cm,width=7.cm]{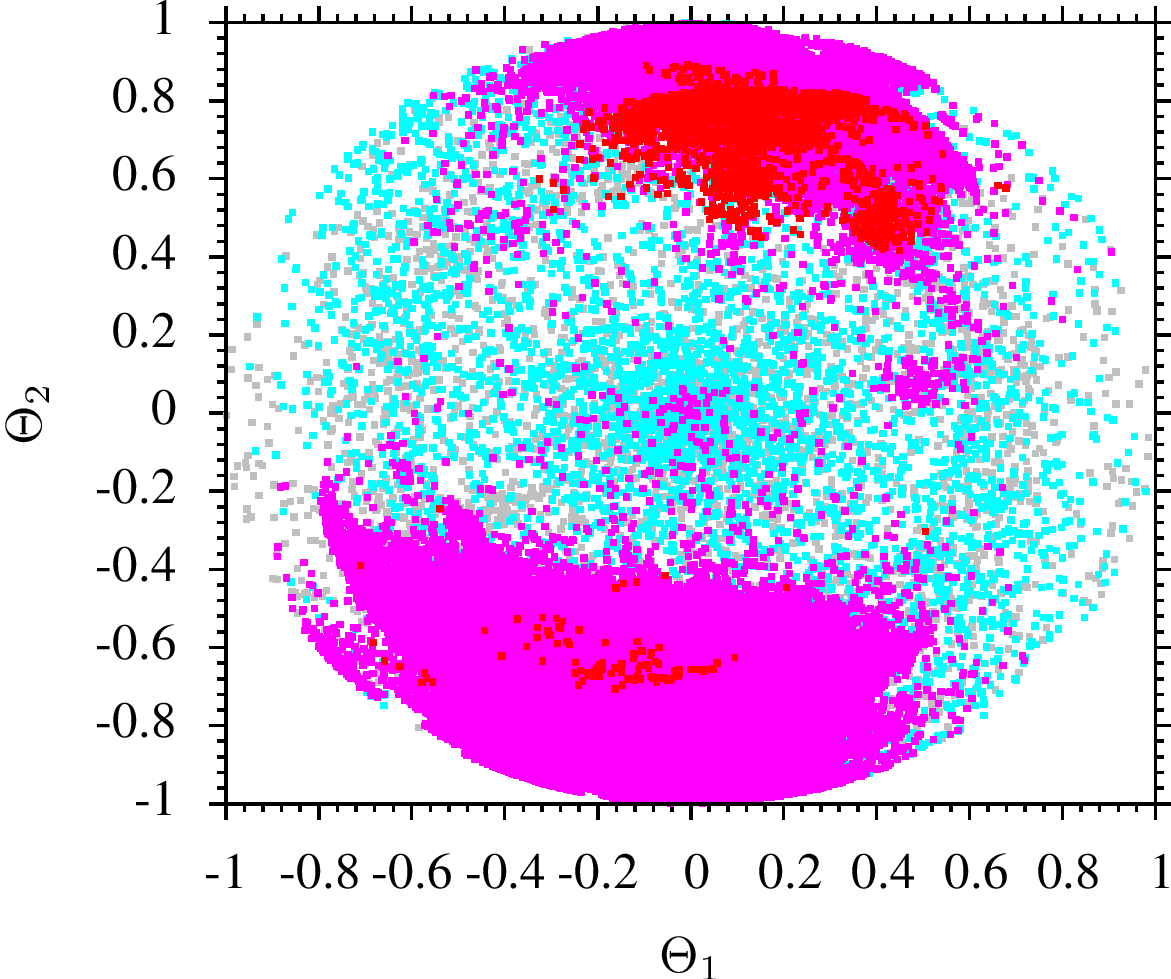}
}
\subfigure{
\includegraphics[totalheight=5.5cm,width=7.cm]{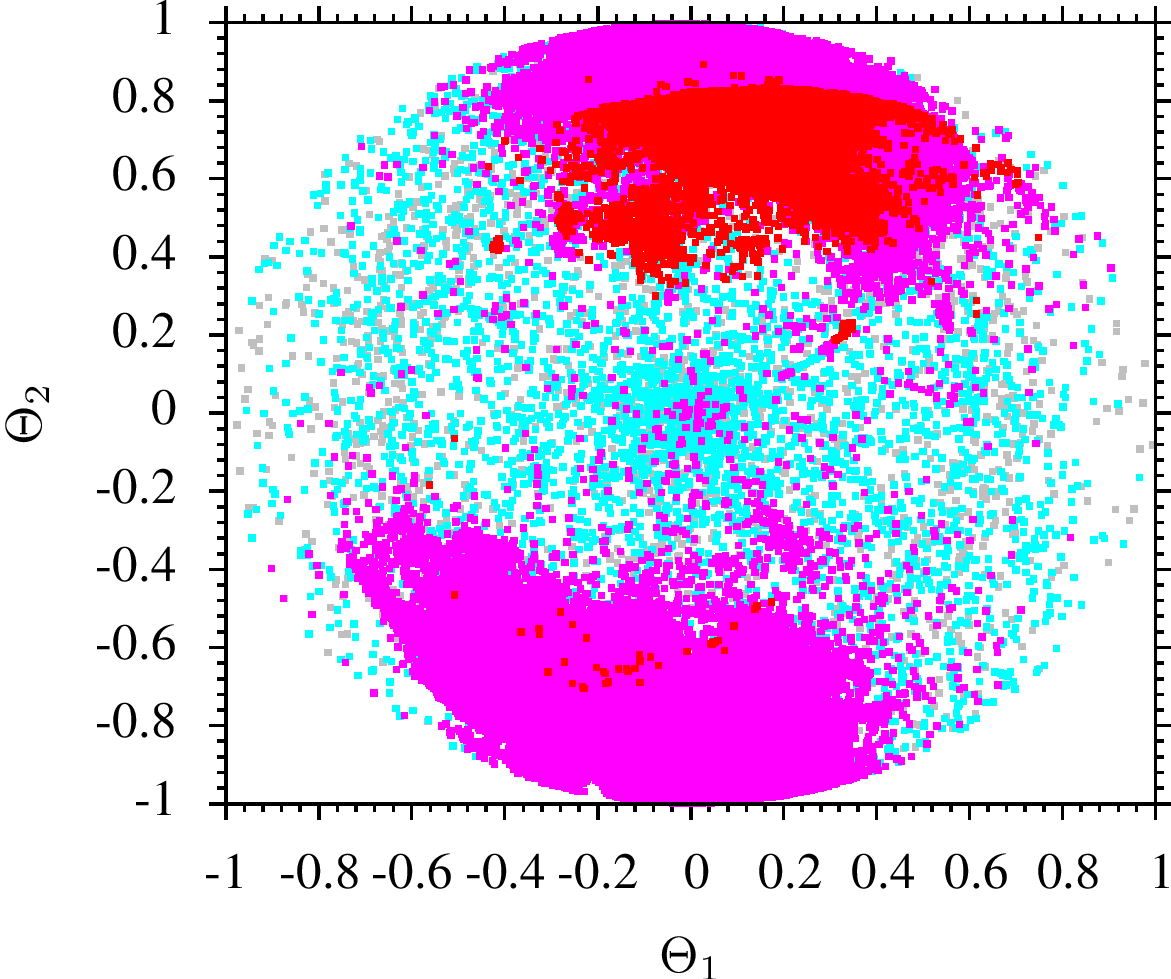}
}
\subfigure{
\includegraphics[totalheight=5.5cm,width=7.cm]{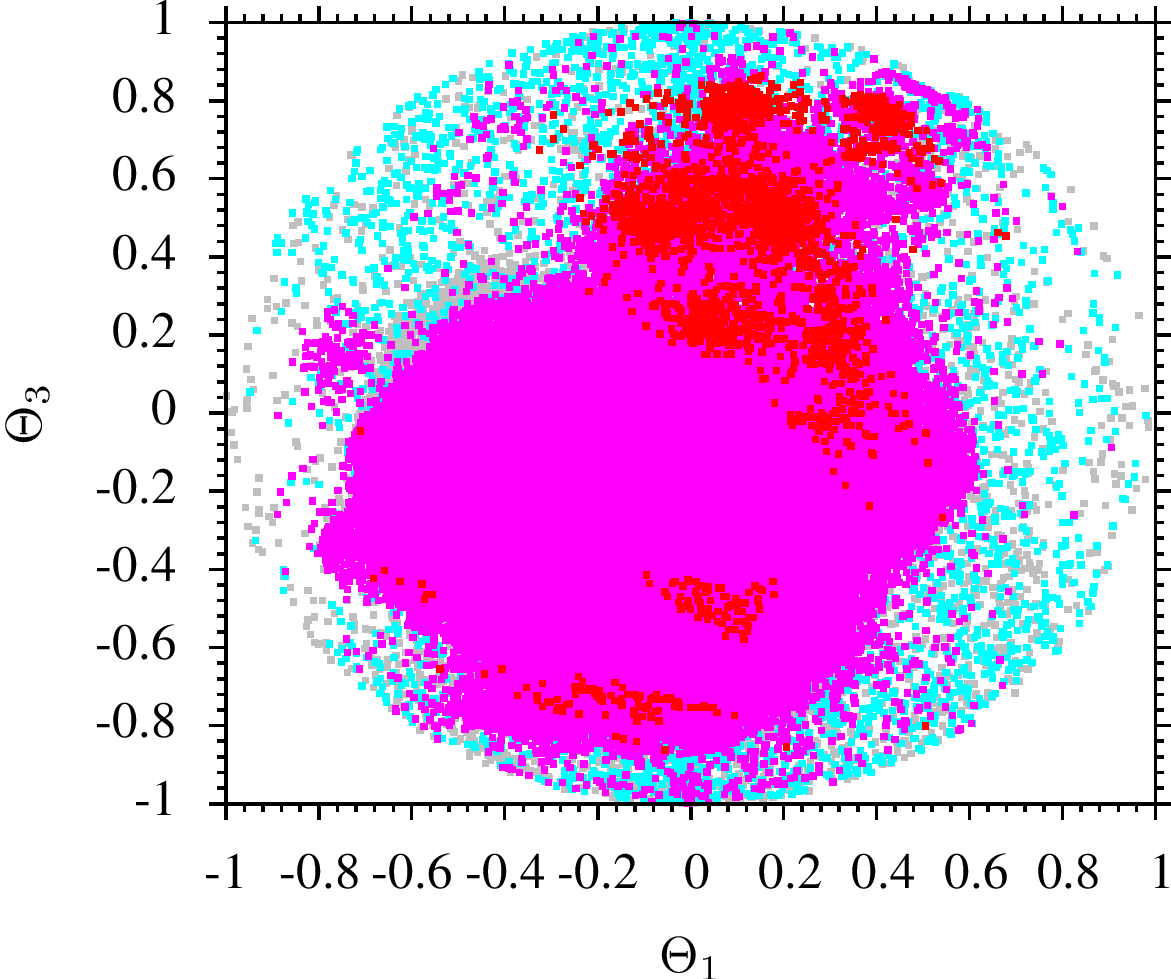}
}
\subfigure{
\includegraphics[totalheight=5.5cm,width=7.cm]{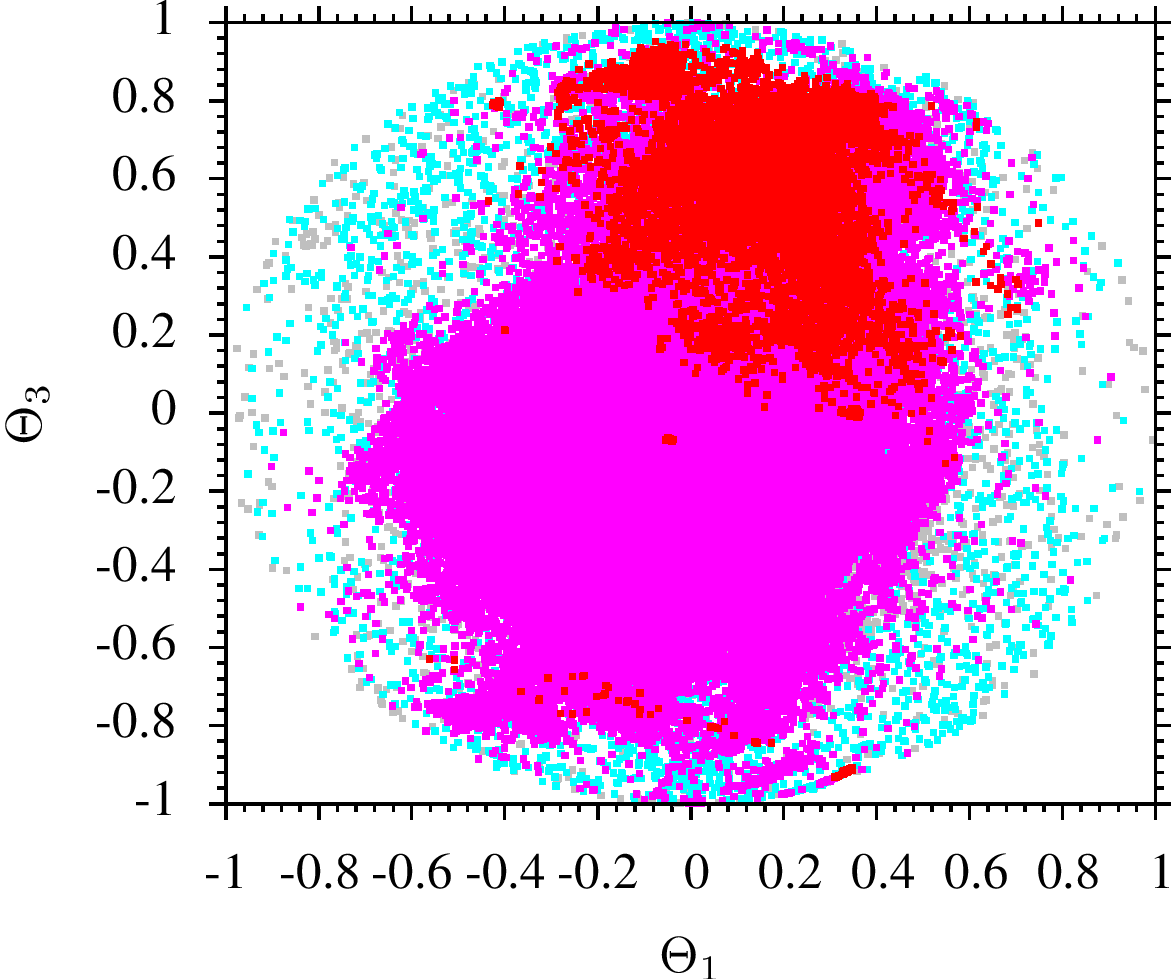}
}
\subfigure{
\includegraphics[totalheight=5.5cm,width=7.cm]{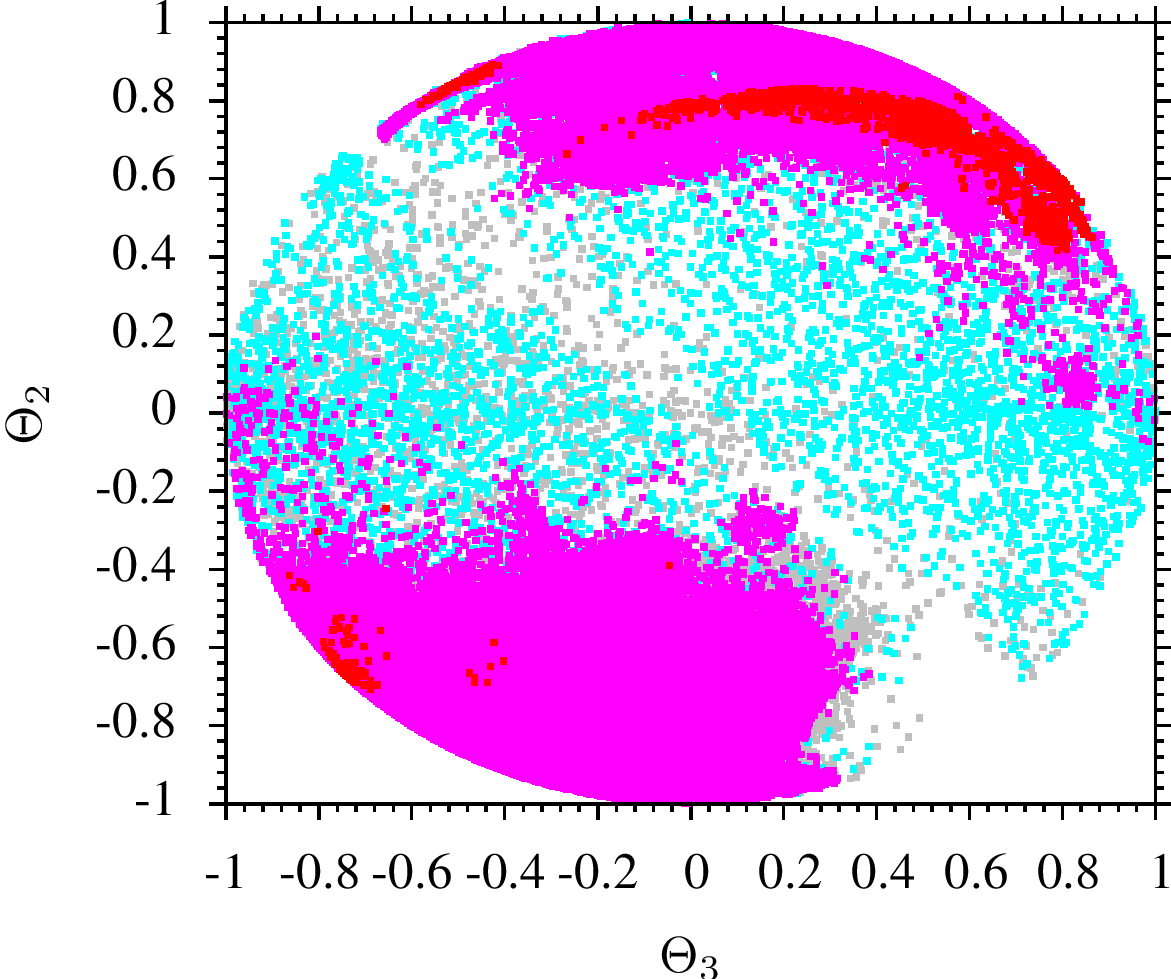}
}
\subfigure{
\includegraphics[totalheight=5.5cm,width=7.cm]{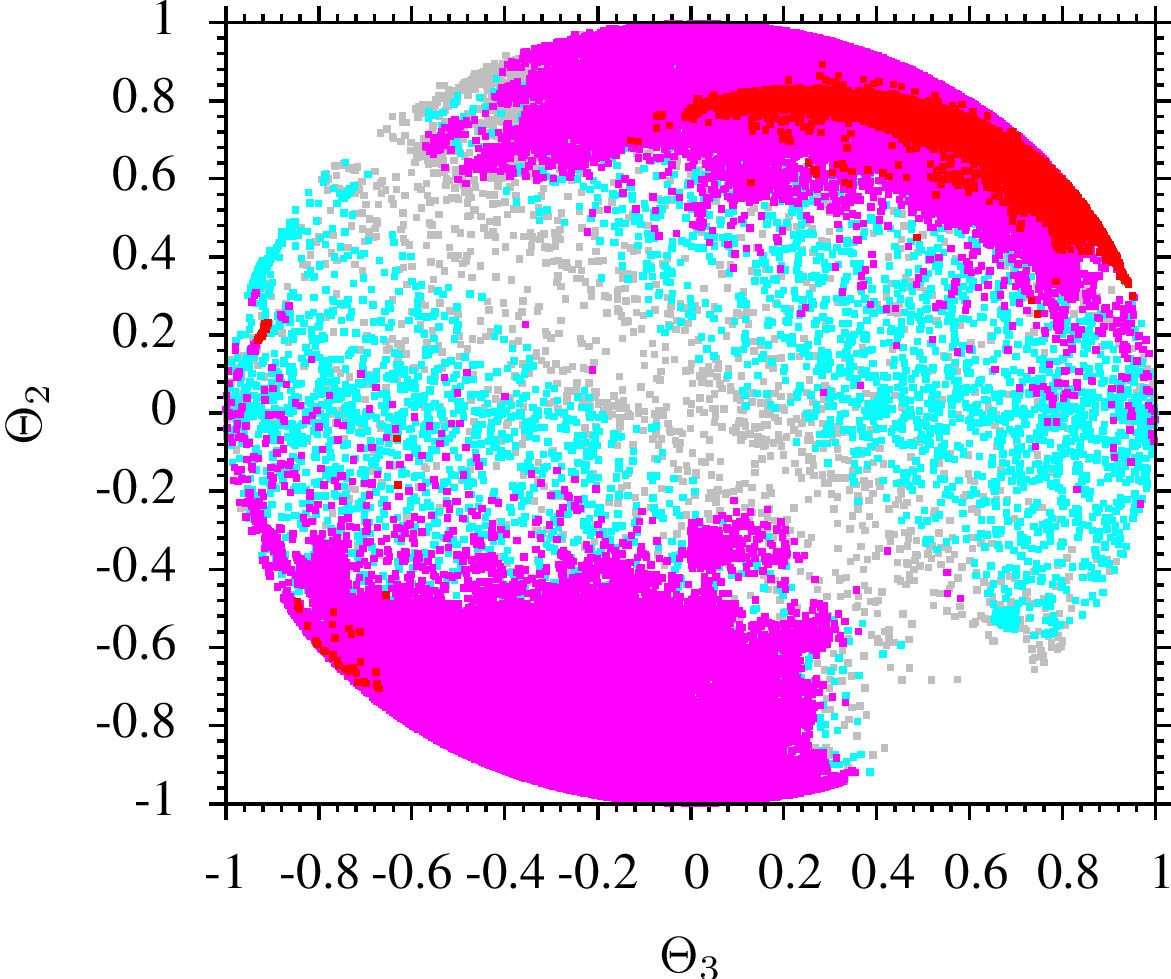}
}

\caption{
Plots in $\Theta_1-\Theta_2$, $\Theta_1-\Theta_3$ and $\Theta_3-\Theta_2$ 
planes for $\mu <0$ (left panels) and $\mu > 0$ (right panels). Grey points satisfy the 
REWSB and yield LSP neutralino. Aqua points satisfy all the 
mass bounds and B-physics bounds. Magenta points are subset of aqua points and also represent 
$123~{\rm GeV}\leqslant m_h\leqslant 127~{\rm GeV}$. 
Red points are subset of magenta points and also satisfy WMAP9 5$\sigma$ bounds.
}
\label{input_params1}
\end{figure}

\begin{figure}[htp!]
\centering
\subfiguretopcaptrue

\subfigure{
\includegraphics[totalheight=5.5cm,width=7.cm]{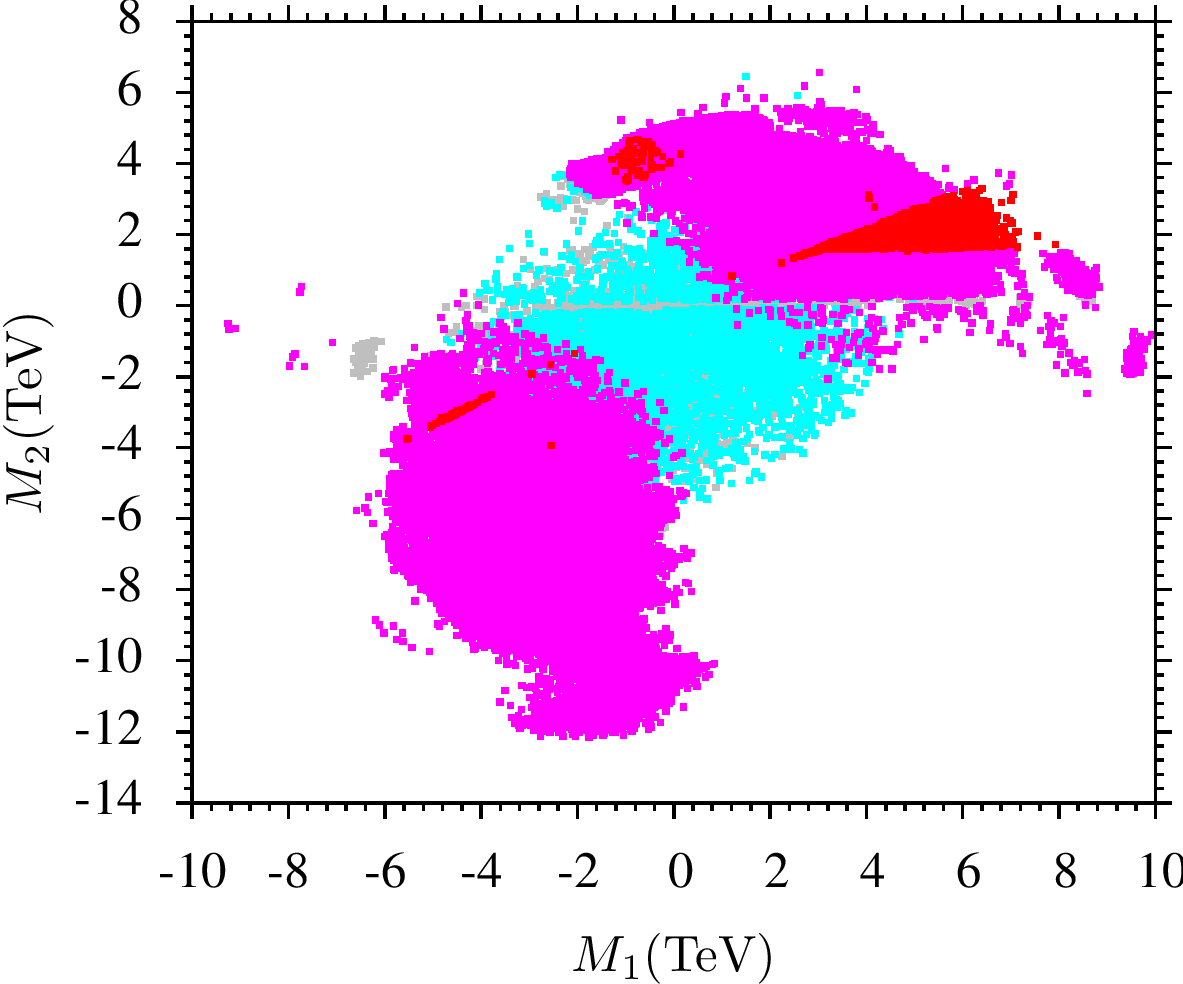}
}
\subfigure{
\includegraphics[totalheight=5.5cm,width=7.cm]{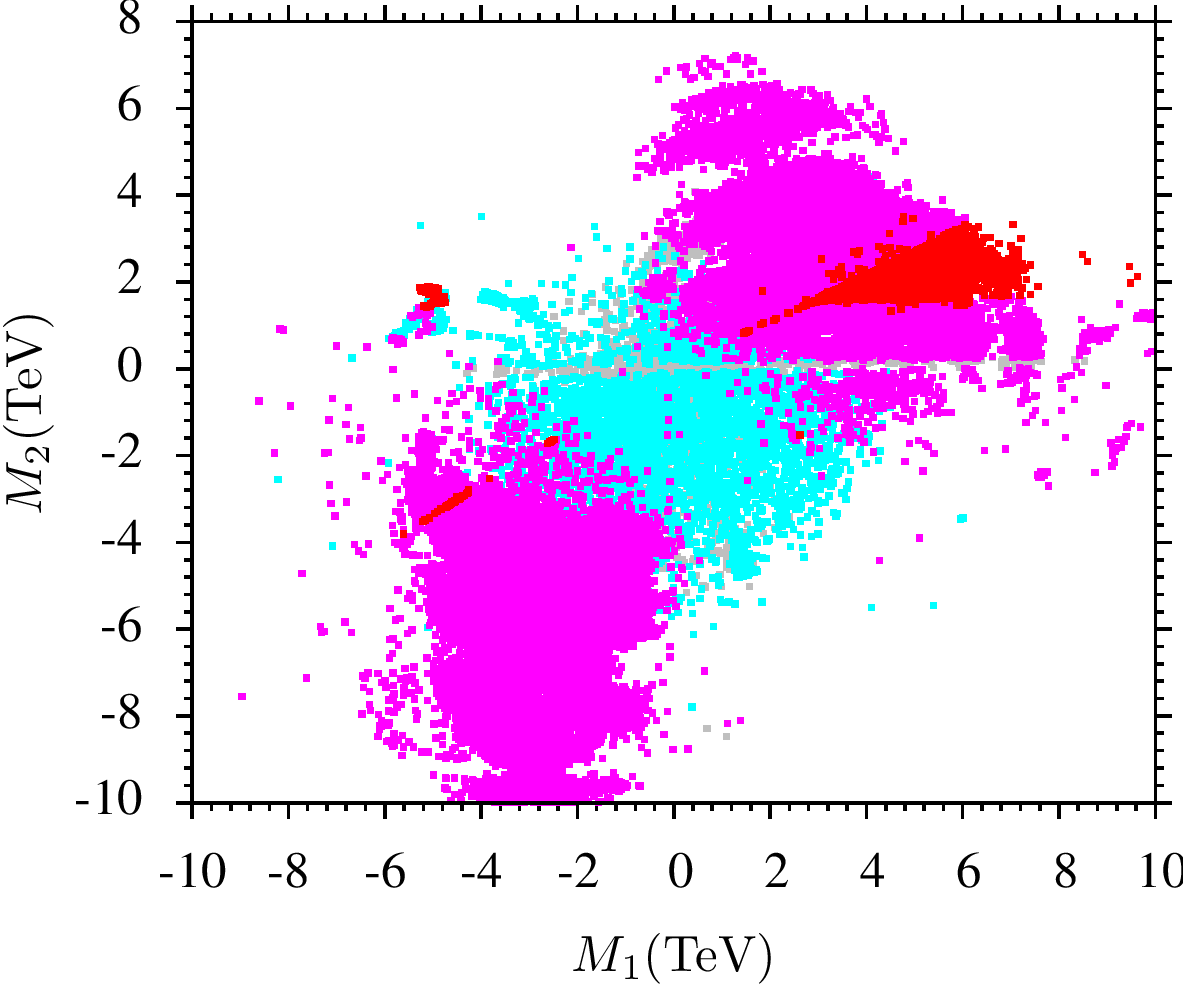}
}
\subfigure{
\includegraphics[totalheight=5.5cm,width=7.cm]{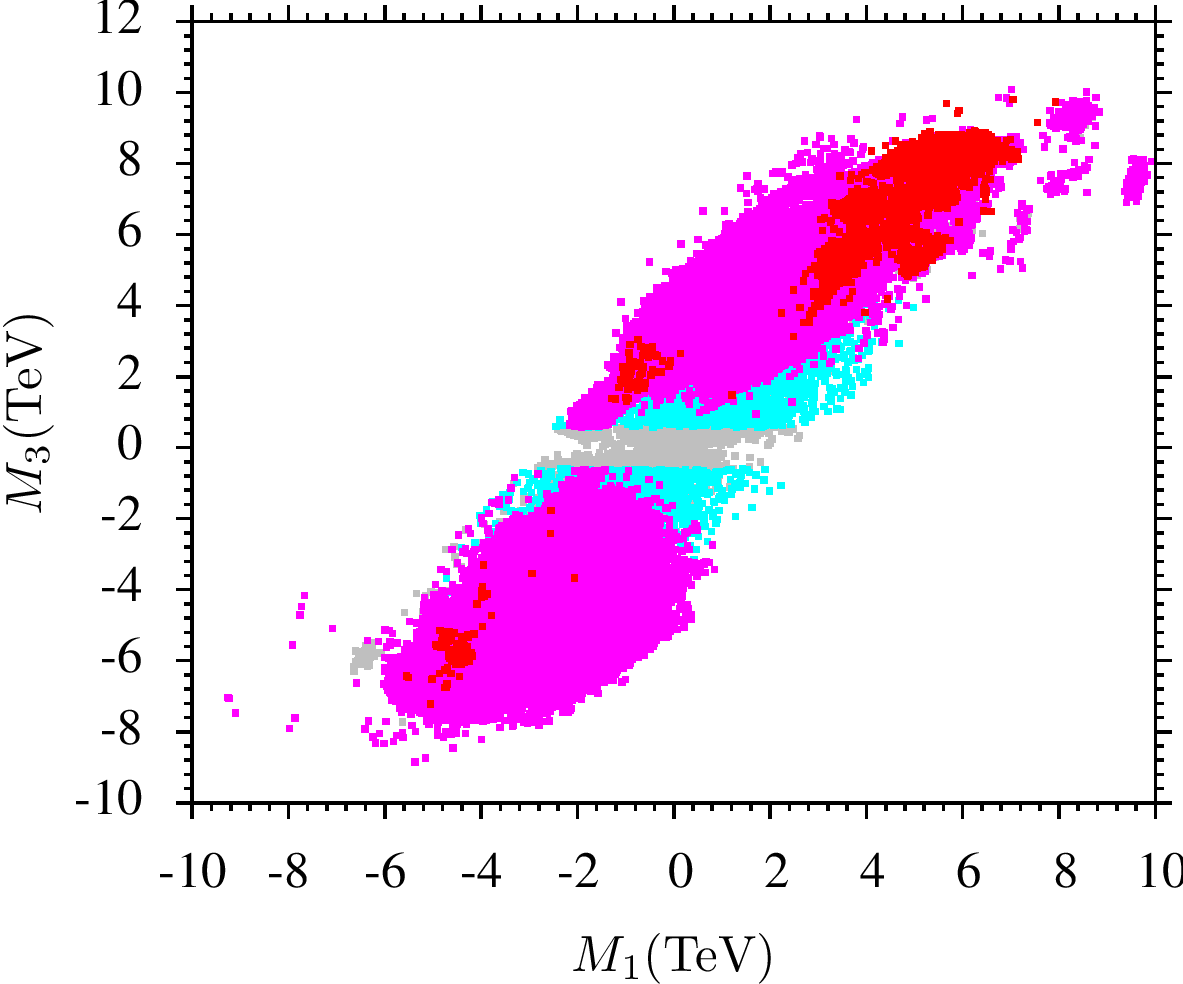}
}
\subfigure{
\includegraphics[totalheight=5.5cm,width=7.cm]{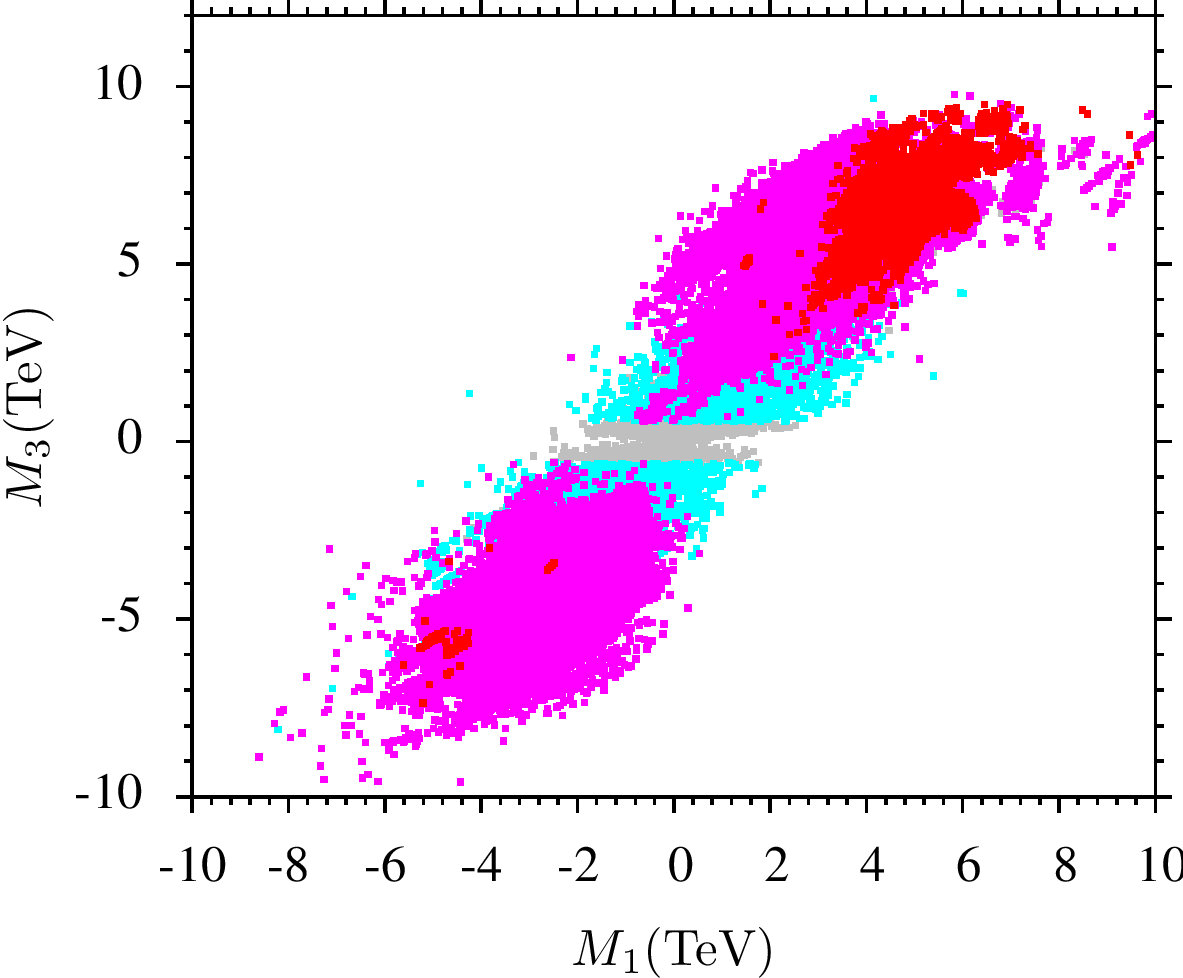}
}
\subfigure{
\includegraphics[totalheight=5.5cm,width=7.cm]{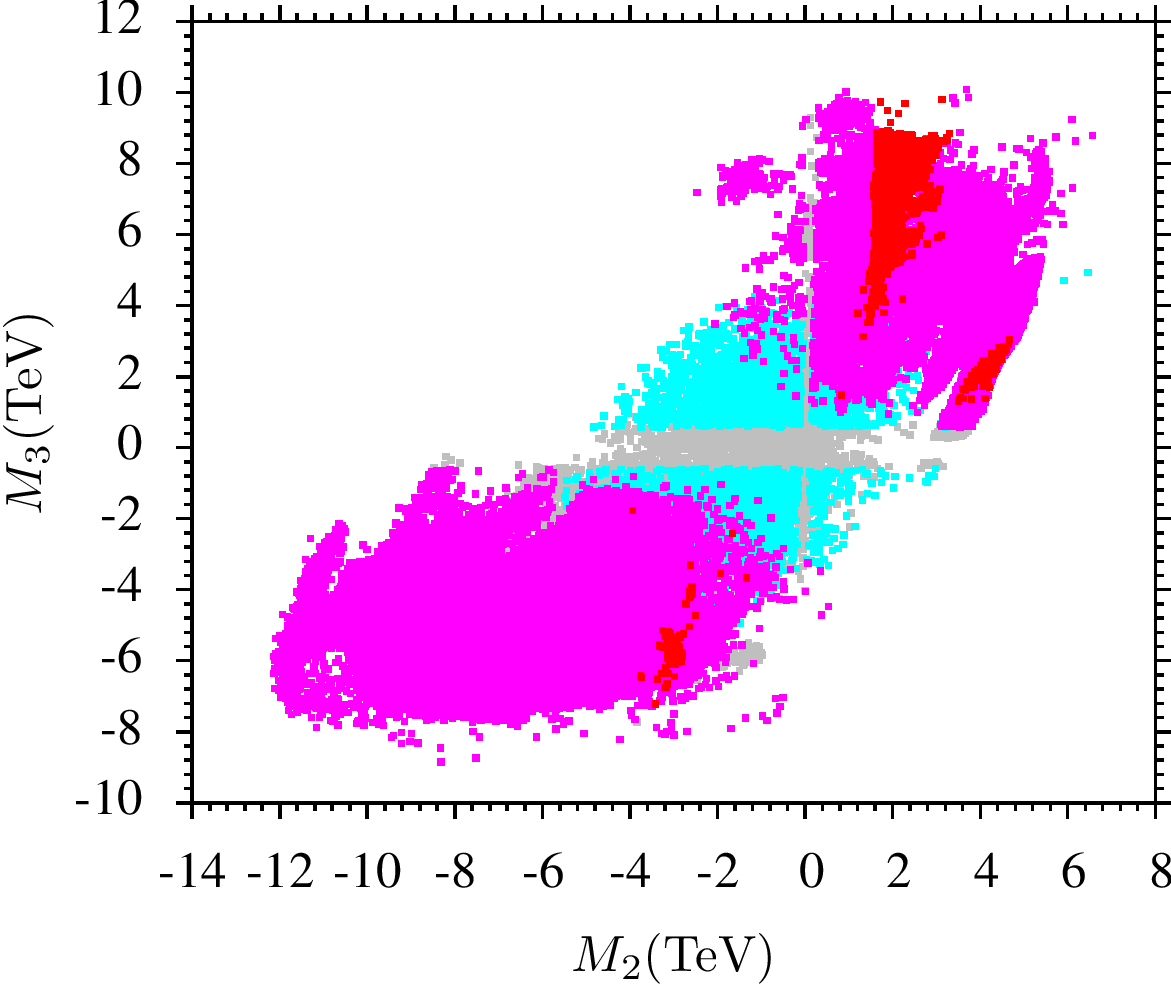}
}
\subfigure{
\includegraphics[totalheight=5.5cm,width=7.cm]{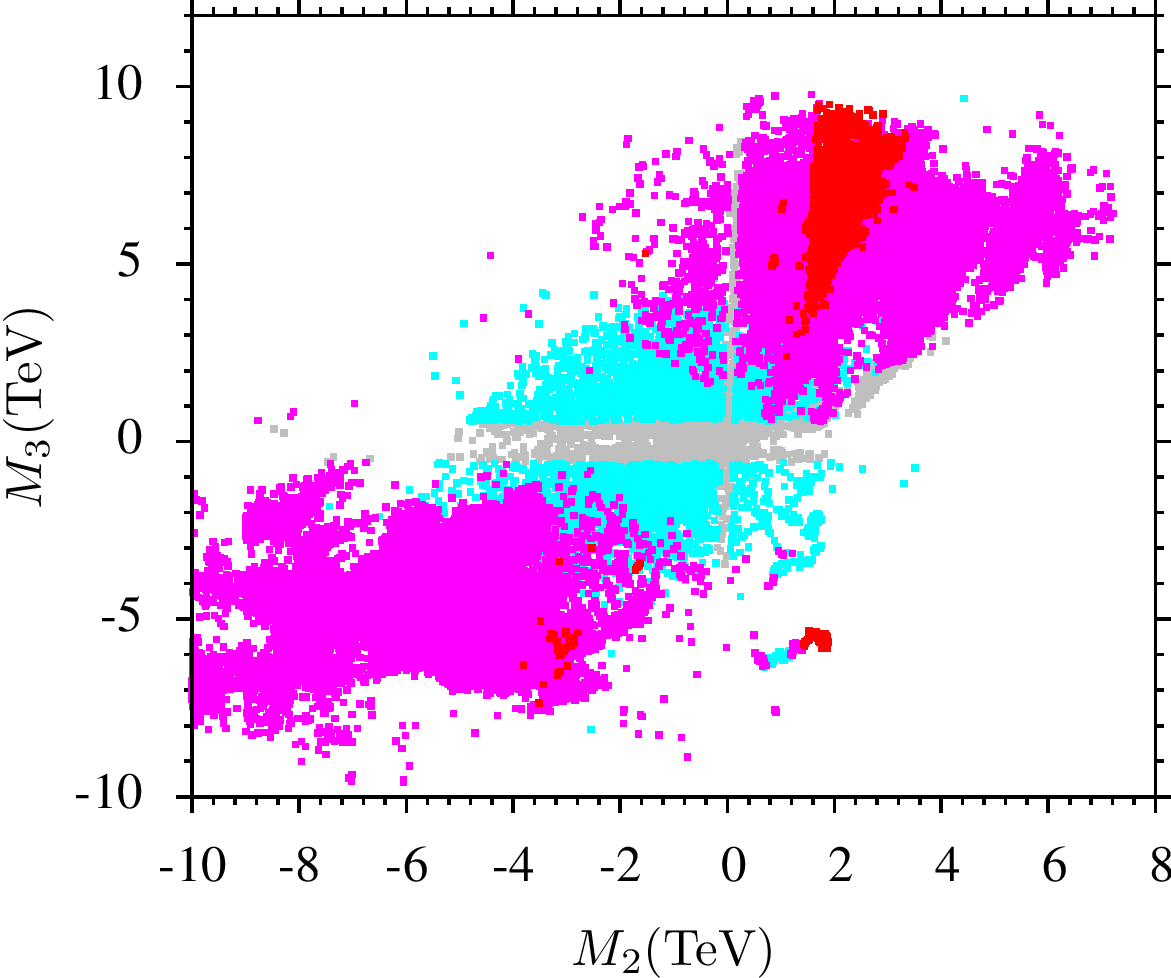}
}

\caption{Plots in $M_1-M_2$, $M_1-M_3$ and $M_3-M_2$ planes. 
Color coding and panel description are same as in Fig.~\ref{input_params1}.
}
\label{funda_params2}
\end{figure}
\begin{figure}[htp!]
\centering
\subfiguretopcaptrue

\subfigure{
\includegraphics[totalheight=5.5cm,width=7.cm]{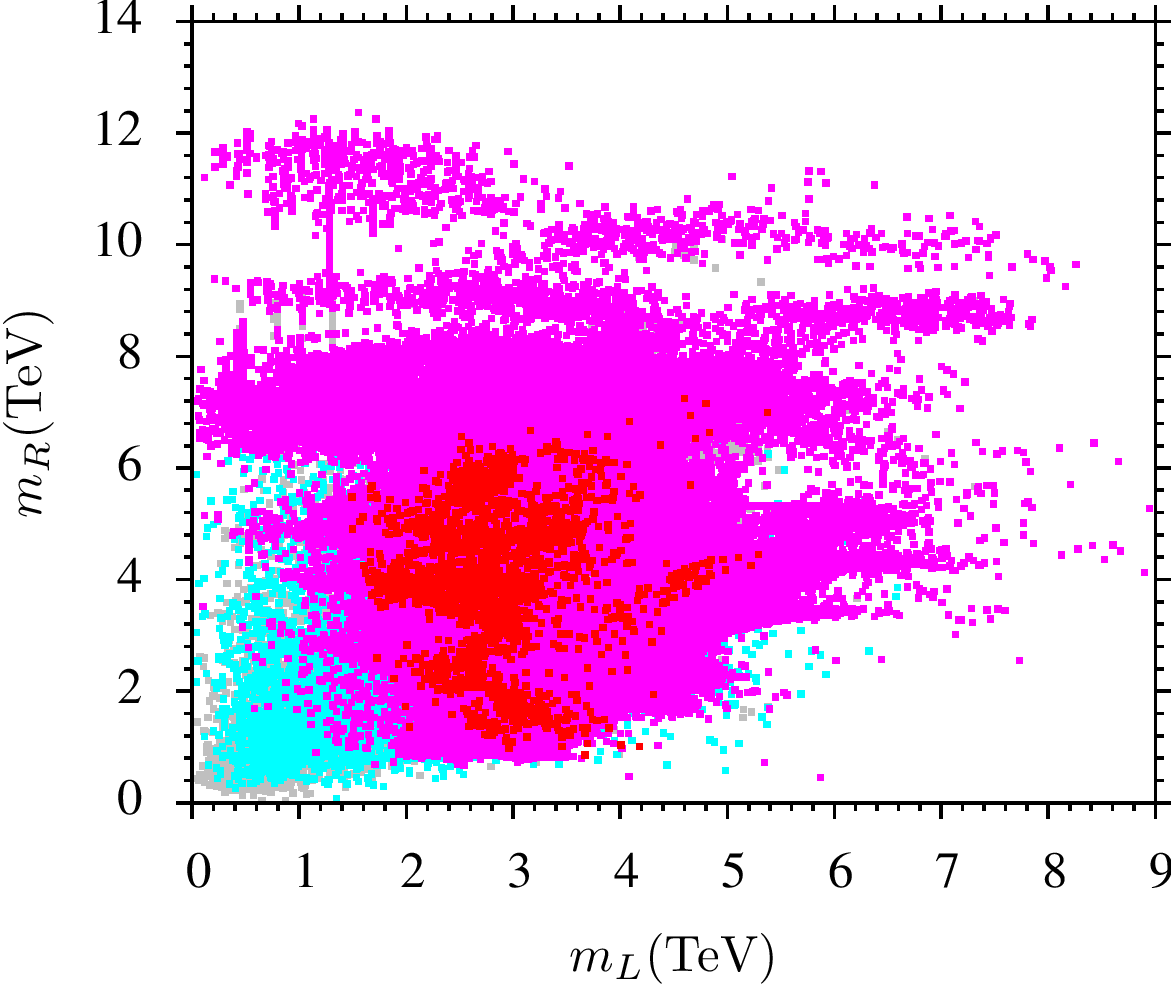}
}
\subfigure{
\includegraphics[totalheight=5.5cm,width=7.cm]{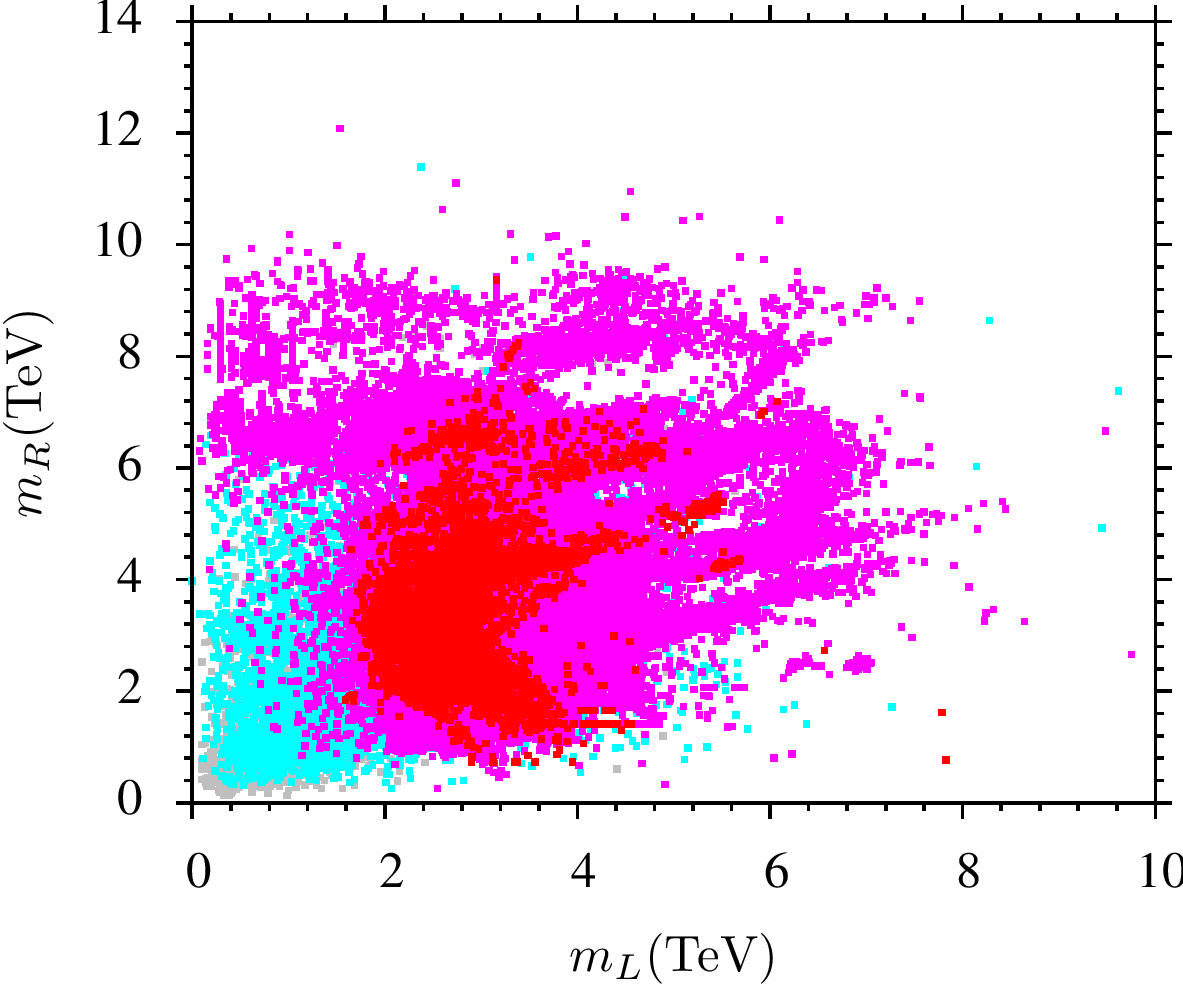}
}
\subfigure{
\includegraphics[totalheight=5.5cm,width=7.cm]{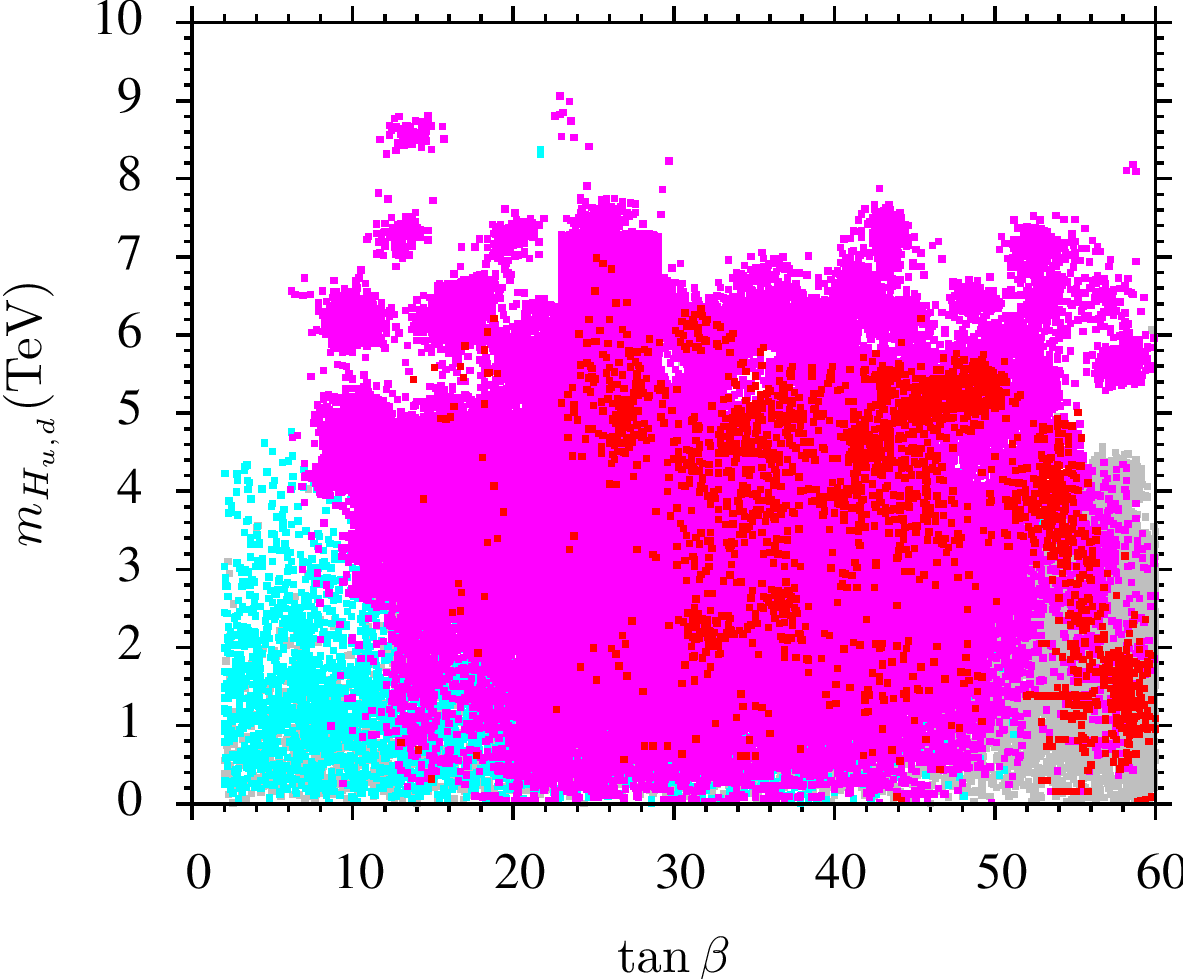}
}
\subfigure{
\includegraphics[totalheight=5.5cm,width=7.cm]{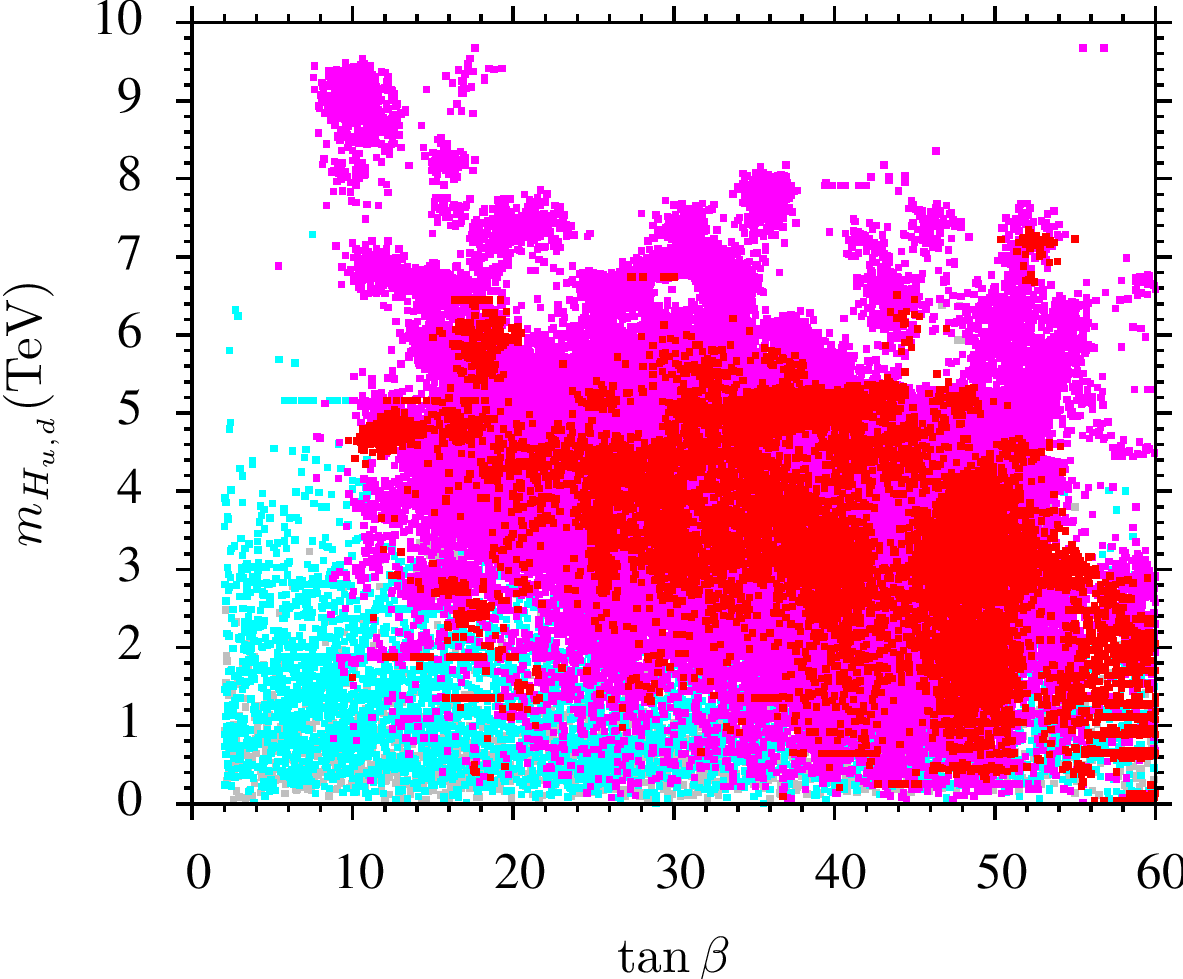}
}
\subfigure{
\includegraphics[totalheight=5.5cm,width=7.cm]{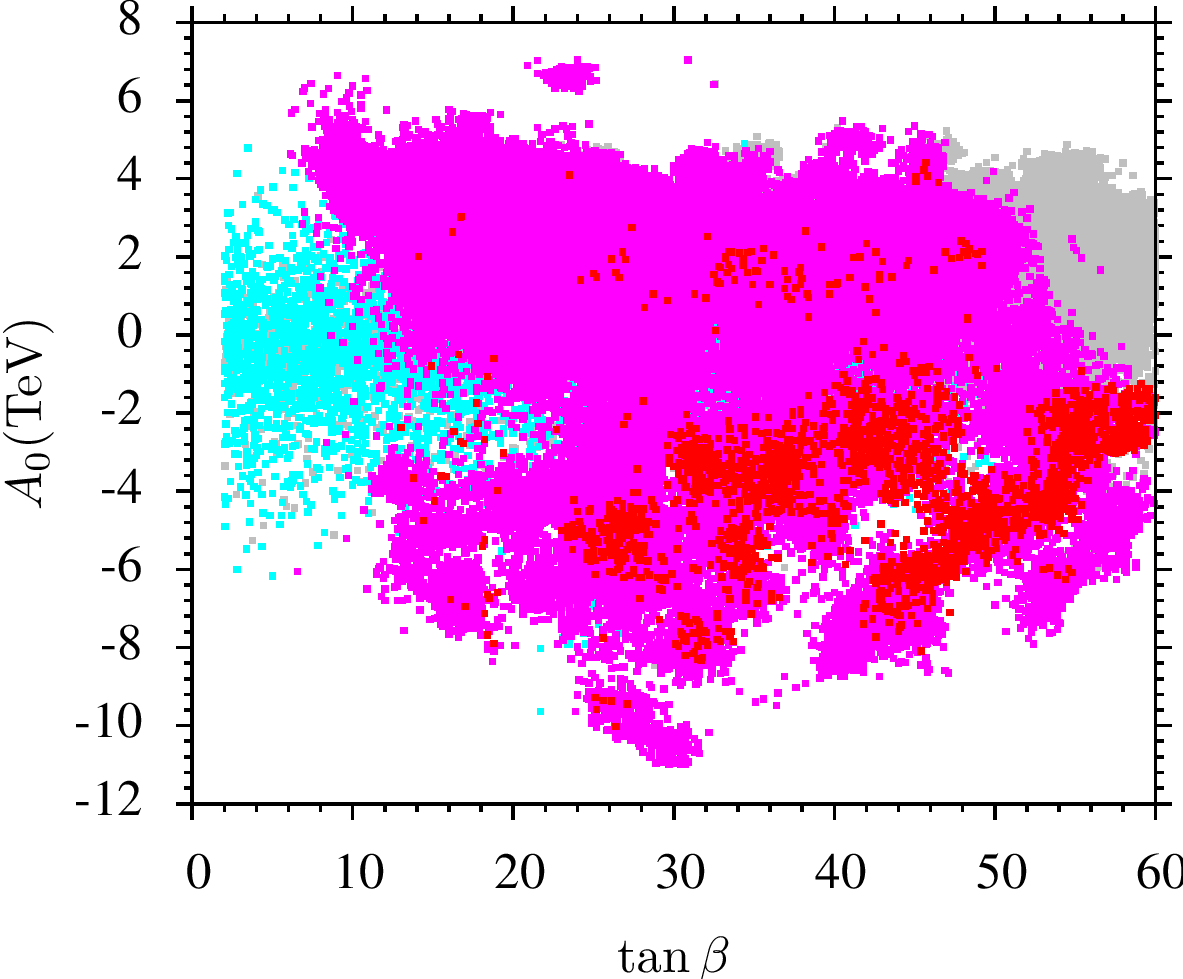}
}
\subfigure{
\includegraphics[totalheight=5.5cm,width=7.cm]{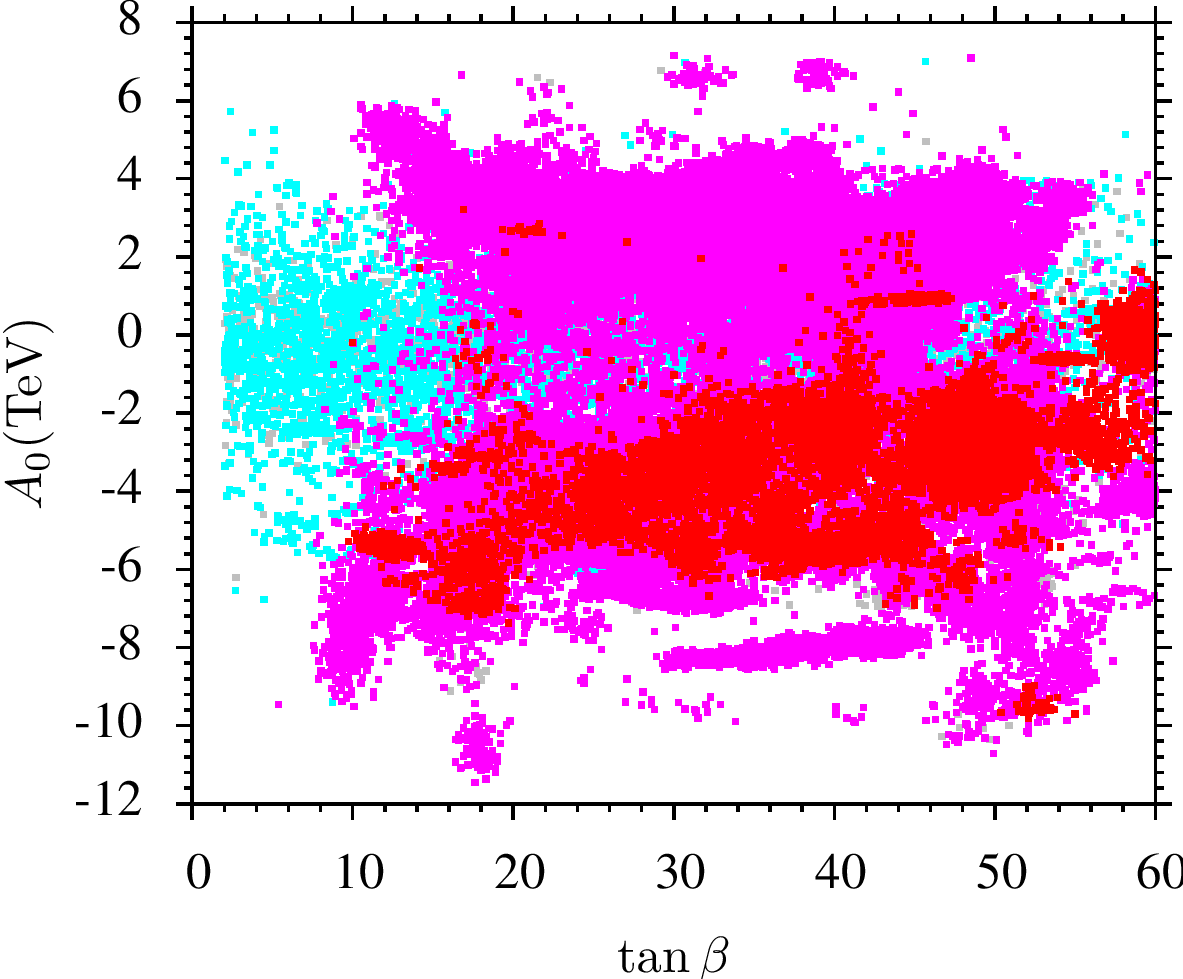}
}
\caption{Plots in $m_L-m_R$, $\tan\beta-m_{H_{u,d}}$ and $\tan\beta-A_0$ planes.
Color coding and panel description are same as in Fig.~\ref{input_params1}.
}
\label{funda_params3}
\end{figure}
\begin{figure}[htp!]
\centering
\subfiguretopcaptrue

\subfigure{
\includegraphics[totalheight=5.5cm,width=7.cm]{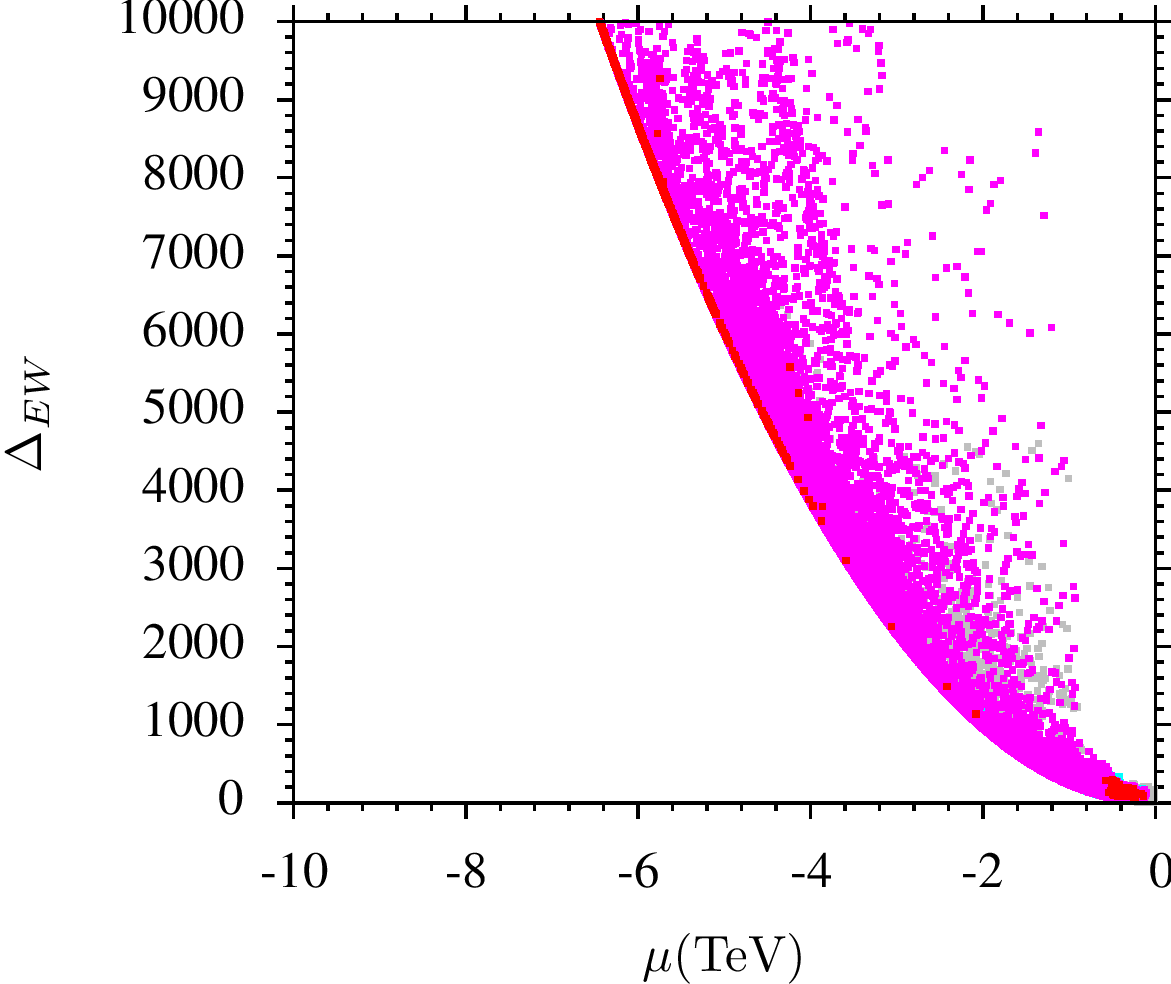}
}
\subfigure{
\includegraphics[totalheight=5.5cm,width=7.cm]{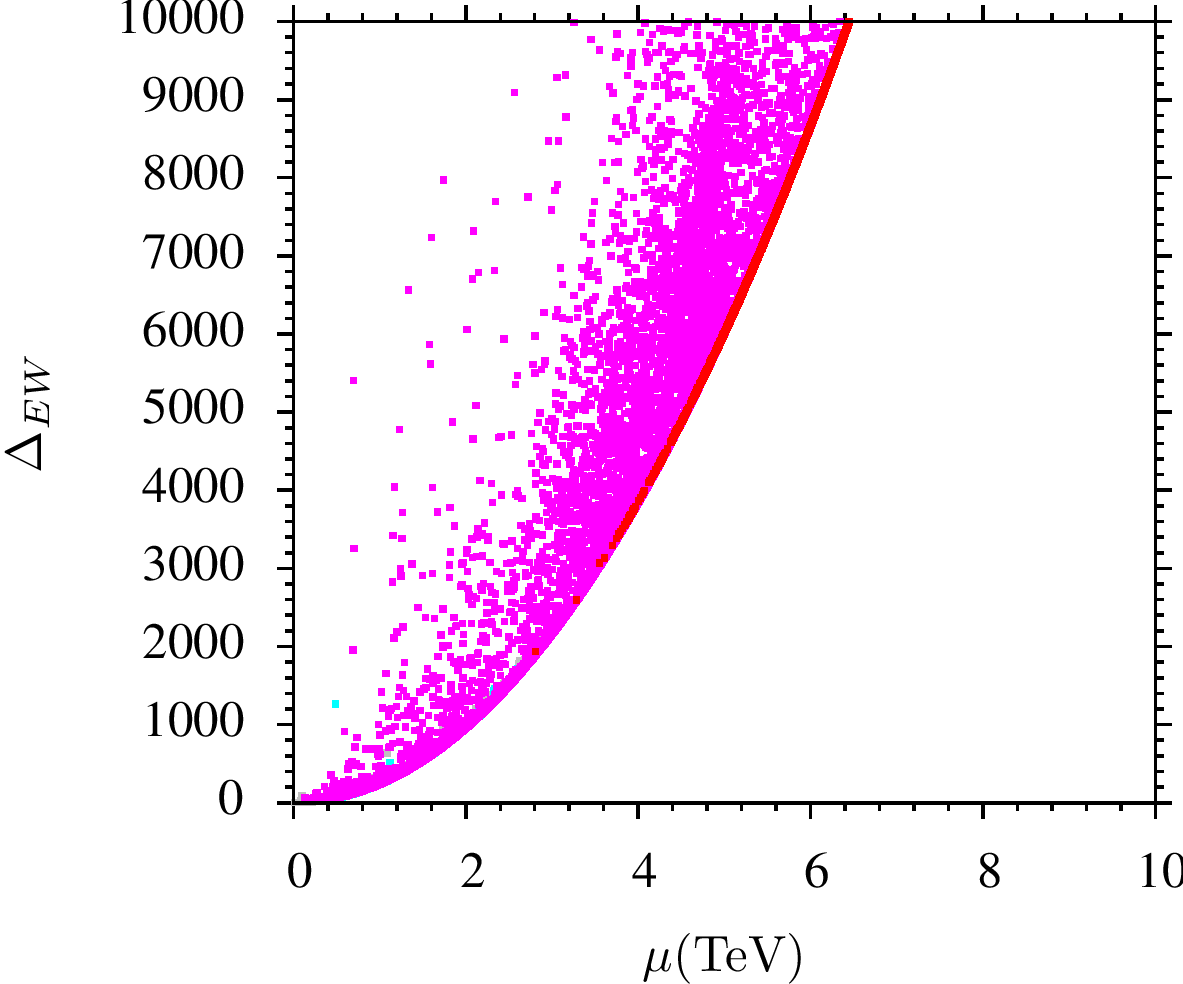}
}
\subfigure{
\includegraphics[totalheight=5.5cm,width=7.cm]{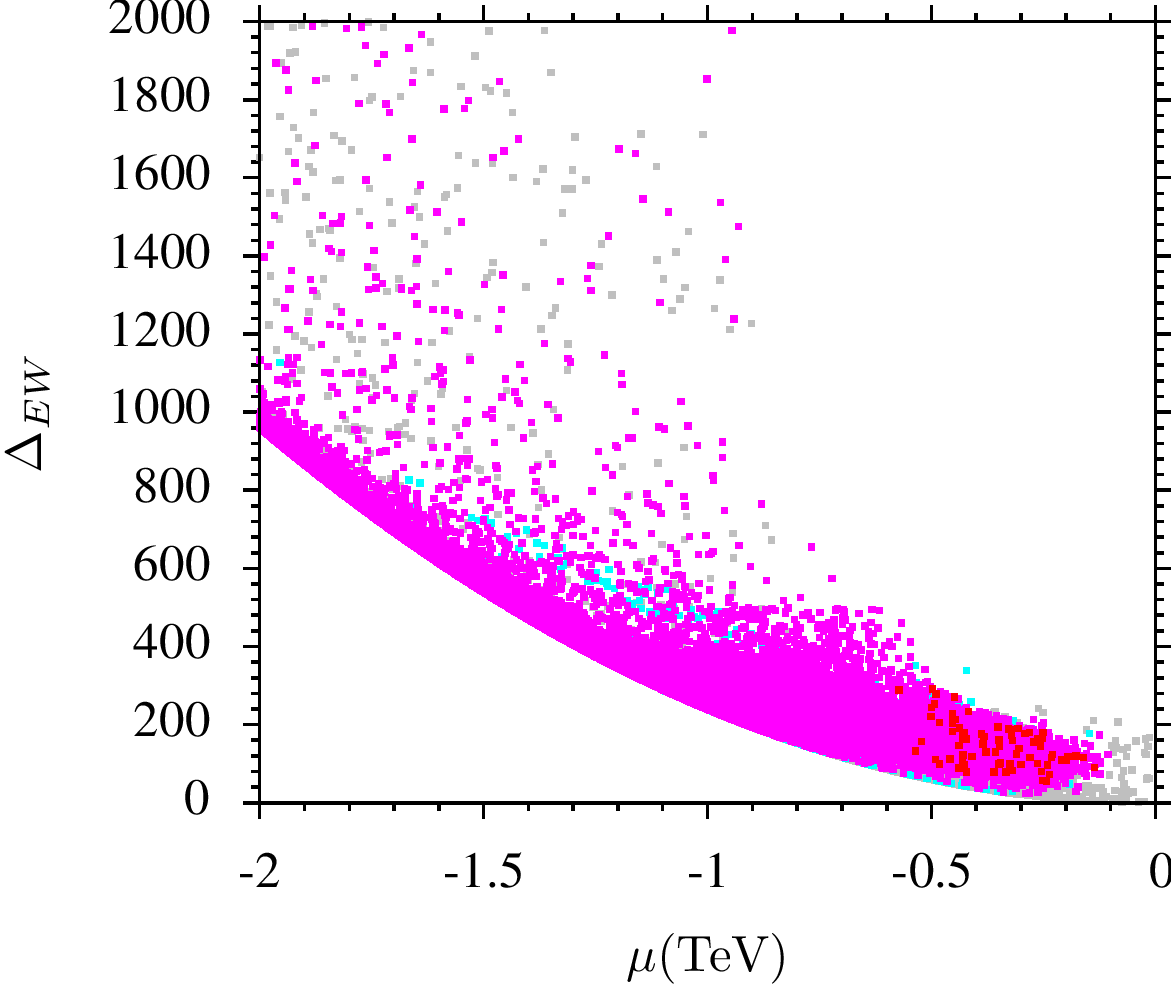}
}
\subfigure{
\includegraphics[totalheight=5.5cm,width=7.cm]{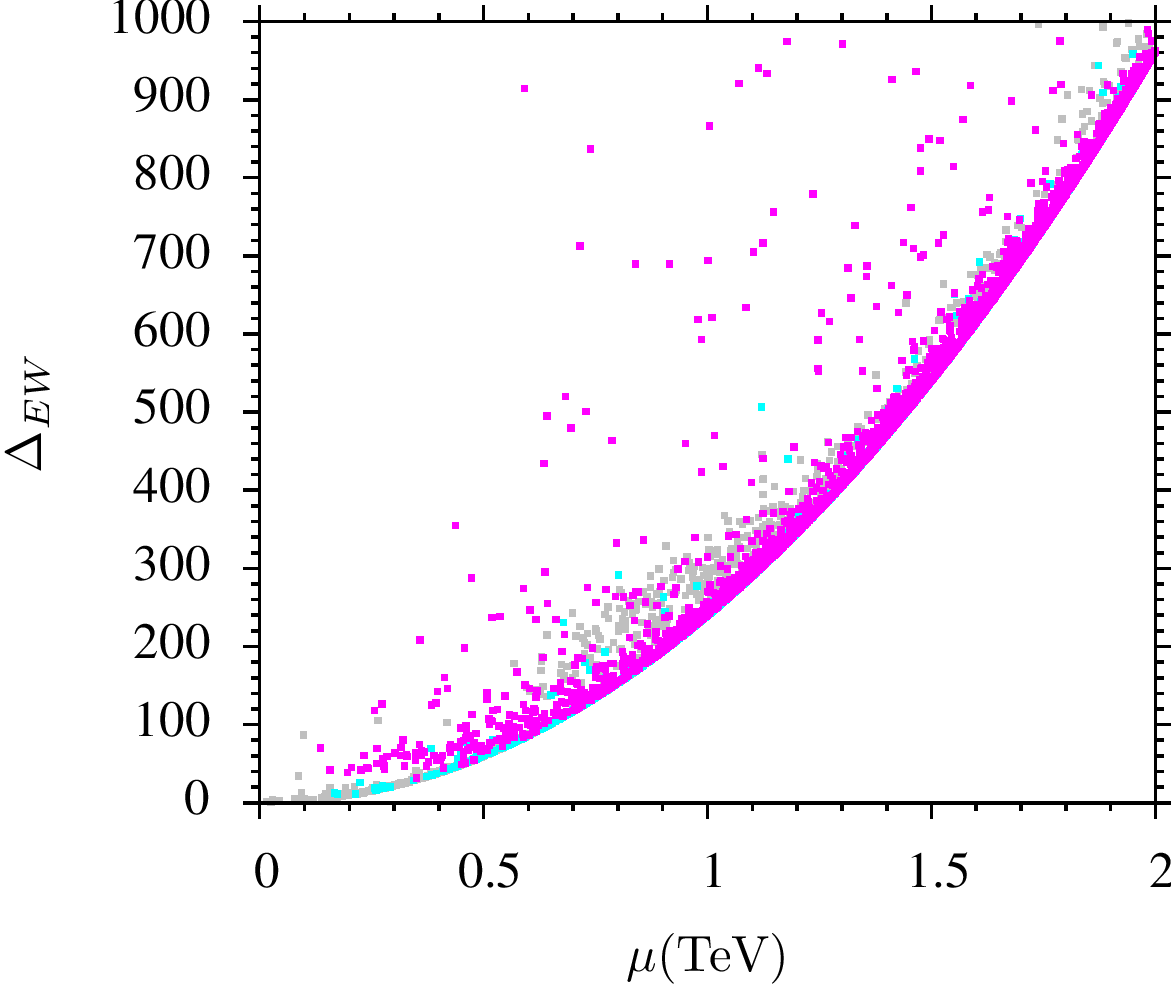}
}
\subfigure{
\includegraphics[totalheight=5.5cm,width=7.cm]{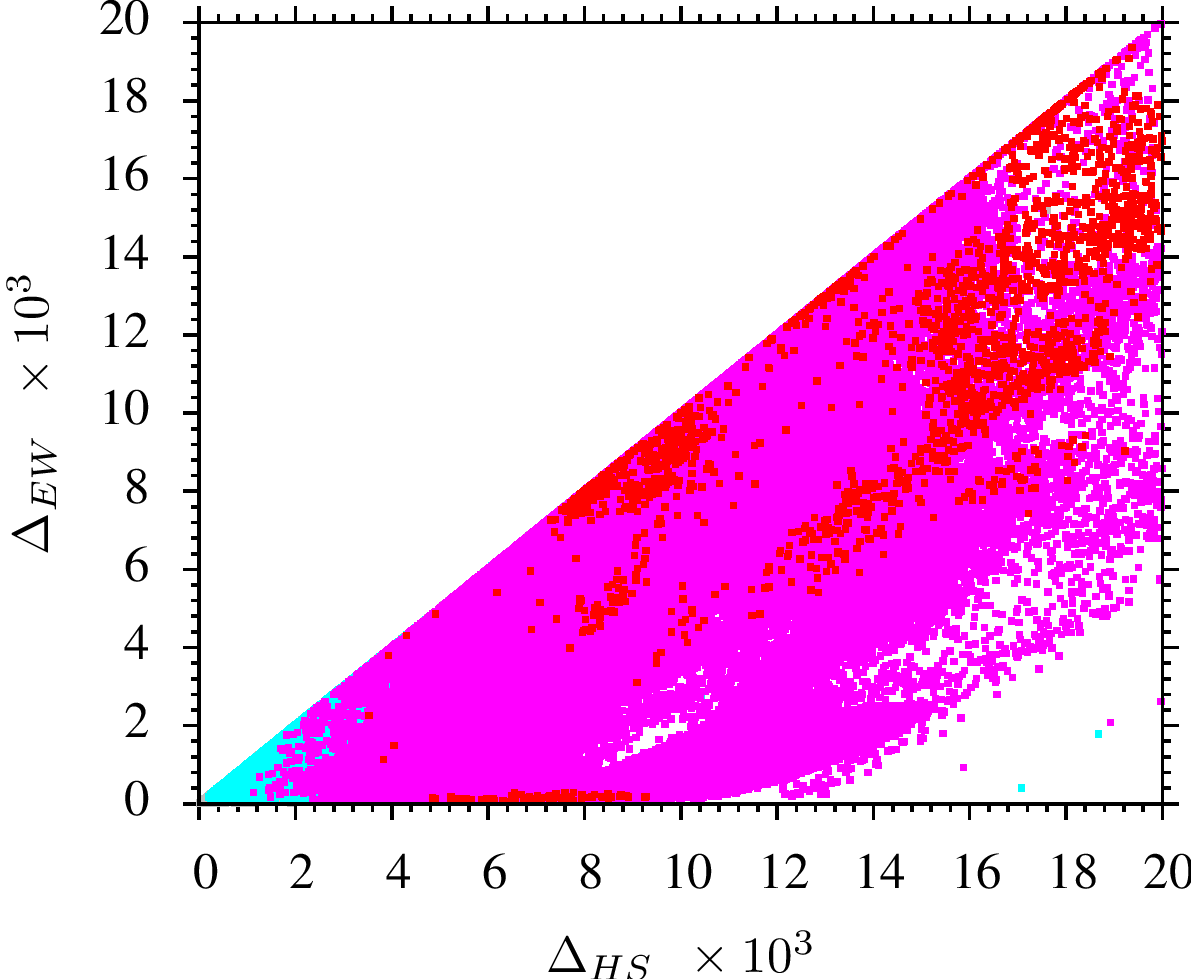}
}
\subfigure{
\includegraphics[totalheight=5.5cm,width=7.cm]{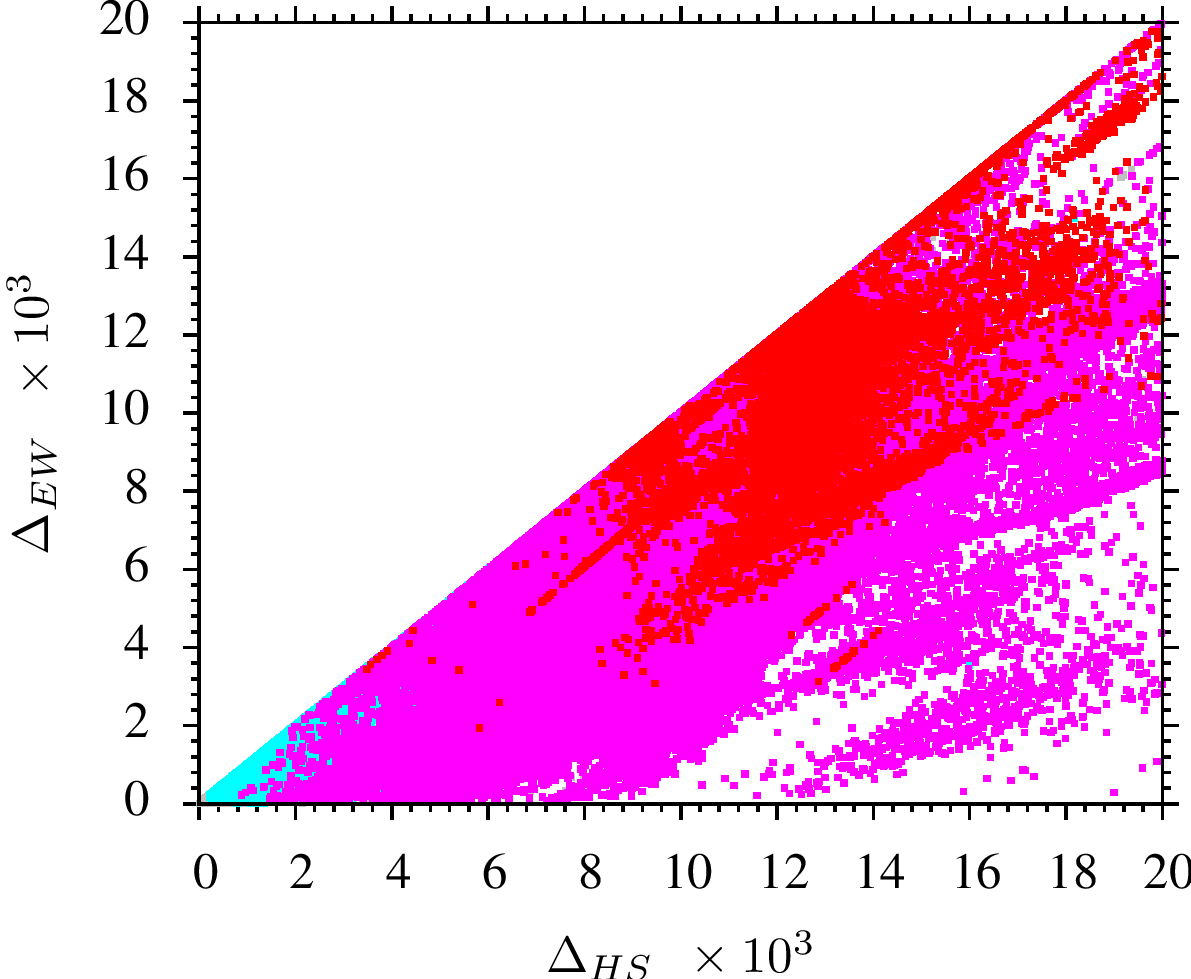}
}
\caption{Plots in $\mu-\Delta_{EW}$ and $\Delta_{HS}-\Delta_{EW}$ planes.
Color coding and panel description are same as in Fig.~\ref{input_params1}.
}
\label{delew}
\end{figure}

\begin{figure}[htp!]
\centering
\subfiguretopcaptrue
\subfigure{
\includegraphics[totalheight=5.5cm,width=7.cm]{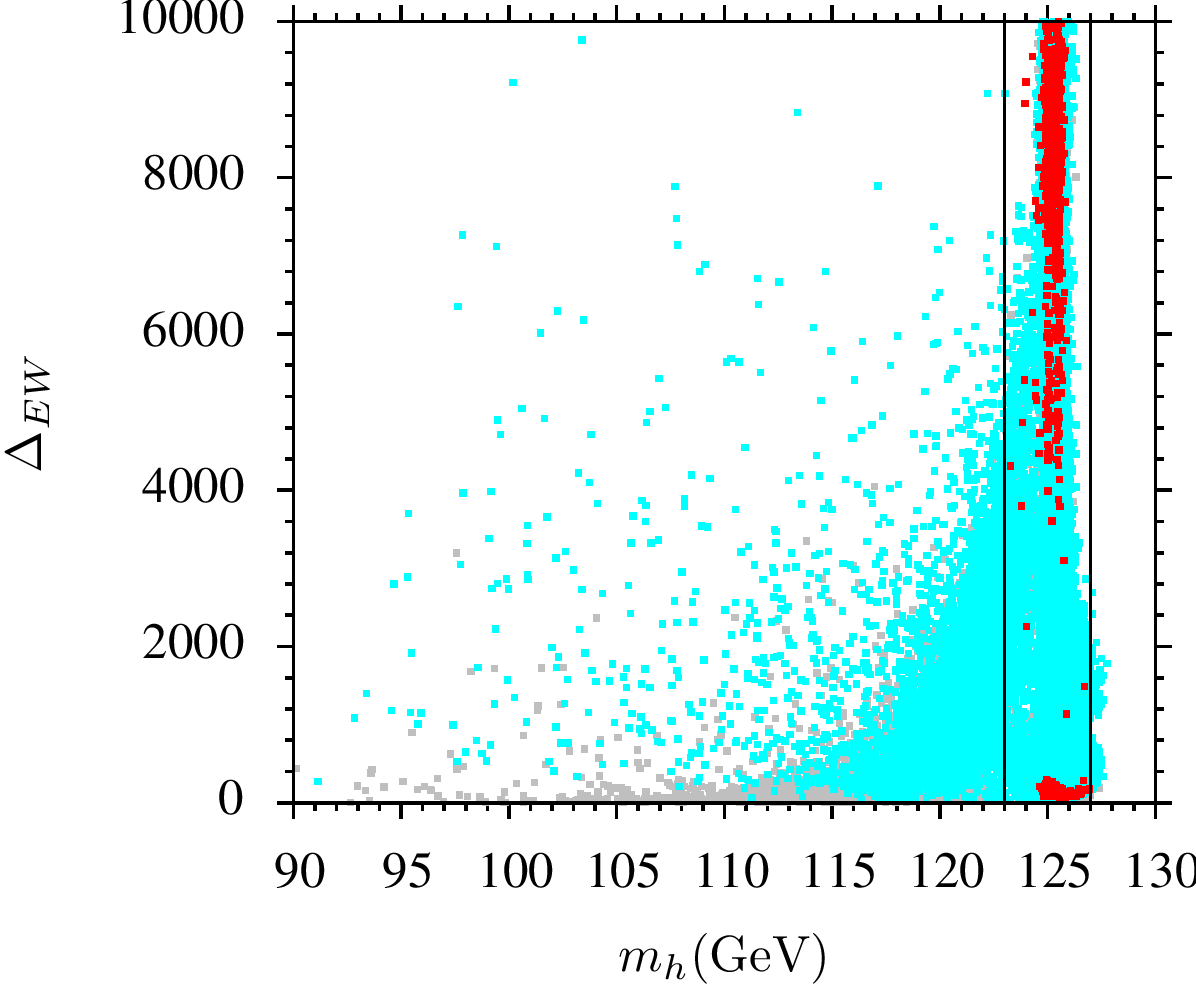}
}
\subfigure{
\includegraphics[totalheight=5.5cm,width=7.cm]{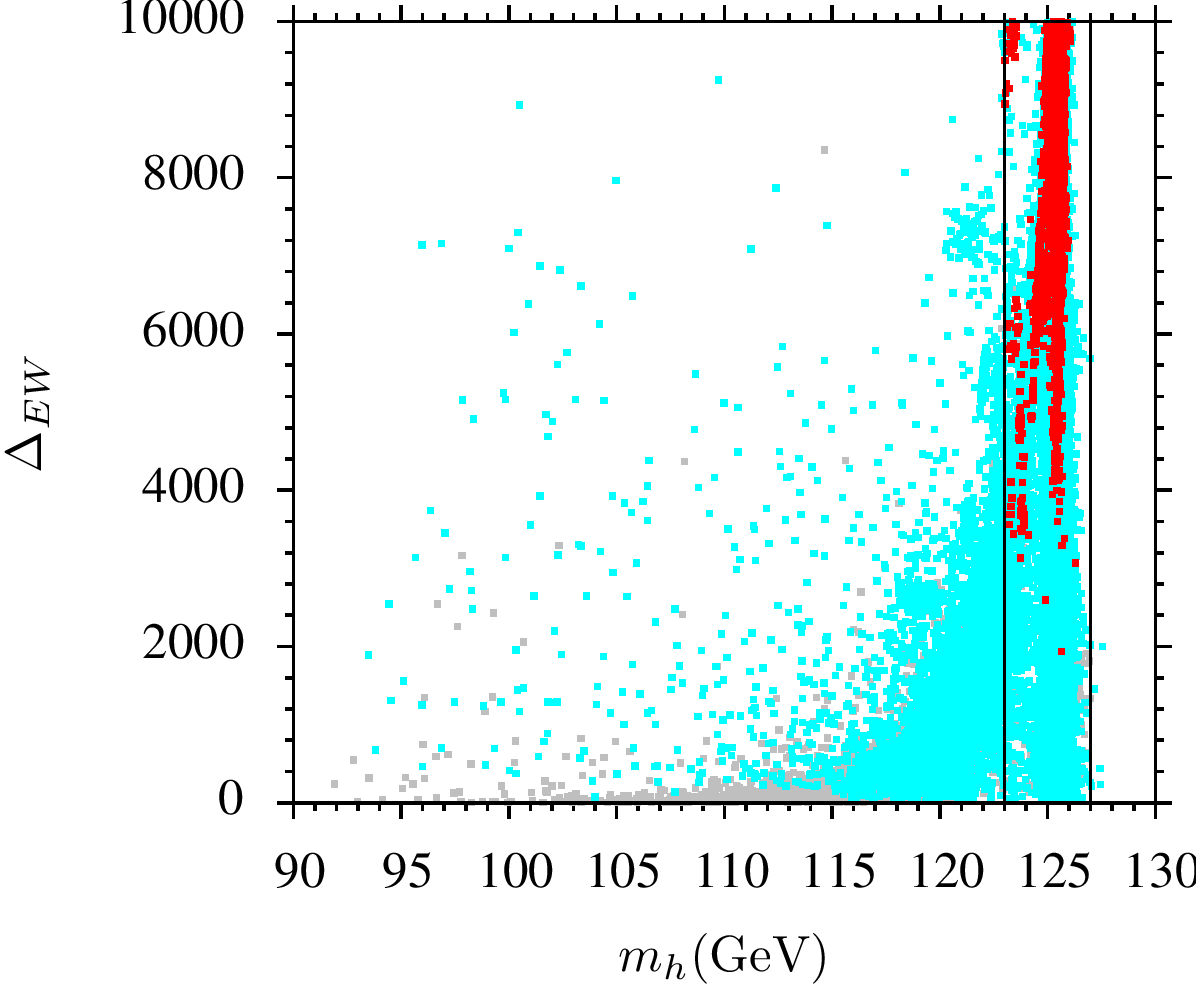}
}
\subfigure{
\includegraphics[totalheight=5.5cm,width=7.cm]{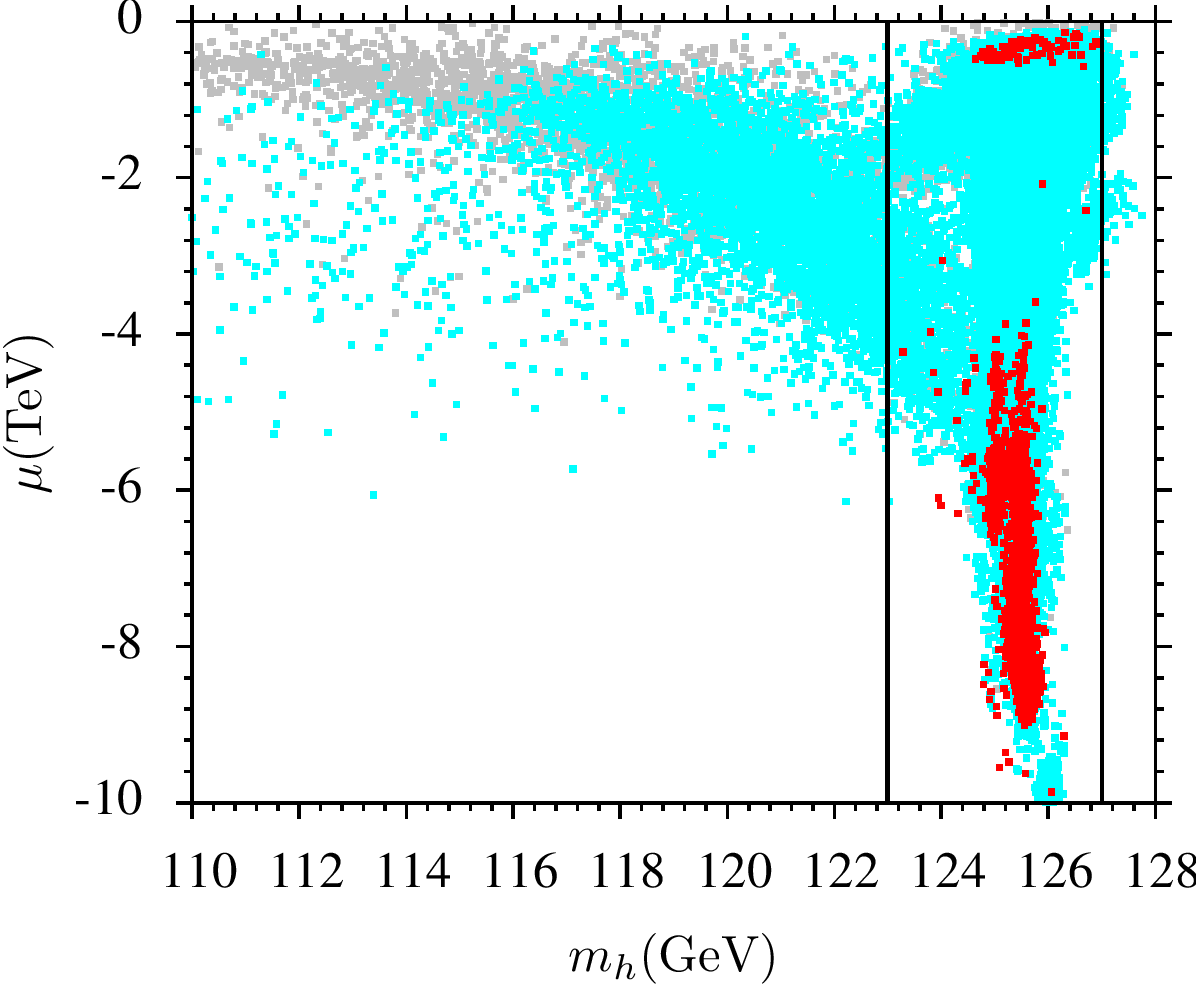}
}
\subfigure{
\includegraphics[totalheight=5.5cm,width=7.cm]{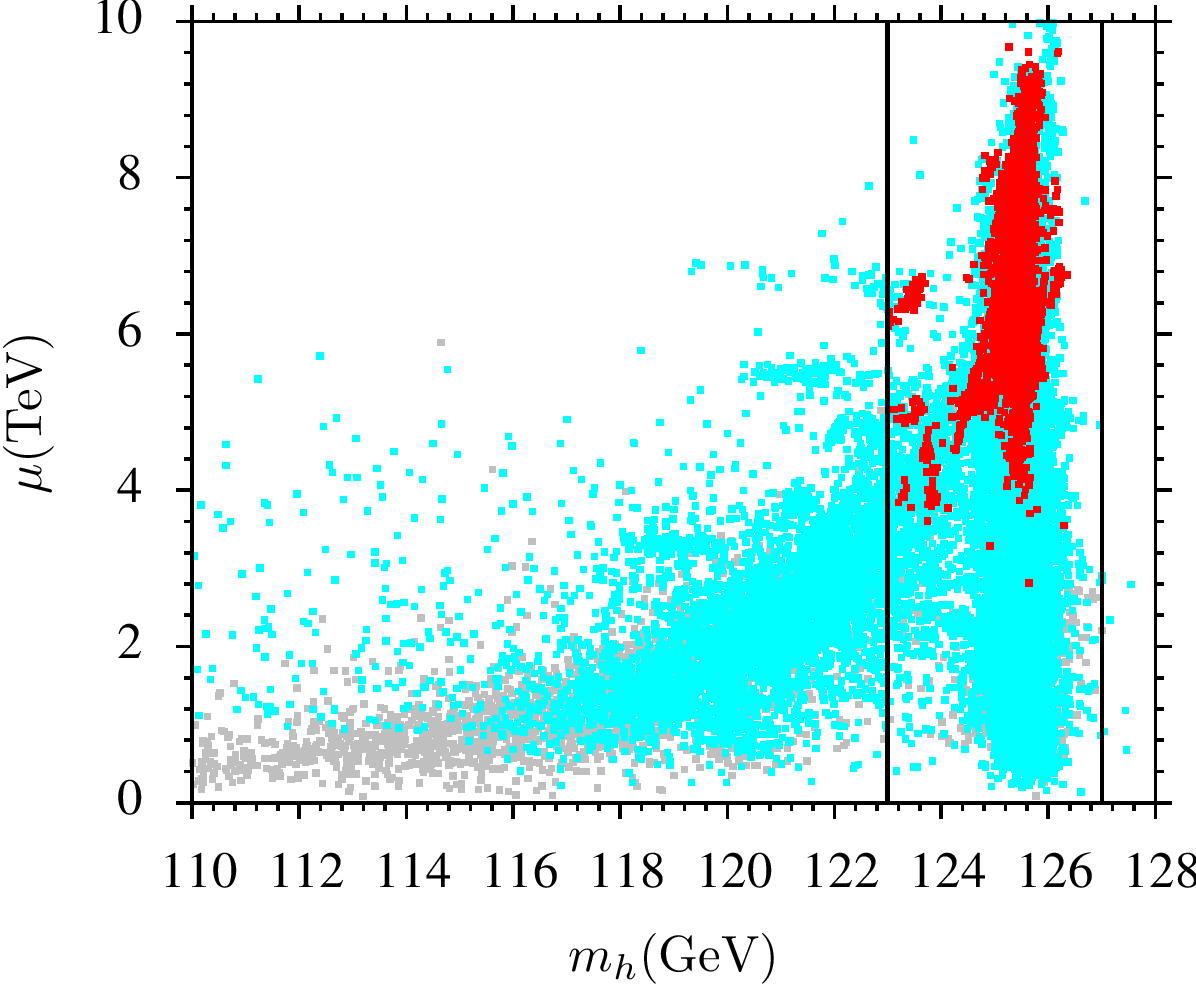}
}
\subfigure{
\includegraphics[totalheight=5.5cm,width=7.cm]{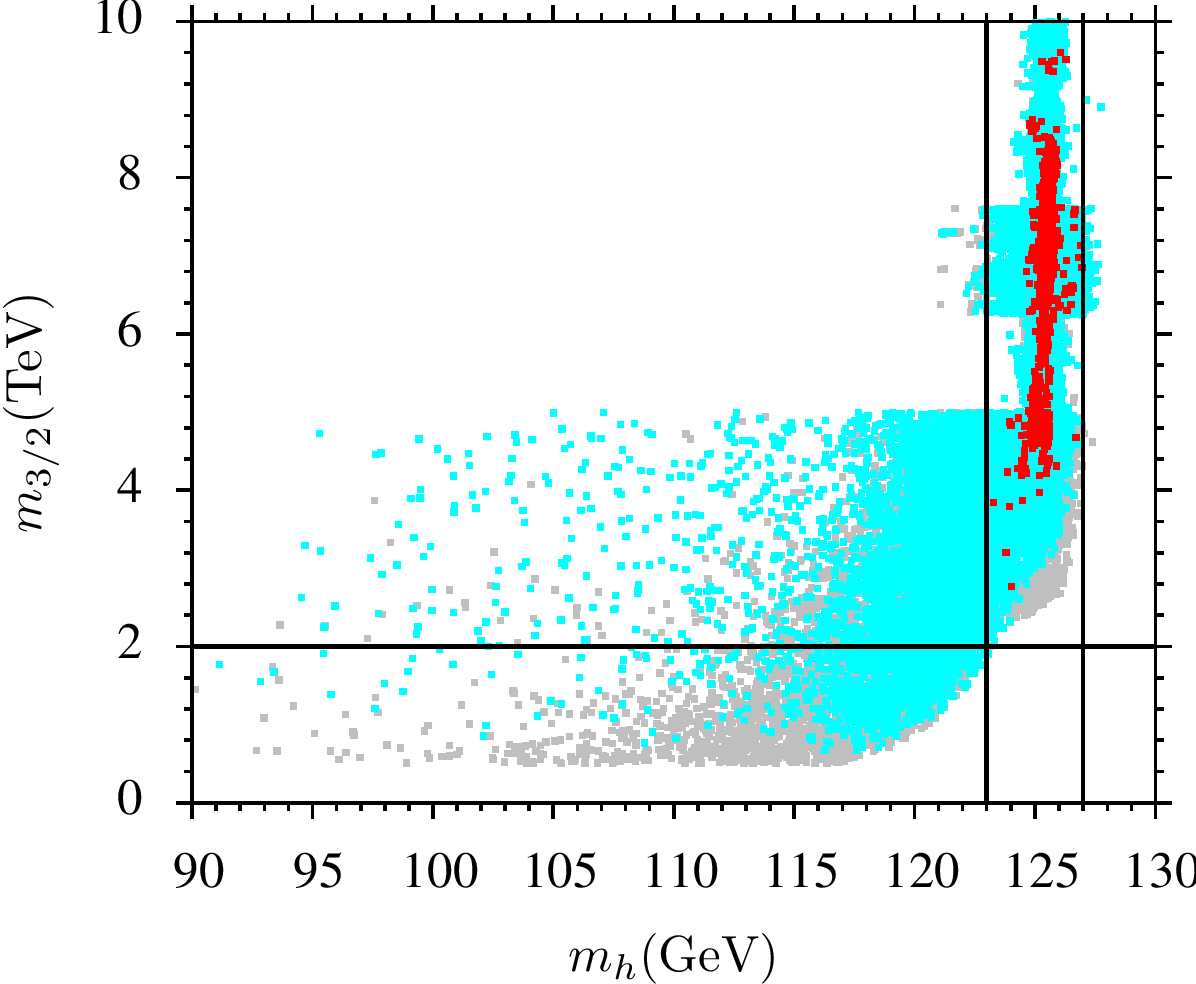}
}
\subfigure{
\includegraphics[totalheight=5.5cm,width=7.cm]{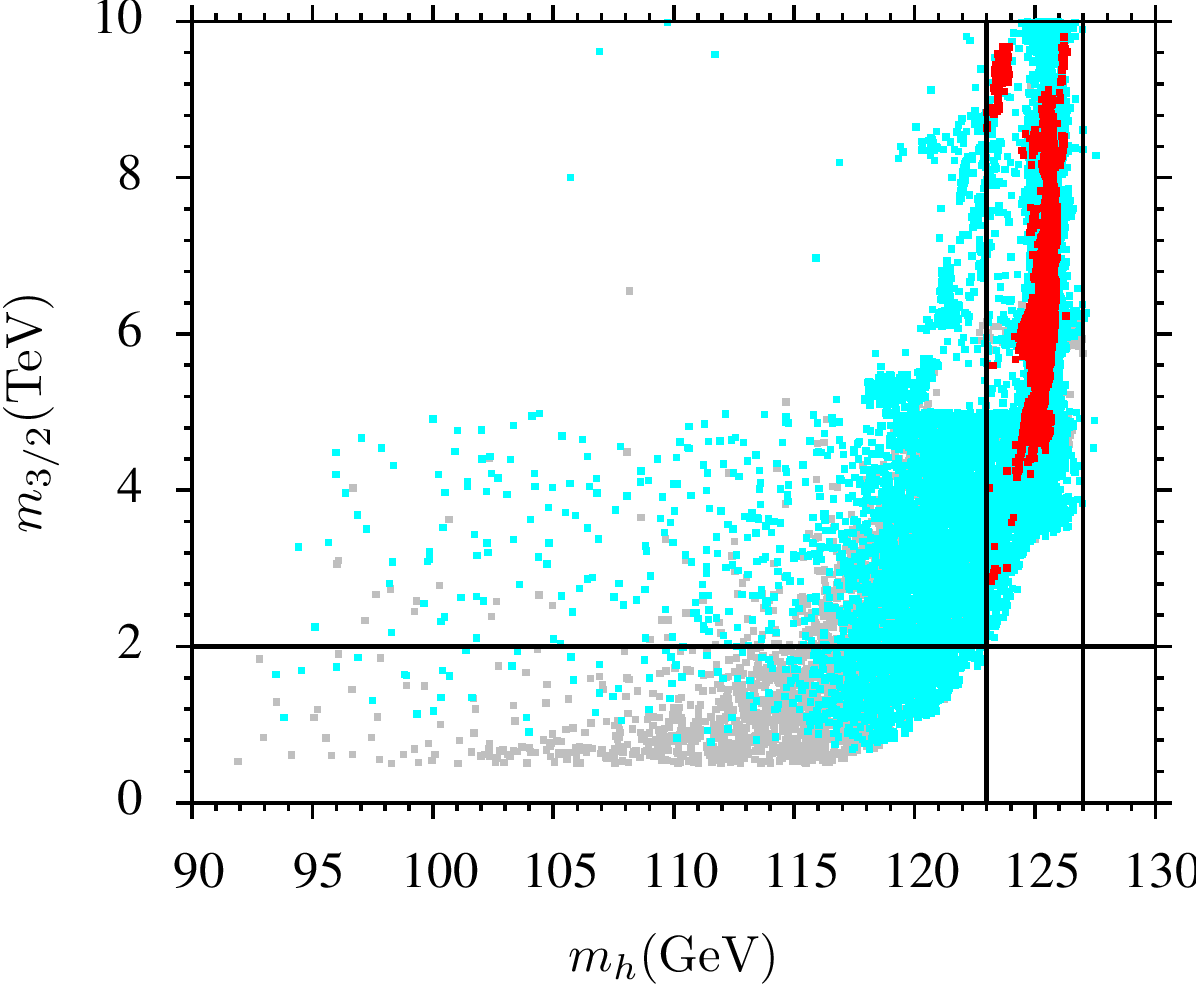}
}

\caption{Plots in $m_h-\Delta_{EW}$, $m_h-\mu$ and $m_h-m_{3/2}$ planes. 
 Grey points satisfy REWSB and yield LSP neutralino. Aqua points
satisfy all the mass bounds and B-physics bounds. 
Red points are subset of aqua points and also satisfy the WMAP9 5$\sigma$ bounds.
}
\label{m32}
\end{figure}
\begin{figure}[htp!]
\centering
\subfiguretopcaptrue

\subfigure{
\includegraphics[totalheight=5.5cm,width=7.cm]{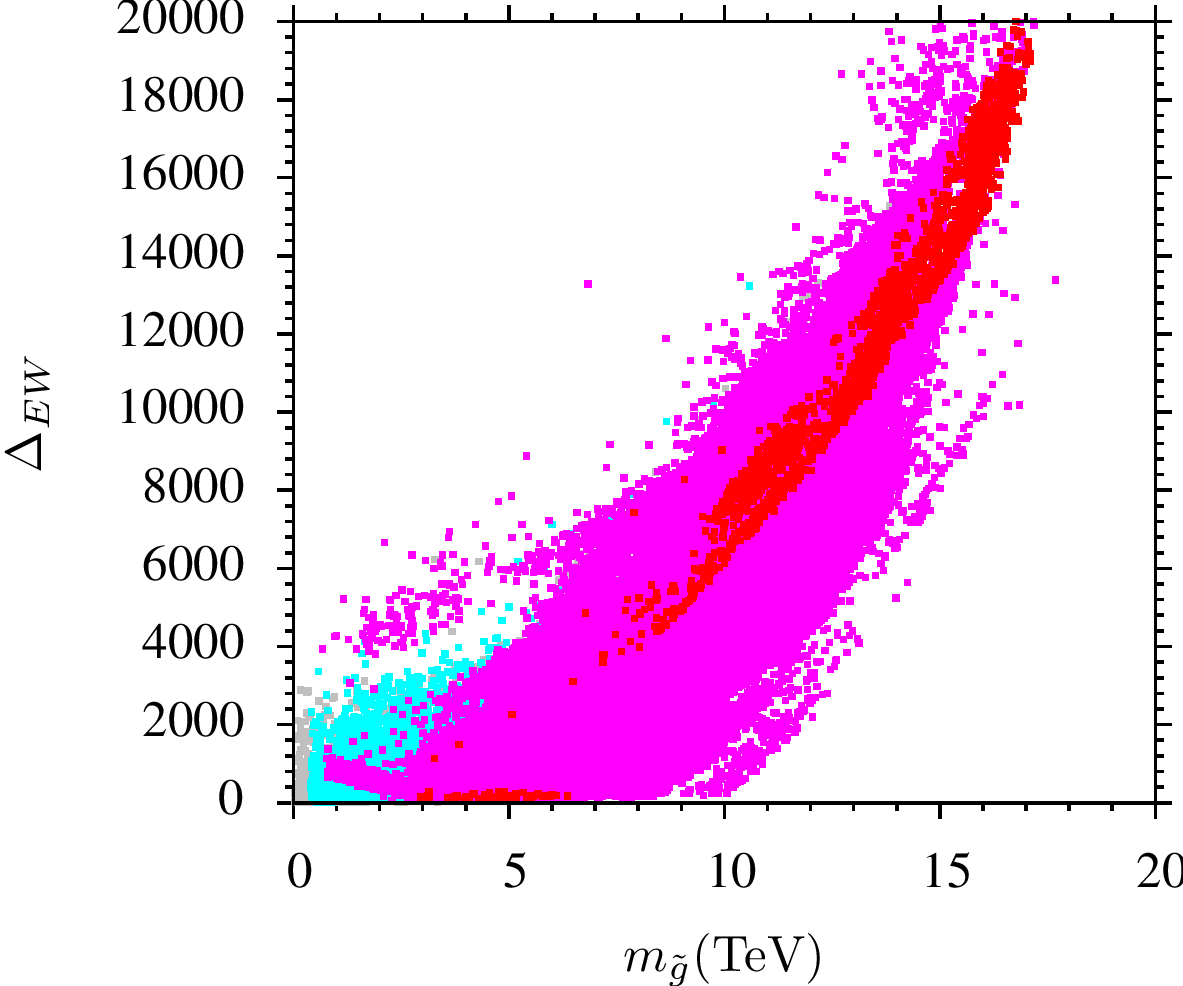}
}
\subfigure{
\includegraphics[totalheight=5.5cm,width=7.cm]{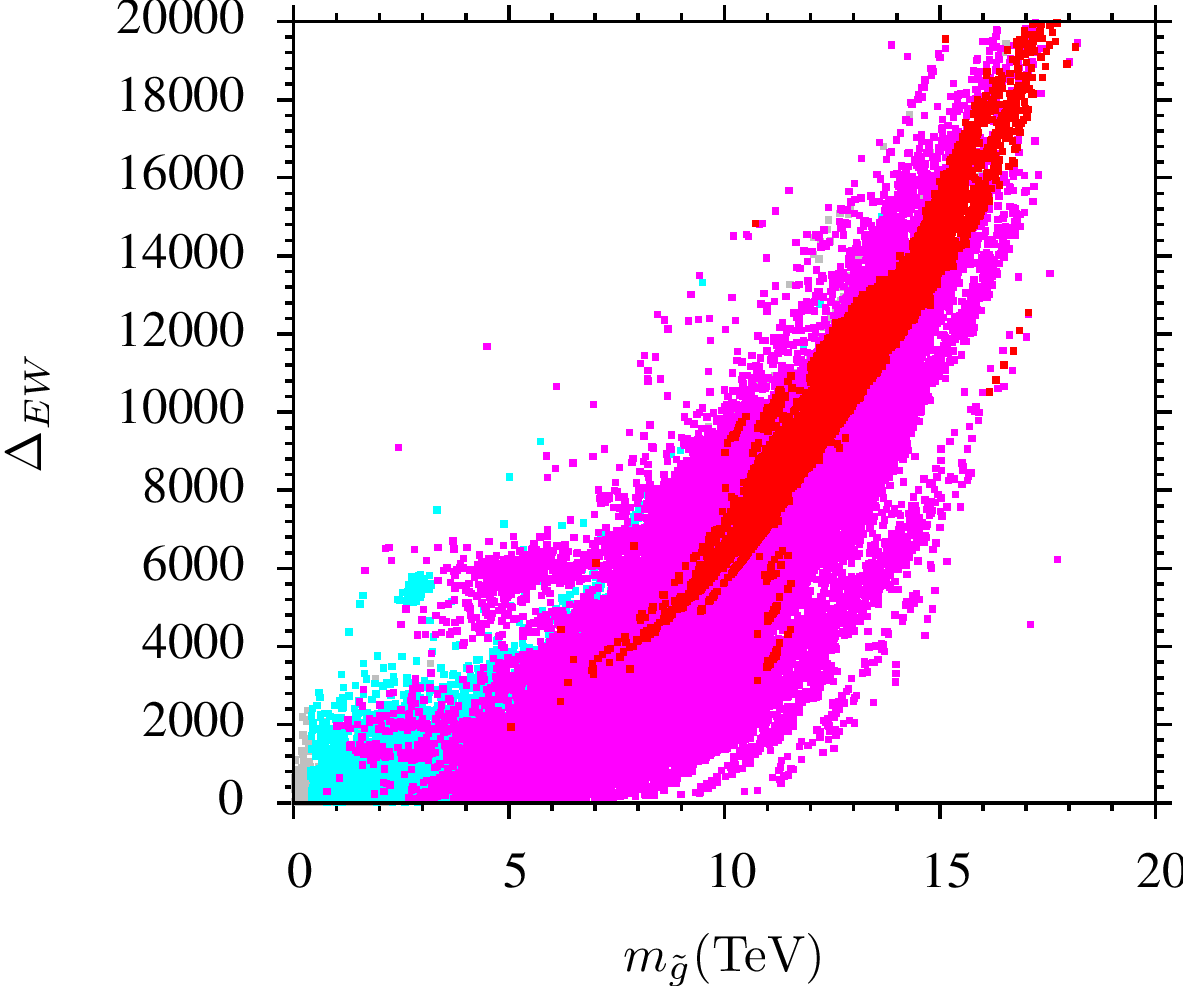}
}
\subfigure{
\includegraphics[totalheight=5.5cm,width=7.cm]{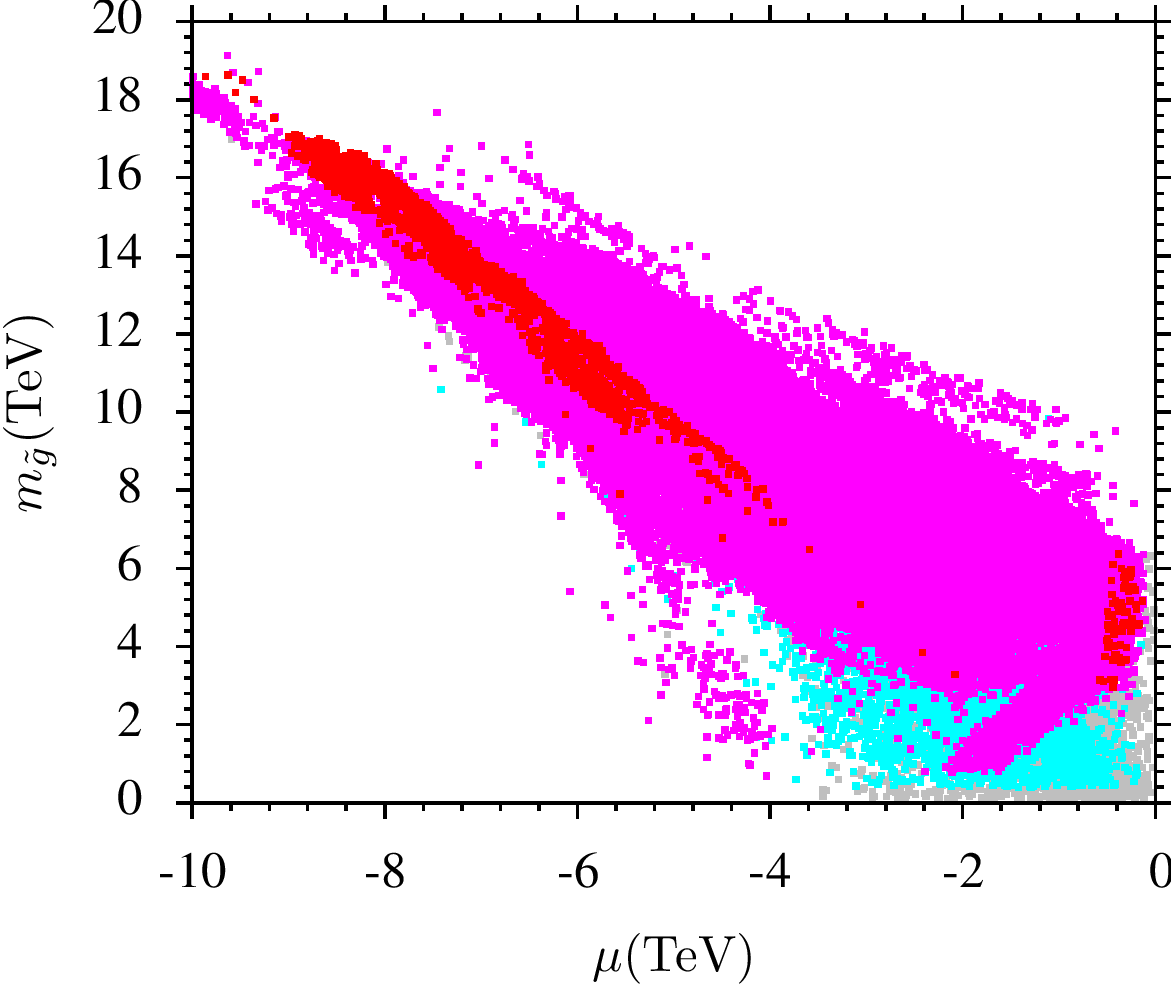}
}
\subfigure{
\includegraphics[totalheight=5.5cm,width=7.cm]{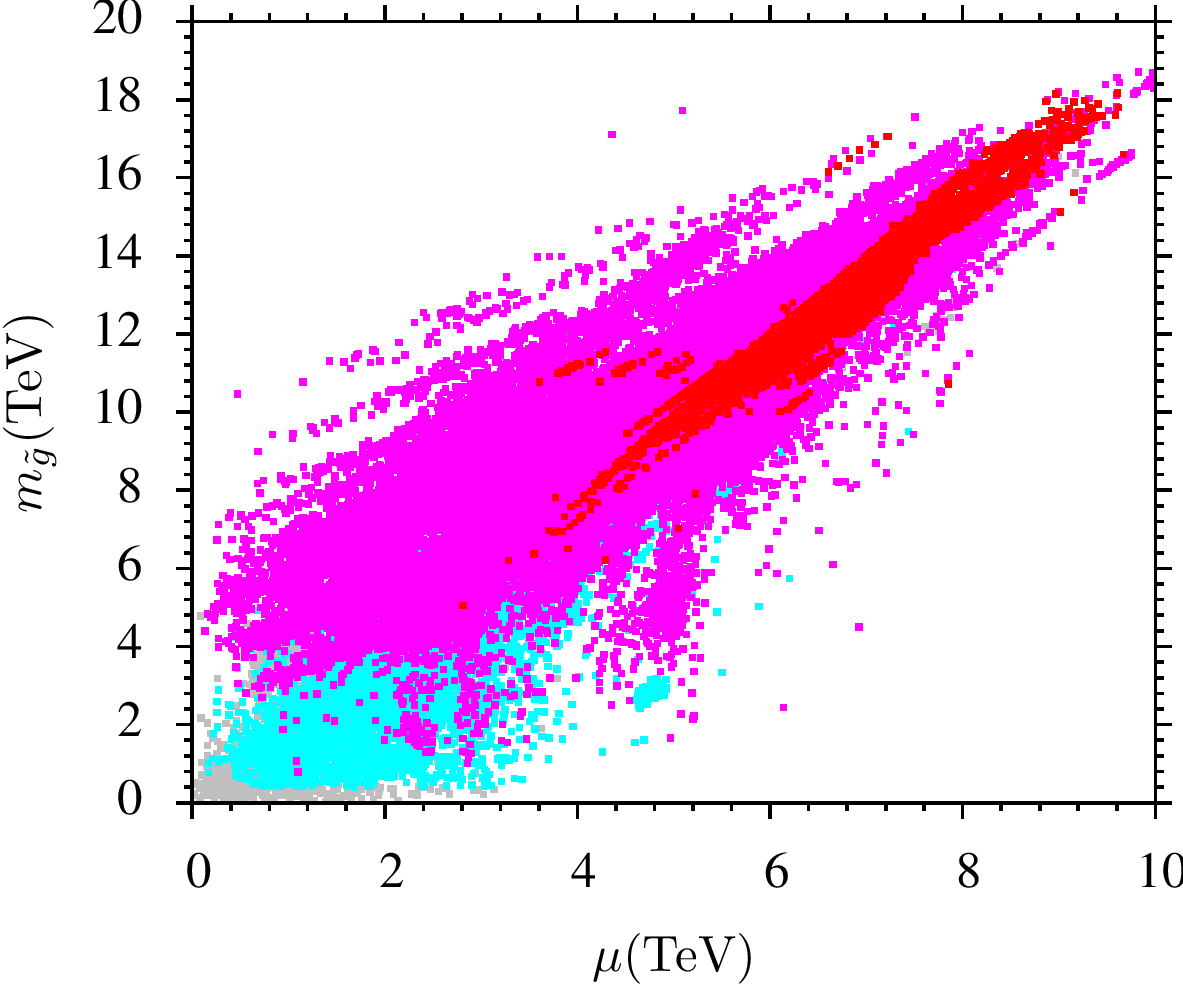}
}

\caption{Plots in $\Delta_{EW}-m_{\tilde g}$ and $\mu-m_{\tilde g}$ planes. 
Color coding and panel description are same as in Fig.~\ref{input_params1}, 
except we do not apply gluino bounds mentioned in Section~\ref{sec:scan}.
}
\label{glumu}
\end{figure}
\begin{figure}[htp!]
\centering
\subfiguretopcaptrue

\subfigure{
\includegraphics[totalheight=5.5cm,width=7.cm]{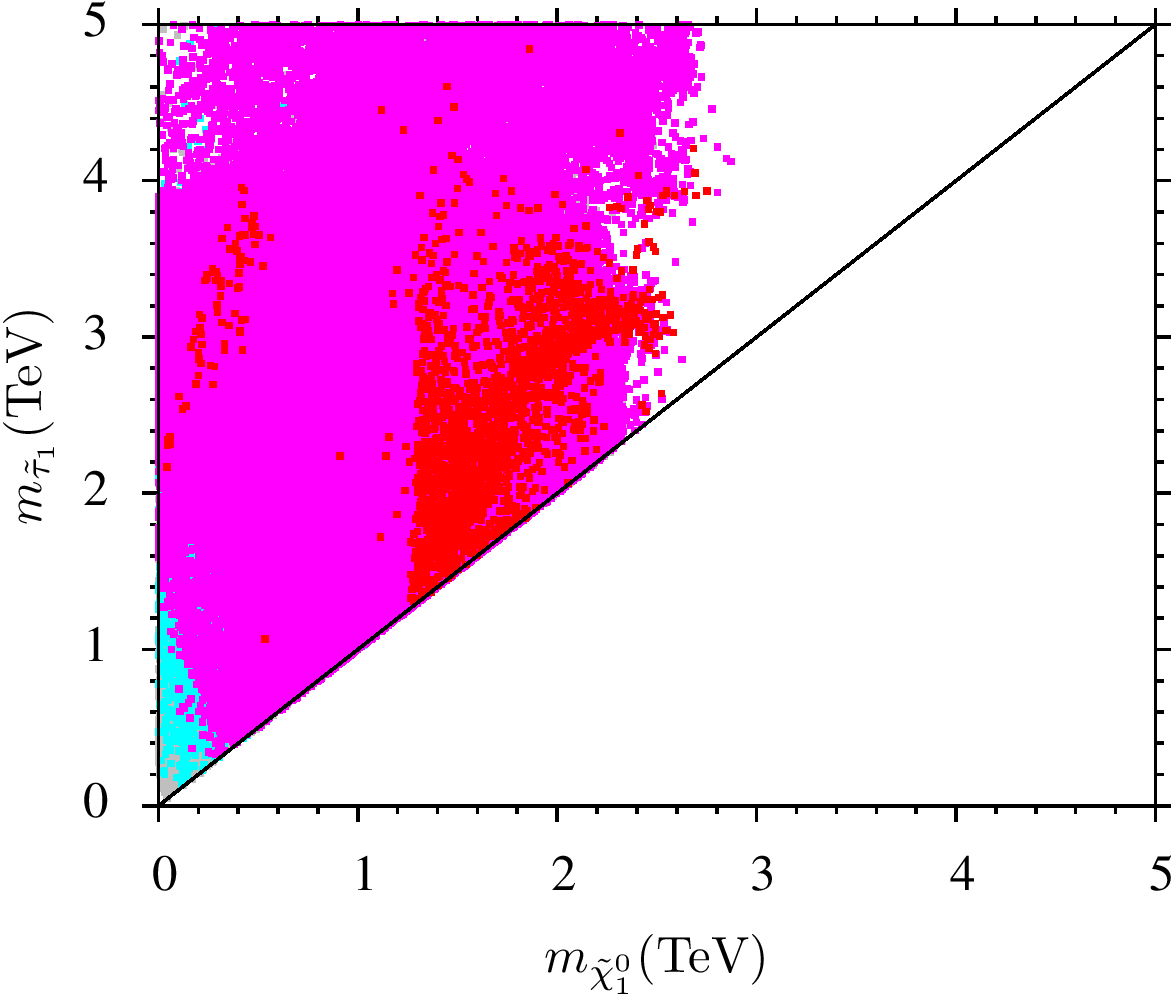}
}
\subfigure{
\includegraphics[totalheight=5.5cm,width=7.cm]{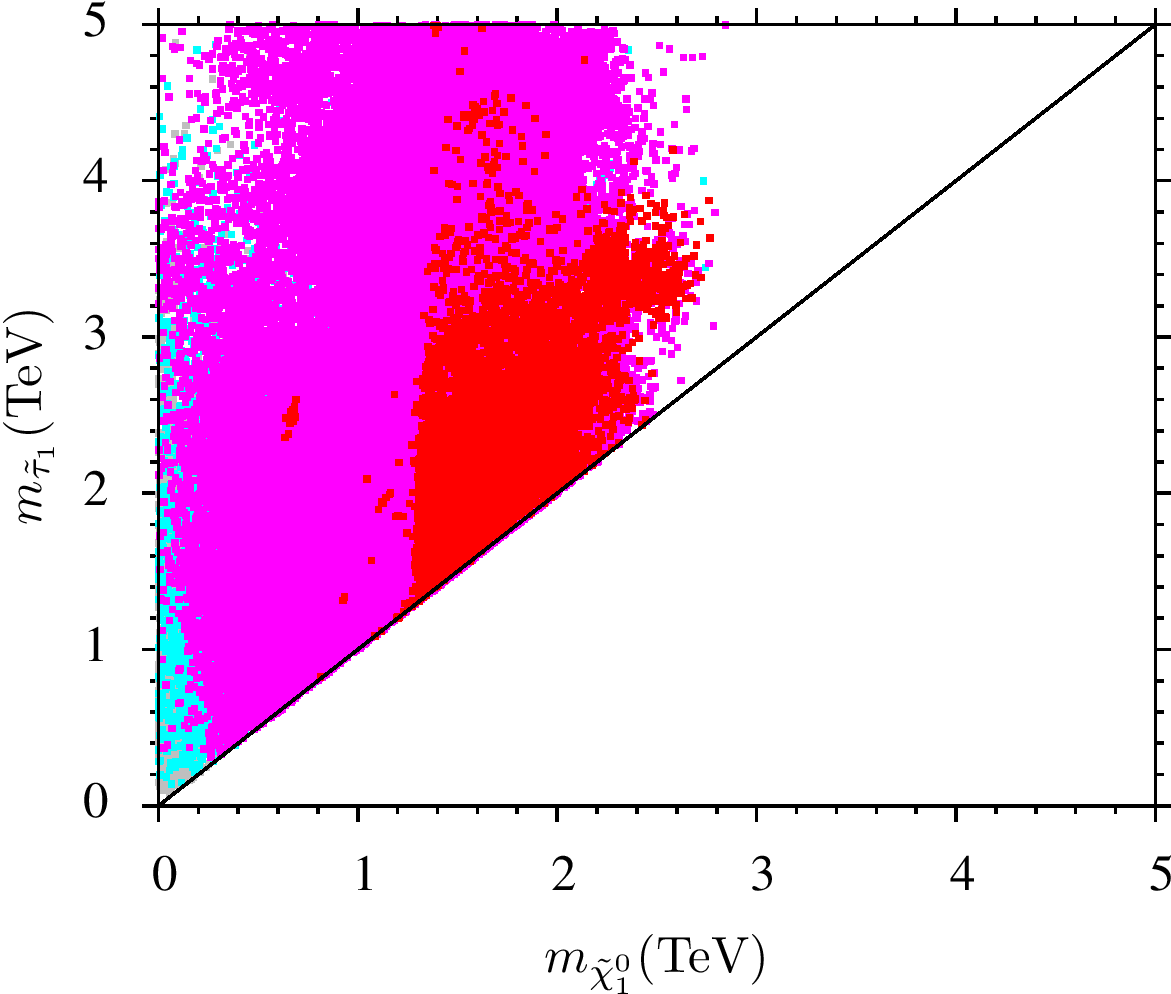}
}
\subfigure{
\includegraphics[totalheight=5.5cm,width=7.cm]{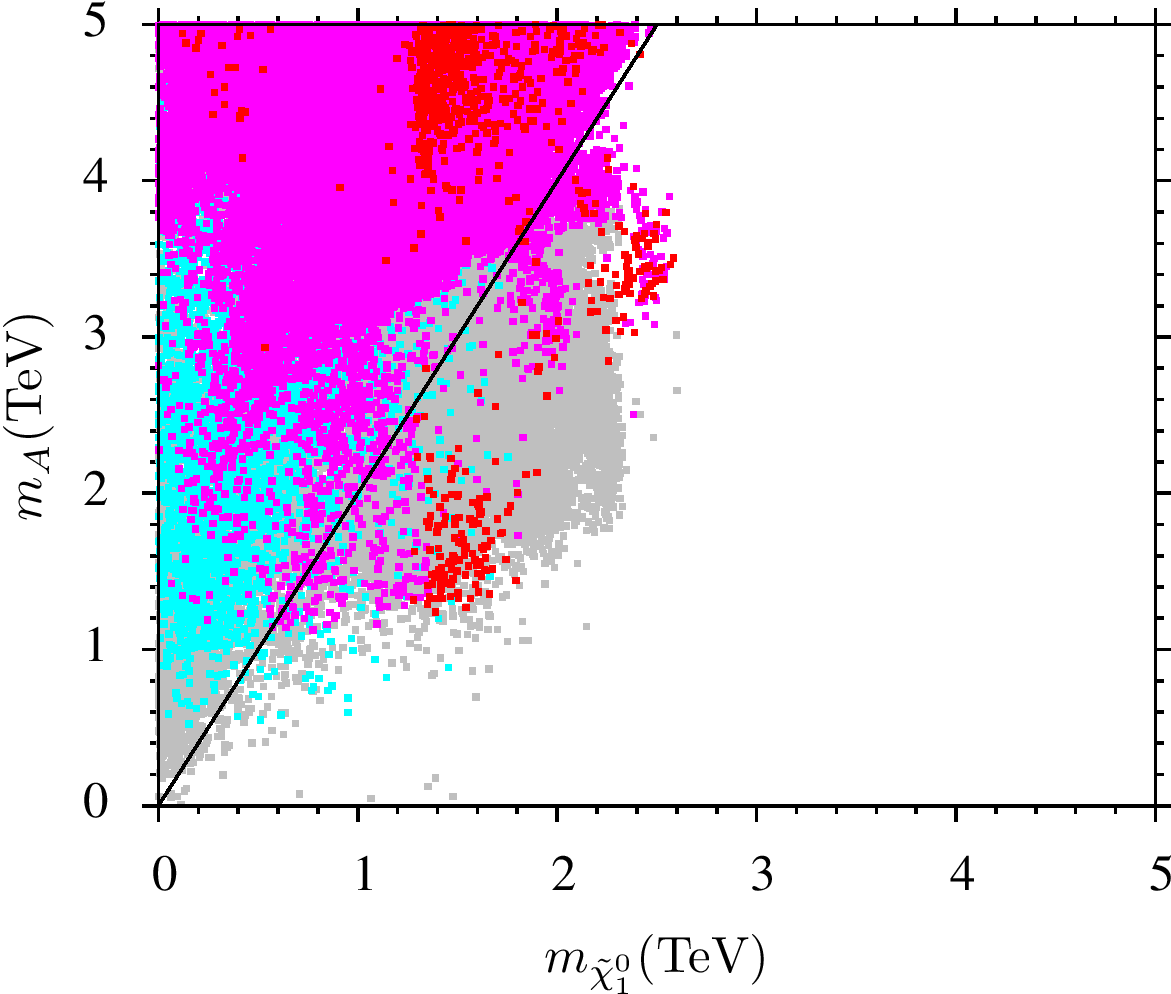}
}
\subfigure{
\includegraphics[totalheight=5.5cm,width=7.cm]{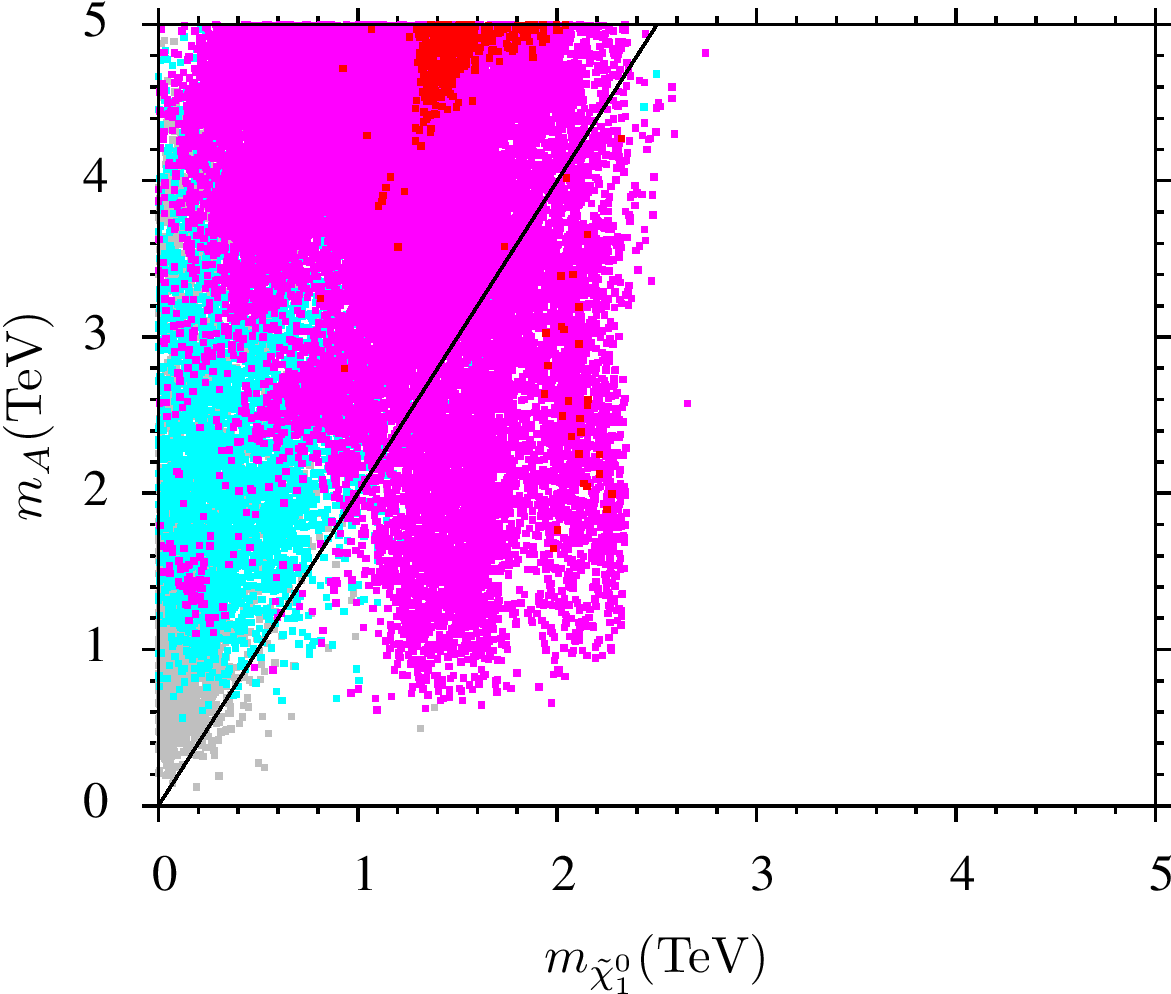}
}

\caption{Plots in $m_{\tilde \chi_{1}^{0}}-m_{\tilde \tau}$ and $m_{\tilde \chi_{1}^{0}}-m_{A}$ planes. 
Color coding and panel description are same as in Fig.~\ref{input_params1}.
}
\label{spectrum1}
\end{figure}
\begin{figure}[htp!]
\centering
\subfiguretopcaptrue

\subfigure{
\includegraphics[totalheight=5.5cm,width=7.cm]{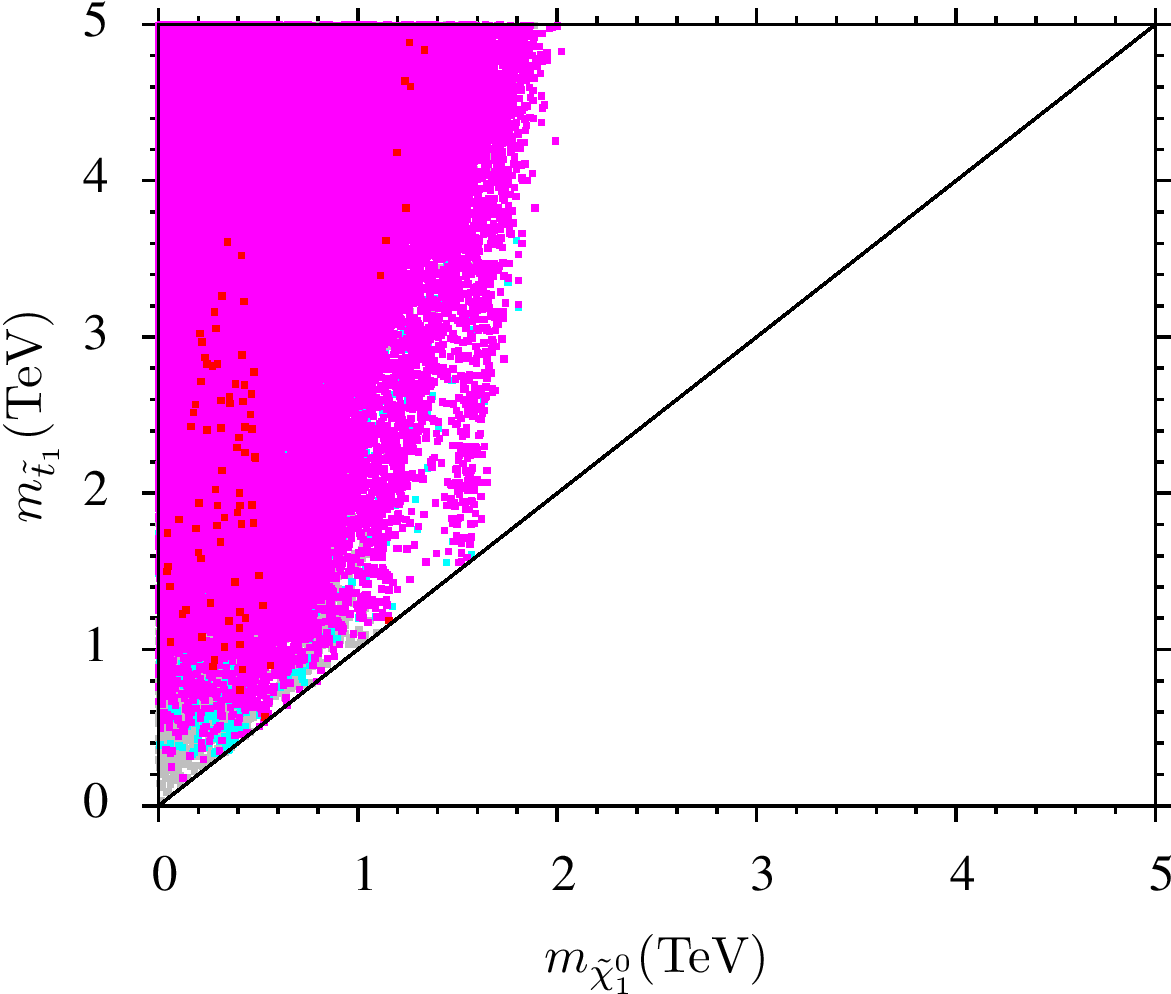}
}
\subfigure{
\includegraphics[totalheight=5.5cm,width=7.cm]{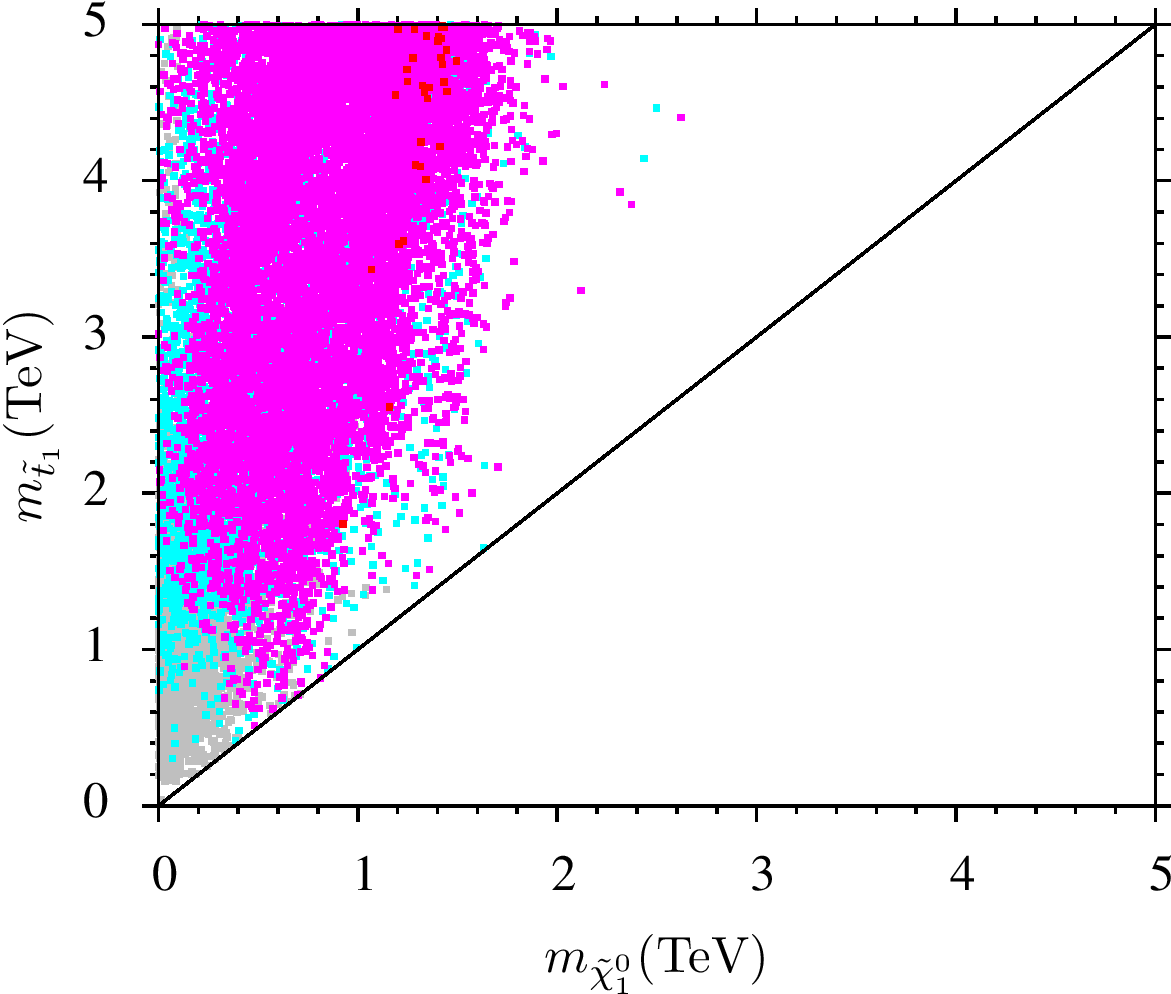}
}
\subfigure{
\includegraphics[totalheight=5.5cm,width=7.cm]{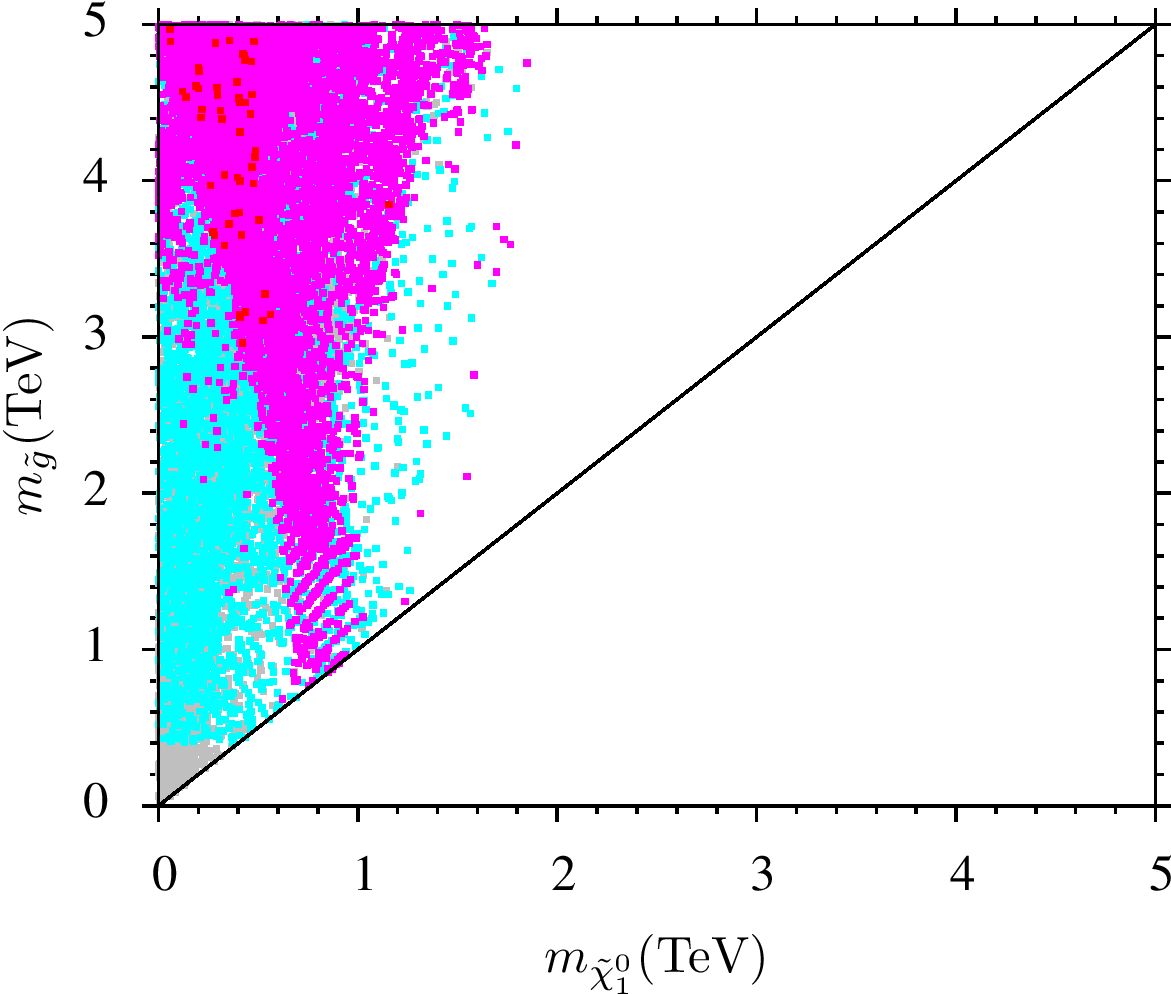}
}
\subfigure{
\includegraphics[totalheight=5.5cm,width=7.cm]{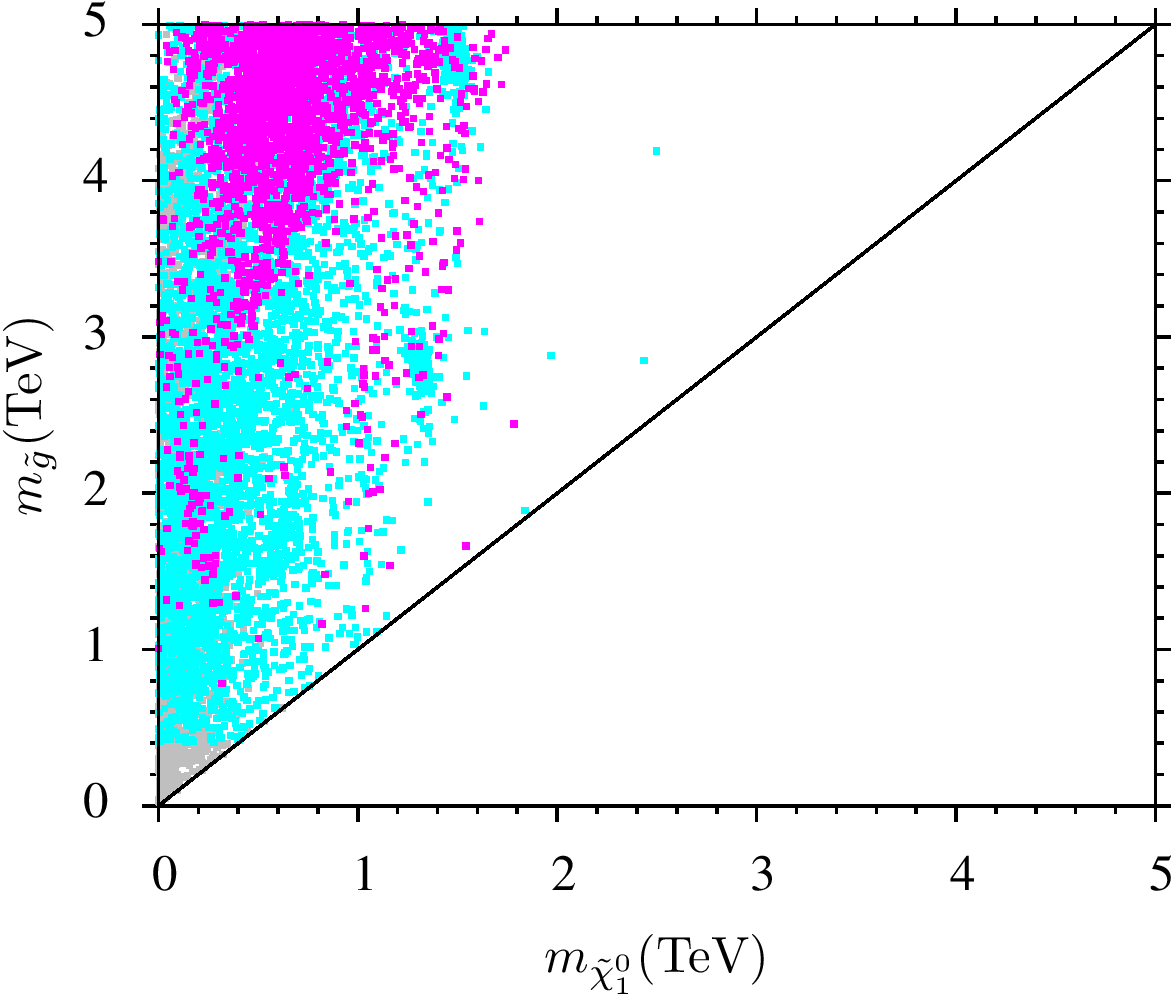}
}
\subfigure{
\includegraphics[totalheight=5.5cm,width=7.cm]{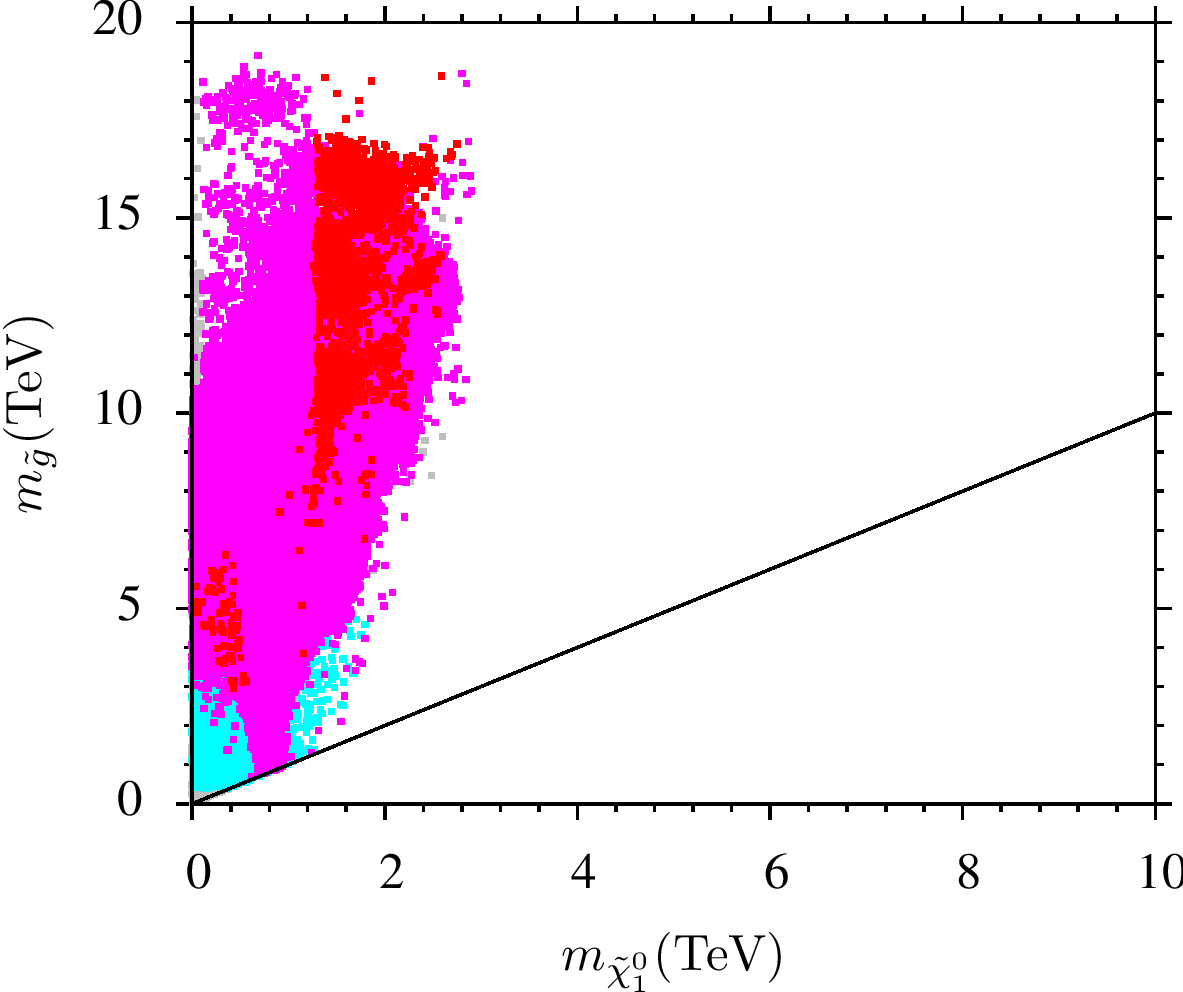}
}
\subfigure{
\includegraphics[totalheight=5.5cm,width=7.cm]{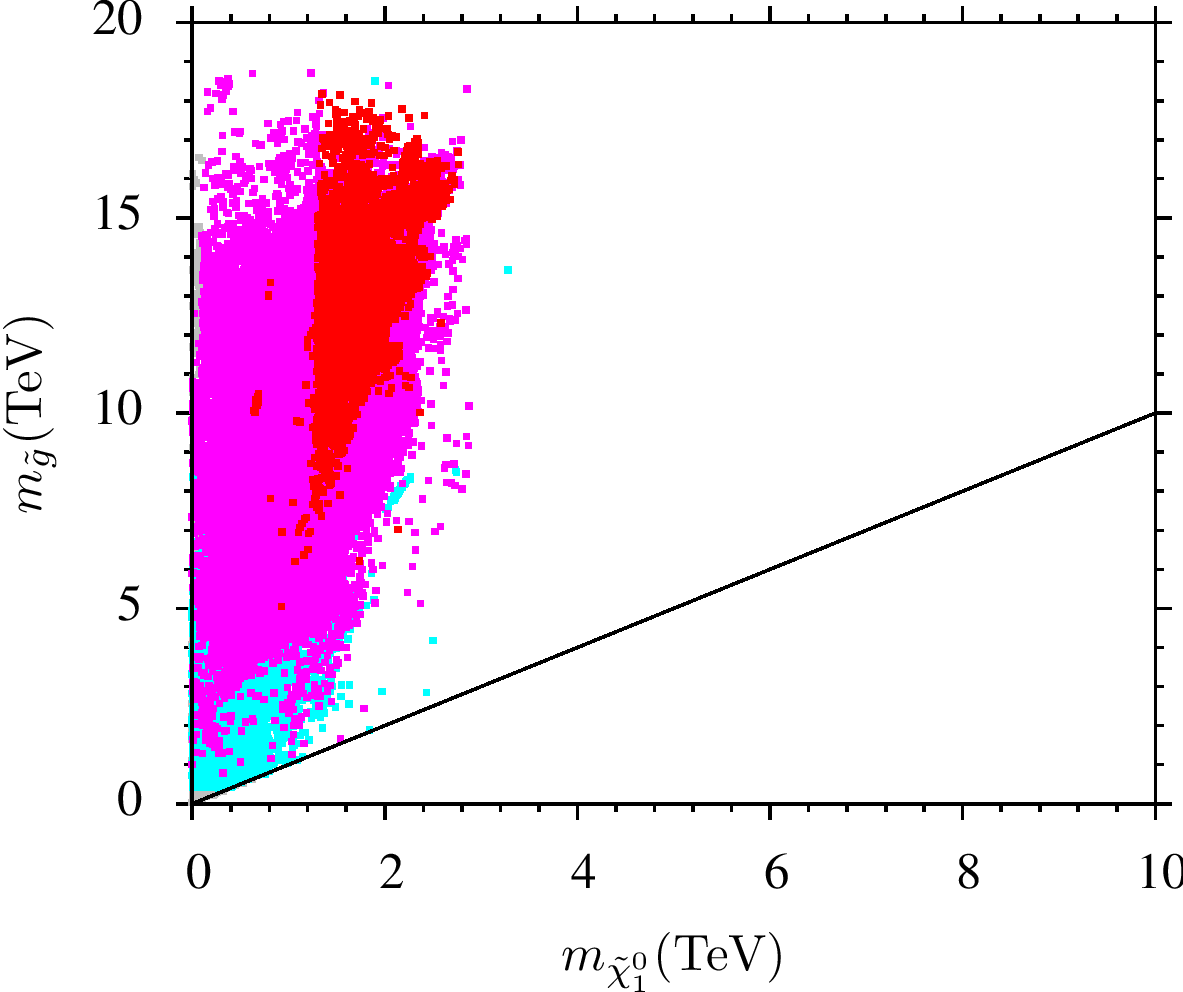}
}

\caption{Plots in $m_{\tilde \chi_{1}^{0}}-m_{\tilde t_1}$ and $m_{\tilde \chi_{1}^{0}}-m_{\tilde g}$ planes.
Color coding and panel description are same as in Fig.~\ref{input_params1}, except in middle 
and bottom panels we do not apply gluino bounds mentioned in Section~\ref{sec:scan}.
}
\label{spectrum2}
\end{figure}

\begin{figure}[htp!]
\centering
\subfiguretopcaptrue

\subfigure{
\includegraphics[totalheight=5.5cm,width=7.cm]{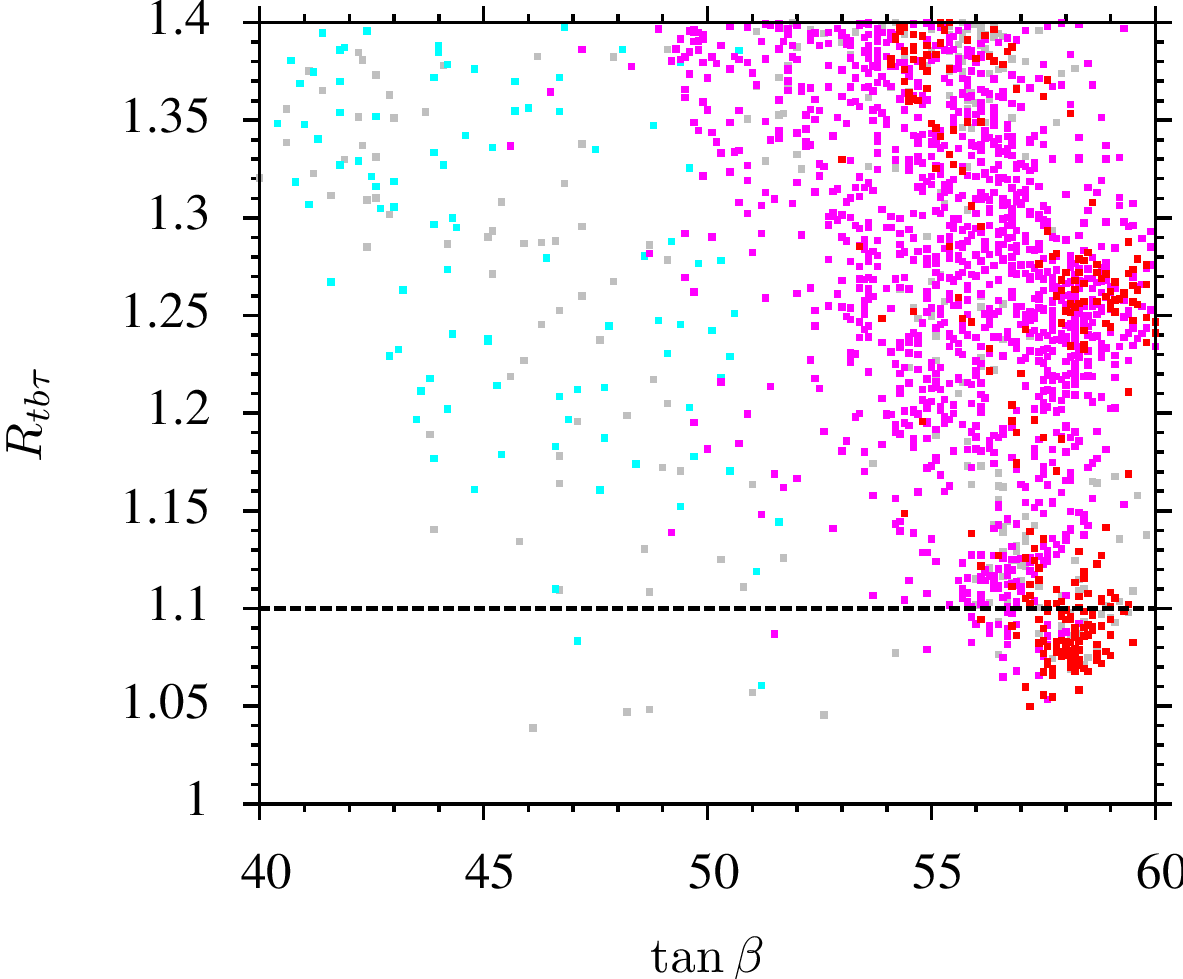}
}
\subfigure{
\includegraphics[totalheight=5.5cm,width=7.cm]{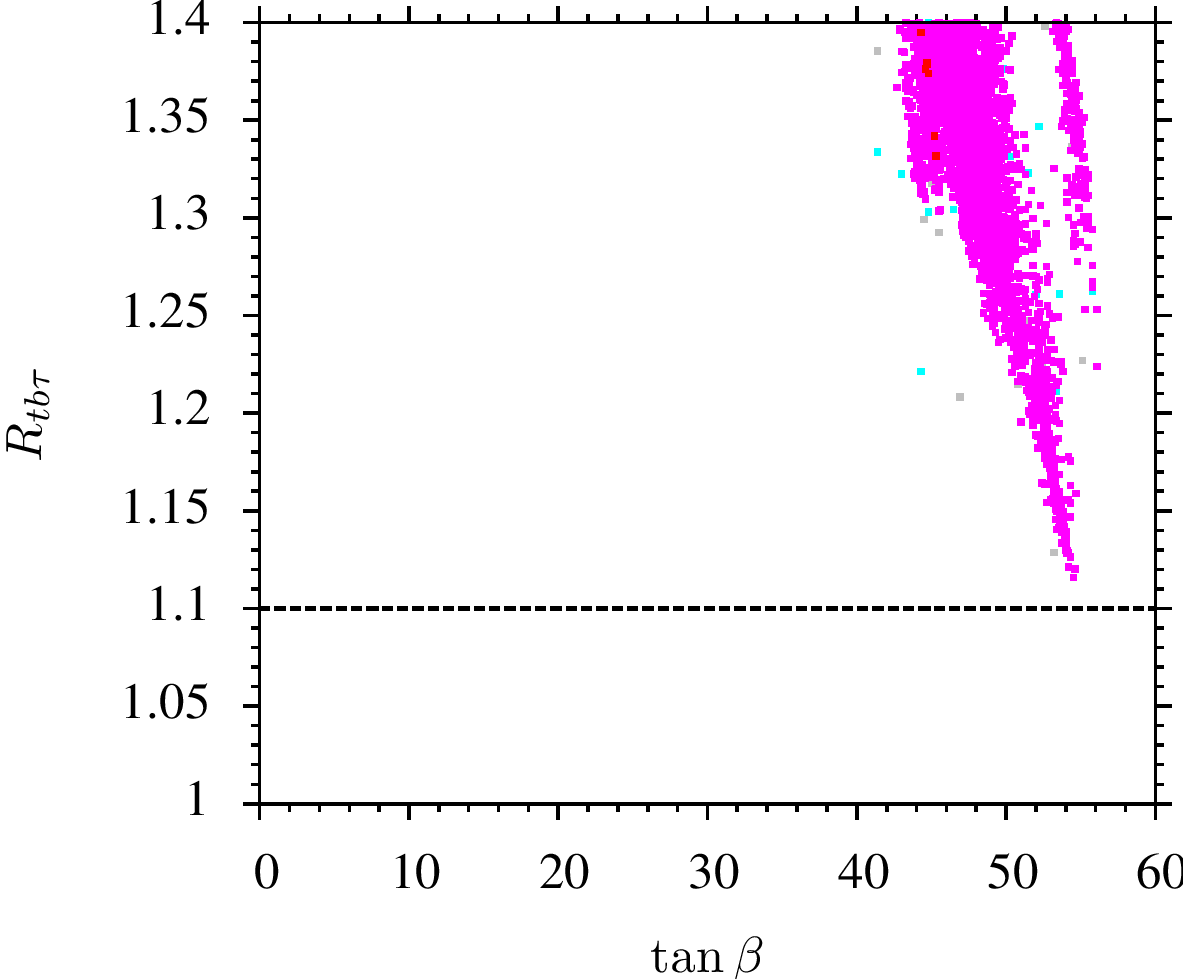}
}
\subfigure{
\includegraphics[totalheight=5.5cm,width=7.cm]{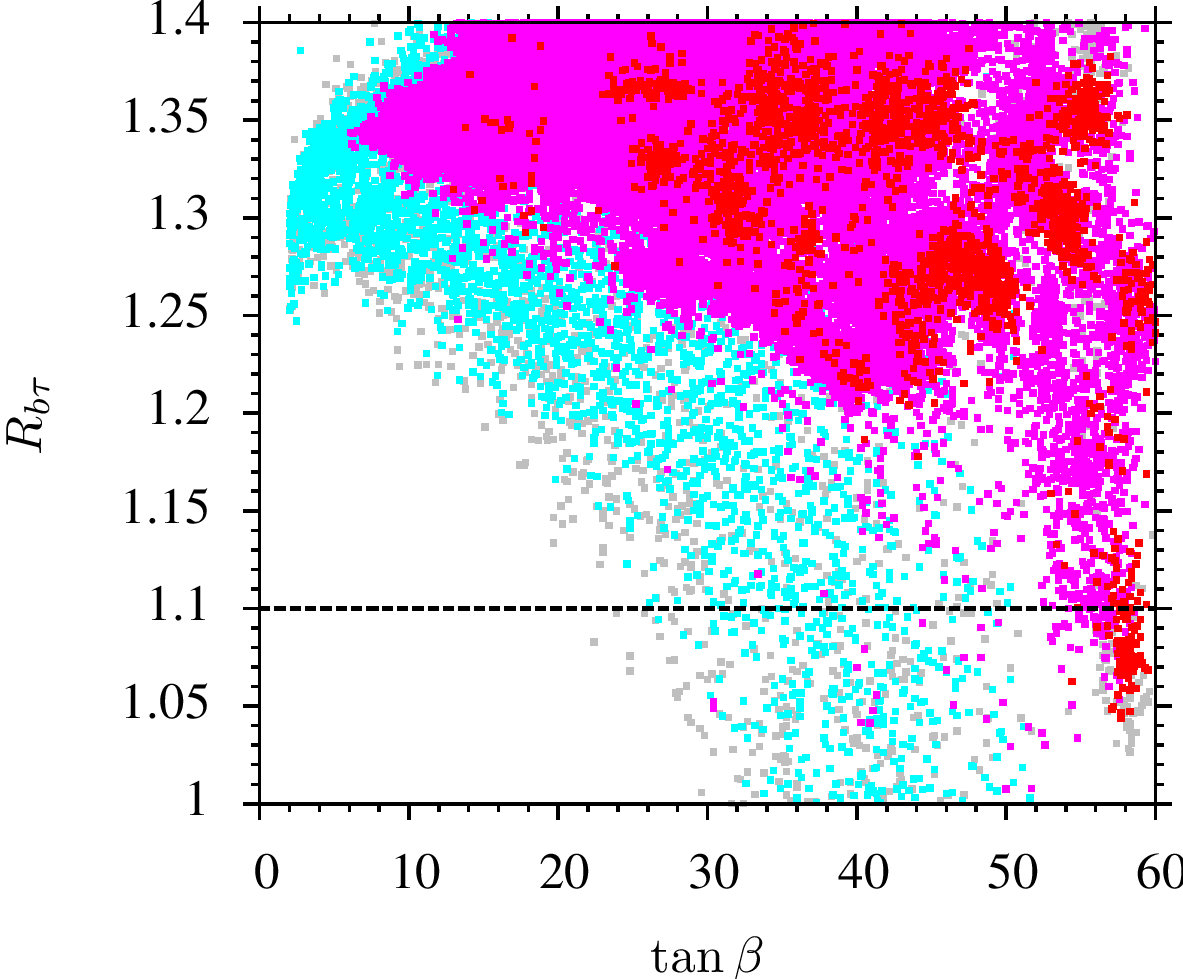}
}
\subfigure{
\includegraphics[totalheight=5.5cm,width=7.cm]{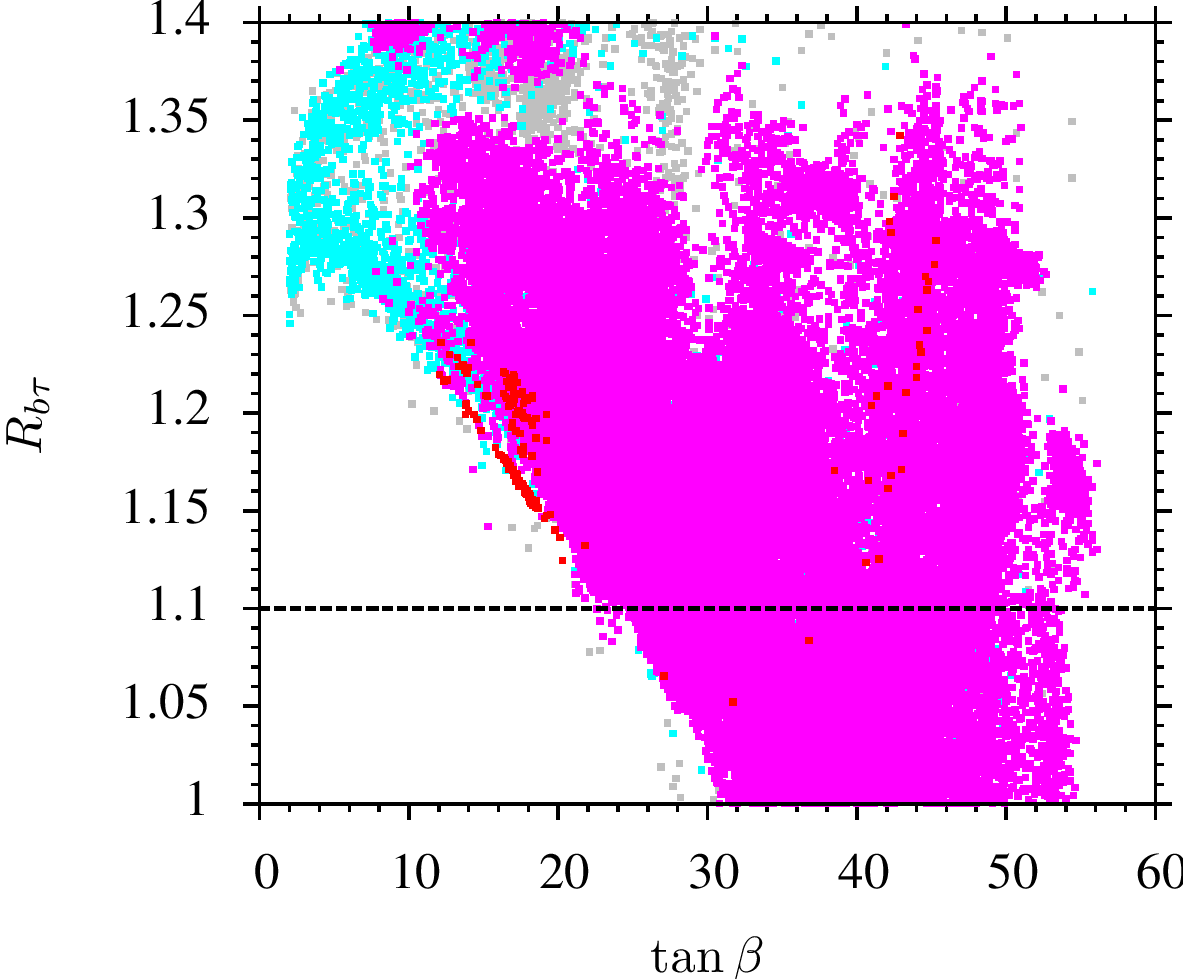}
}

\caption{Plots in $\tan\beta-R_{tb\tau}$ and $\tan\beta-R_{b\tau}$ planes. 
Color coding and panel description are same as in Fig.~\ref{input_params1}.
}
\label{YU}
\end{figure}
\begin{figure}[htp!]
\centering
\subfiguretopcaptrue

\subfigure{
\includegraphics[totalheight=5.5cm,width=7.cm]{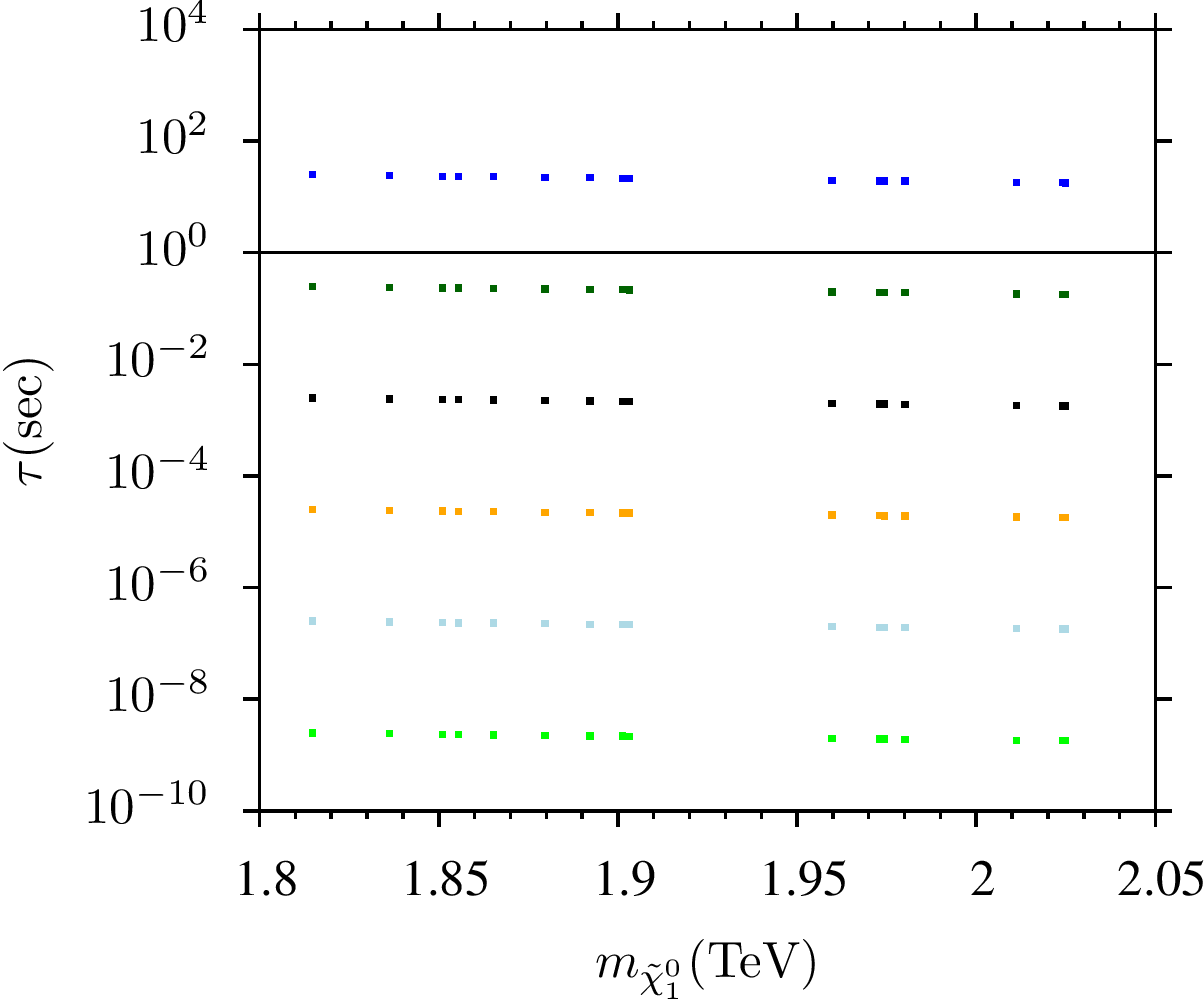}
}
\subfigure{
\includegraphics[totalheight=5.5cm,width=7.cm]{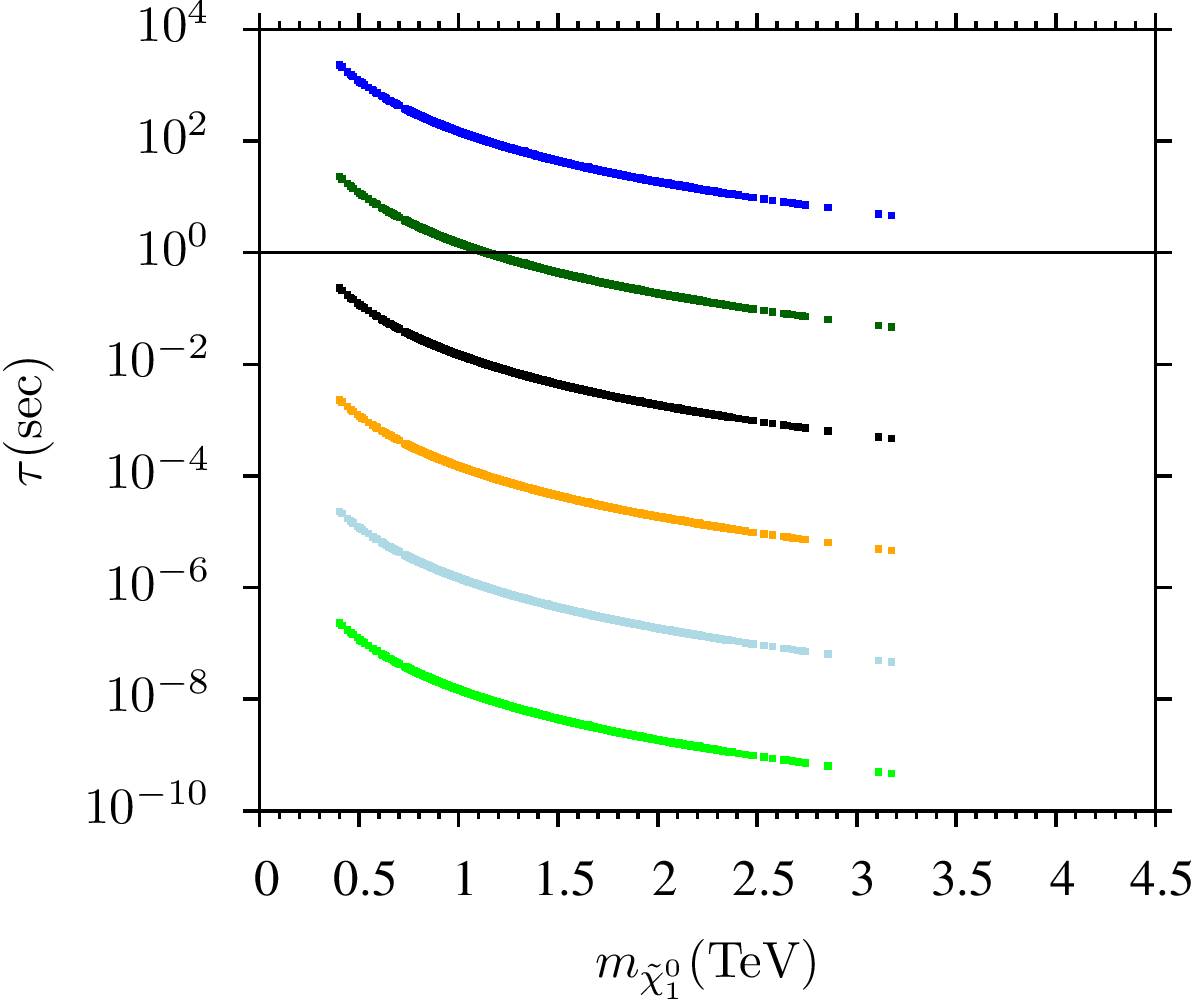}
}

\caption{The NLSP bino-like neutralino mass ($m_{\tilde \chi_{1}^{0}}$) versus lifetime ($\tau$) plot. 
We use $C_{aYY}=8/3$, $N$=6. Here, 
 light-Green, light-blue, orange, black, dark-green, and blue colors 
represent $f_a=10^{10}-10^{15}\,{\rm GeV}$, respectively.
The black solid line shows $\tau=$1 second.
}
\label{LT}
\end{figure}
\begin{figure}[htp!]
\centering
\subfiguretopcaptrue

\subfigure{
\includegraphics[totalheight=5.5cm,width=7.cm]{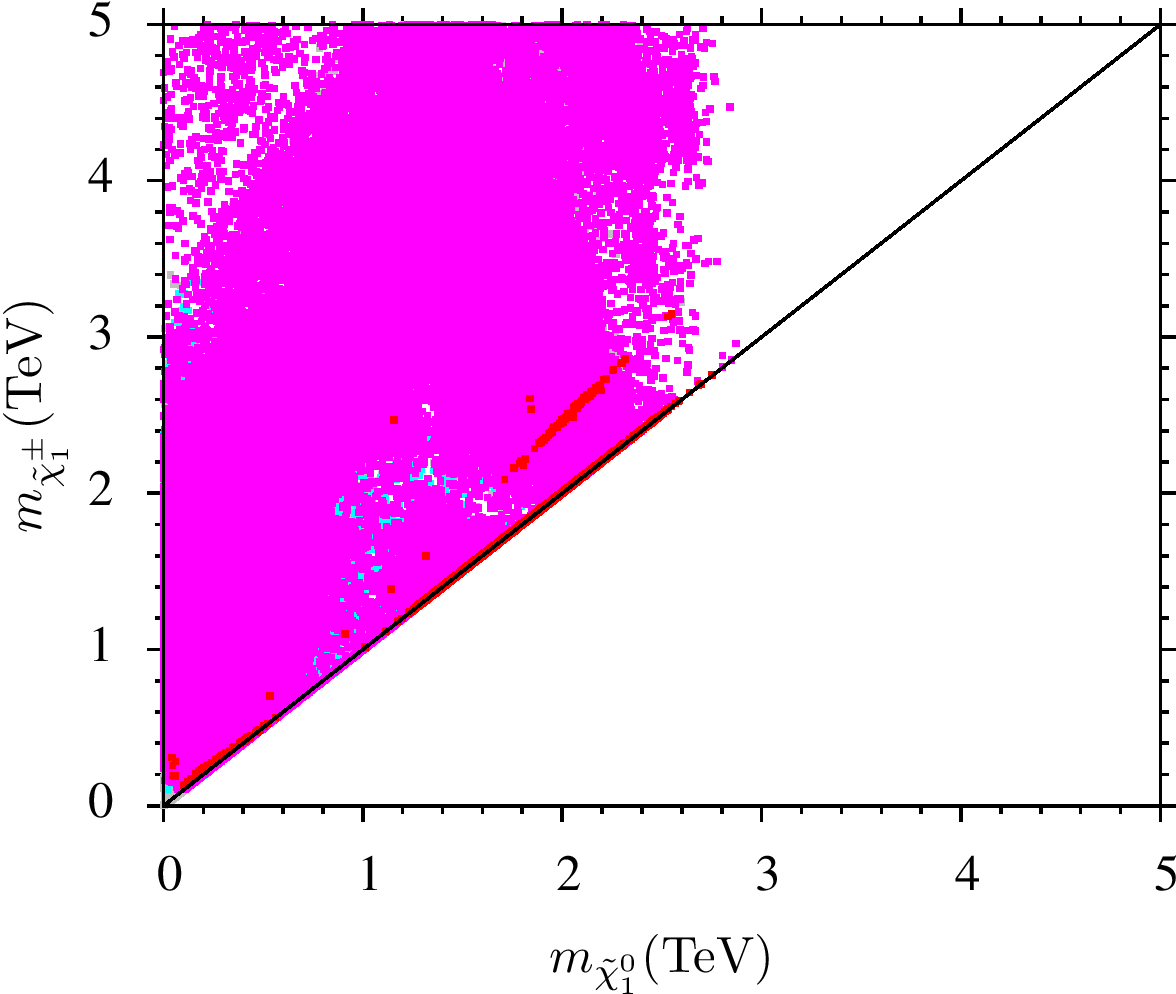}
}
\subfigure{
\includegraphics[totalheight=5.5cm,width=7.cm]{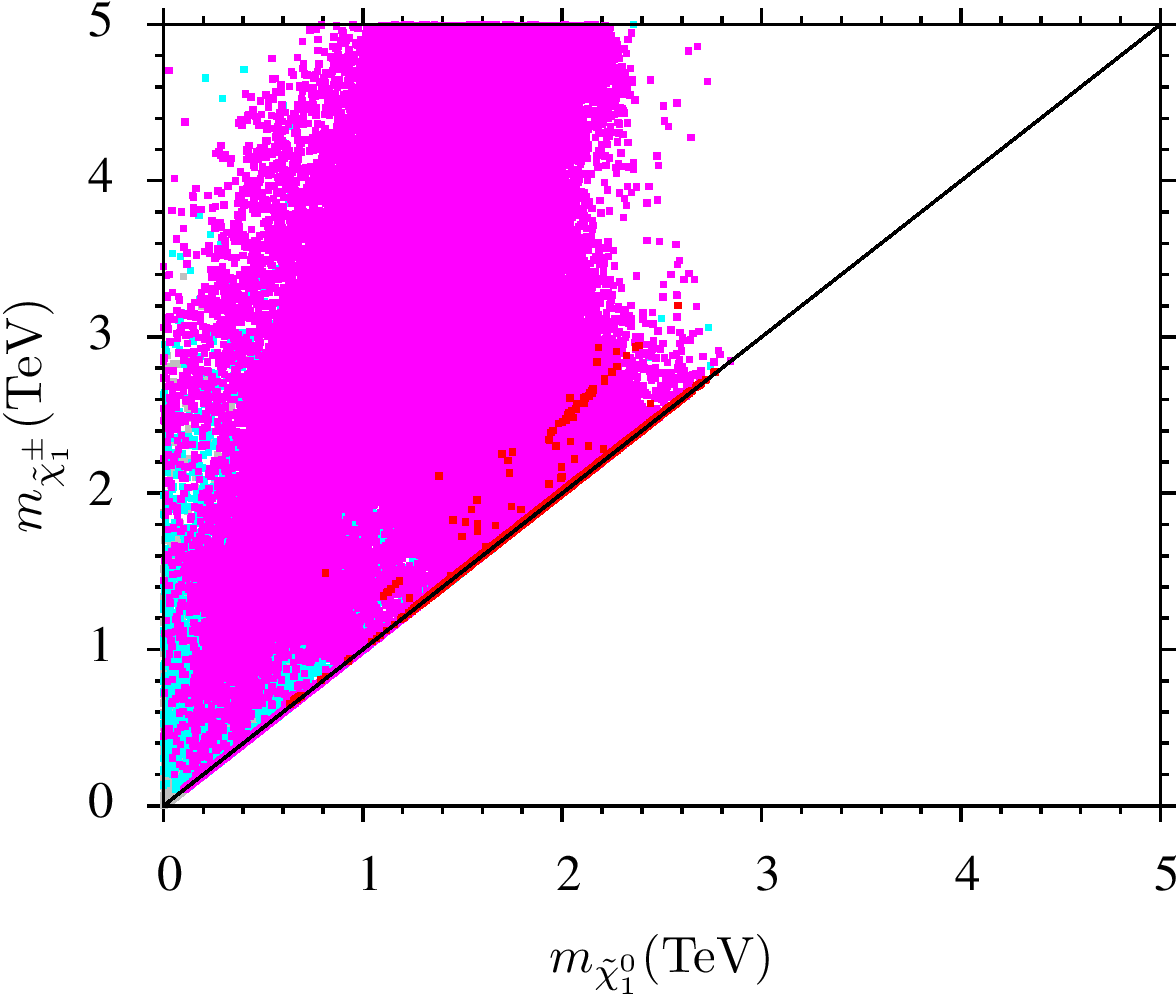}
}
\subfigure{
\includegraphics[totalheight=5.5cm,width=7.cm]{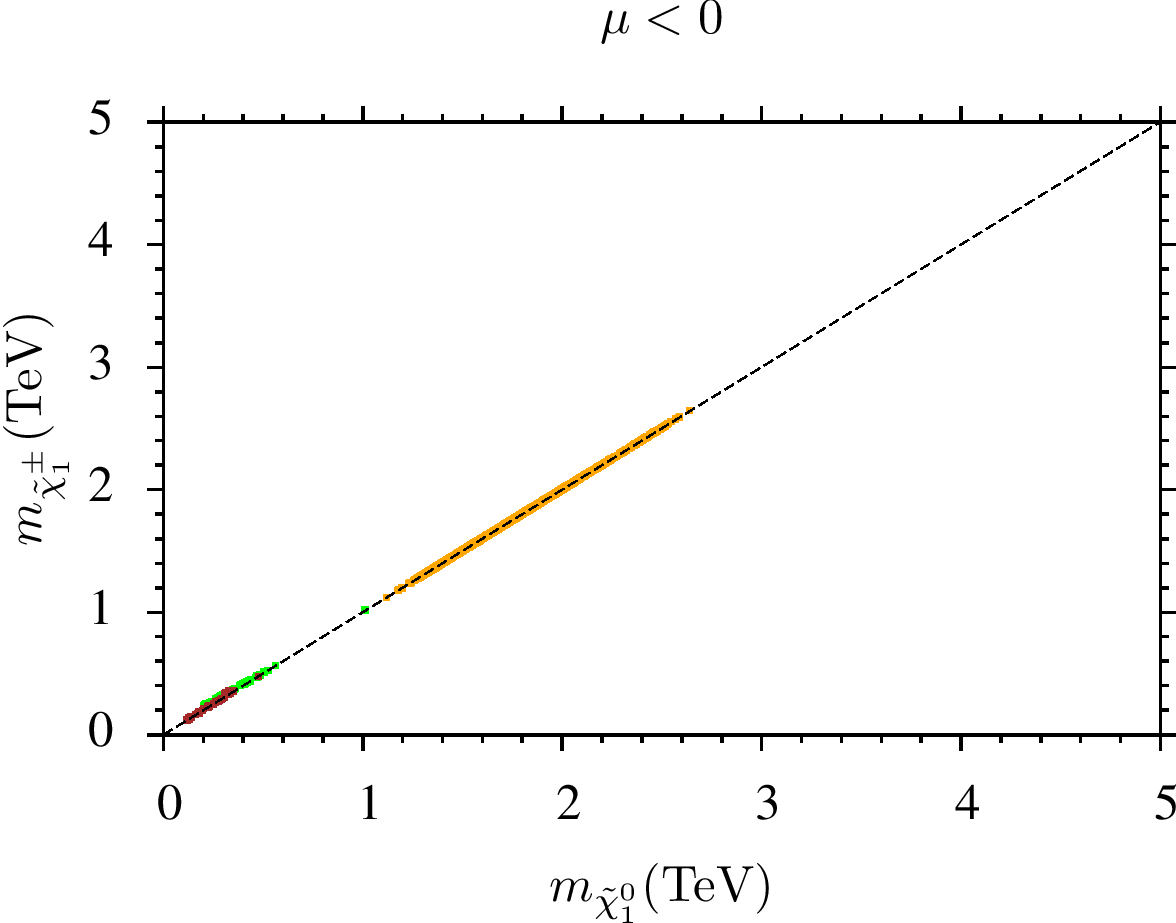}
}
\subfigure{
\includegraphics[totalheight=5.5cm,width=7.cm]{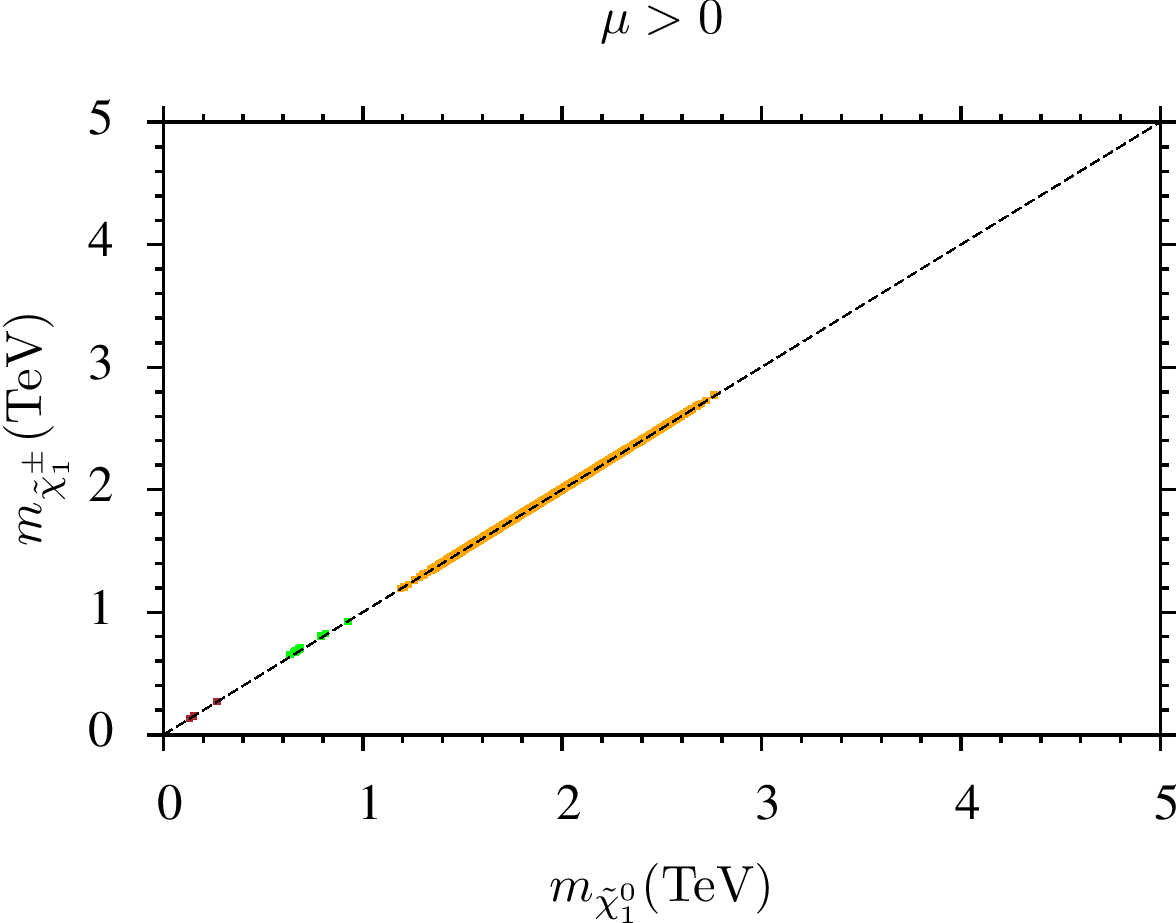}
}

\caption{Plots in $m_{\tilde \chi_{1}^{0}}-m_{\tilde \chi_{1}^{\pm}}$ planes.
Color coding and panel description are same as in Fig.~\ref{input_params1} for plots in top left and
right panels. In bottom left and right plots, the brown, green and orange  points represent higgsino-like, 
bino-like, and
wino-like neutralinos respectively with same panel description as in top panels. 
}
\label{charNeu}
\end{figure}
\begin{figure}[htp!]
\centering
\subfiguretopcaptrue

\subfigure{
\includegraphics[totalheight=6cm,width=8.cm]{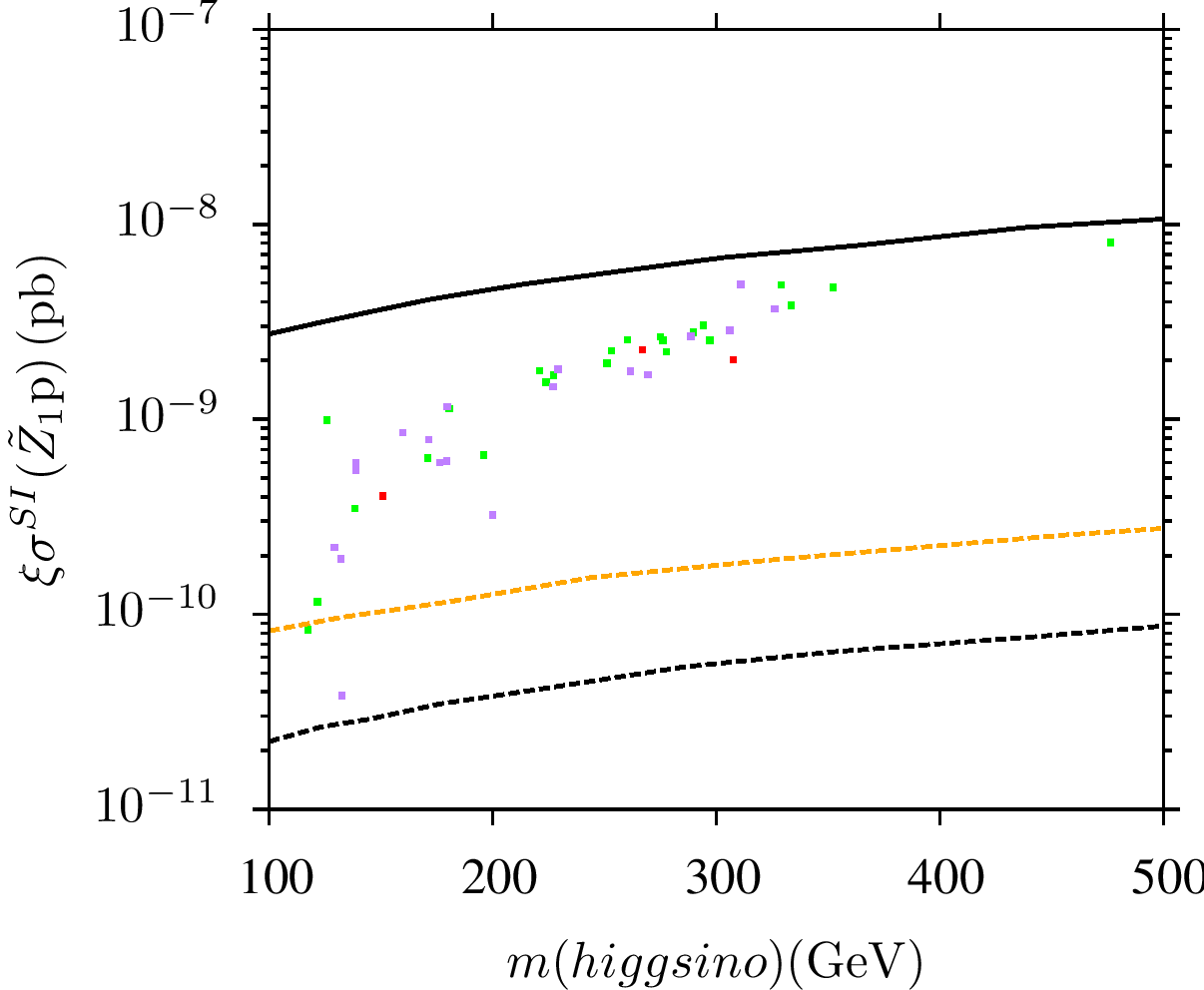}
}
\subfigure{
\includegraphics[totalheight=6cm,width=8.cm]{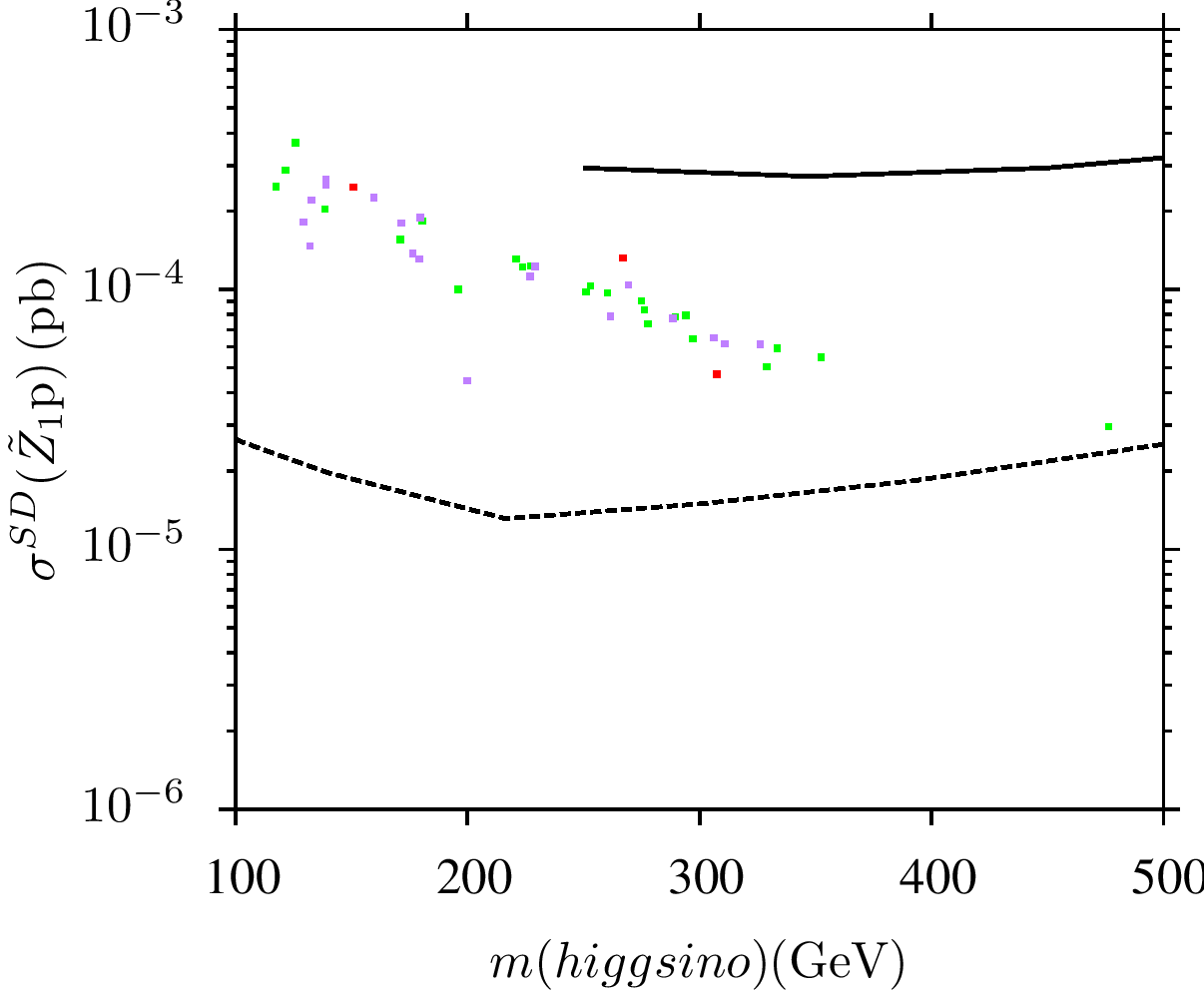}
}

\caption{Plots in rescaled higgsino-like neutralino spin-independent cross section $\xi \sigma^{SI}(\tilde Z_{1}p)$ versus
$m({\rm higgsino})$ and (non-rescaled) higgsino-like neutralino spin-dependent cross section 
$\sigma^{SD}(\tilde Z_{1}p)$ versus $m({\rm higgsino})$.
In the left panel, the orange solid line represents the current
 upper bound set by the CDMS experiment and black solid line depicts the upper bound set by XENON100,
 while the orange and black dashed lines represent respectively the future reach
 of the SuperCDMS and XENON1T experiments.
In the right panel, the IceCube DeepCore (black solid line) bound is shown and future IceCube DeepCore bound 
is depicted by the black dashed line. Green points represent solutions with $\Delta_{EW} \gtrsim$100, 
purple points display solutions with 50$\lesssim \Delta_{EW} \lesssim$100 while red solutions
satisfy $\Delta_{EW} \lesssim$50. Here, we have combined the solutions with $\mu<0$ and $\mu>0$.
}
\label{SISD}
\end{figure}

\begin{table}[b]\hspace{-1.0cm}
\centering
\begin{tabular}{|c|ccccc|}
\hline
\hline
                 & Point 1 & Point 2 & Point 3 & Point 4 & Point 5\\

\hline
$m_{L}$        & 2763.6 & 4077.8   & 1353.6    & 3125.3 & 2599.1\\
$m_{R}$        & 2187.1 & 3343.4   & 1401.1    & 1414.8 & 4991.8\\
$M_{1} $       & -379.47 & -441.48  & -865.62   & 4894.5 &5792.8\\
$M_{2}$        & 2850.2 & 3957.4   & -2681.8   & 1802.4 & 2082.9\\
$M_{3}$        & 1495.1 & 2206.2   & -1194.1   & 4953.4  &8736.4\\
$A_0$          & -3144.2 & -5006.1  & 2127.6    & -2884.4 & -2626.3\\
$\tan\beta$    & 12.8  & 23            & 11.1      & 57.7 & 46.8\\
$m_{H_u}=m_{H_d}$   & 3470  & 5127.6  & 1857    & 1778.4 & 3692.7\\
\hline
$\mu$            &-290   & -244        &   -1110        & -5500 & -8688\\
$\Delta_{EW}$    & 24    & 56          &    297         & 7279 & 18163\\
$\Delta_{HS}$    & 2941  & 6395        &    1125        & 7663 & 20667\\
\hline

\hline
$m_h$            &124   & 125     & 123       & 125 & 126\\
$m_H$            & 3828 & 5146    & 2712      &1369  &6761\\
$m_A$            & 3803 & 5112    & 2695      &1360  &6717\\
$m_{H^{\pm}}$    & 3828 & 5146    & 2713      &1373  &6762\\

\hline
$m_{\tilde{\chi}^0_{1,2}}$
                 & 177, 300 & 202, 255    & 377, 1120    & 1503, 2221 &1697, 2647\\

$m_{\tilde{\chi}^0_{3,4}}$
                 & 310, 2375   & 273, 3314 & 1122, 2209    & 5403, 5403 &8561, 8561\\

$m_{\tilde{\chi}^{\pm}_{1,2}}$
                 & 285, 2348  & 238, 3274 &  1139, 2193    &1506, 5356  &1700, 8478\\
\hline
$m_{\tilde{g}}$  & 3281 & 4725 &  2630  & 9818 &16760\\
\hline
$m_{ \tilde{u}_{L,R}}$
                 & 4239, 3490 & 6112, 5133 & 3069, 2626  & 8934, 8510 & 14308, 15024\\
$m_{\tilde{t}_{1,2}}$
                 & 1240, 3536 & 1933, 4958  & 1273, 2620  & 6955, 7659 & 12604, 12906\\
\hline $m_{ \tilde{d}_{L,R}}$
                 & 4240, 3491 & 6112, 5136  & 3070, 2622  & 8934, 8456 & 14308, 14994\\
$m_{\tilde{b}_{1,2}}$
                 & 3386, 3579 & 4751, 5025  & 2565, 2638  & 7066, 7617 & 12672, 13902\\
\hline
$m_{\tilde{\nu}_{1,2}}$
                 & 3294 & 4767  & 2175  & 3415 &3015\\
$m_{\tilde{\nu}_{3}}$
                 &  3268 & 4624 & 2162  & 3221 &2551\\
\hline
$m_{ \tilde{e}_{L,R}}$
                & 3291, 2187 & 4763, 3342 &2176, 1435  & 3423, 2273 & 3006, 5410  \\
$m_{\tilde{\tau}_{1,2}}$
                & 2094, 3264 & 2898, 4620&  1386, 2161 & 1511, 3218 &2527, 4810 \\
\hline

$\sigma_{SI}({\rm pb})$
                & $2.41\times 10^{-9}$ & $ 8.34\times 10^{-9} $ & $ 5.97\times 10^{-11}$ & $2.78\times 10^{-12} $ & 
$5.52\times 10^{-13} $\\

$\sigma_{SD}({\rm pb})$
                & $3.39\times 10^{-5}$ & $1.29\times 10^{-4}$ & $8.55\times 10^{-8}$ & $1.52\times 10^{-9} $ &
$1.71\times 10^{-10} $\\

$\Omega_{CDM}h^{2}$&  0.731 & 0.101   & 21.3  & 0.126 & 0.115\\
$R_{tb\tau}$     & 7.44     & 4.04    & 8.68  & 1.05 & 1.85\\
$R_{b\tau}$      & 1.35     & 1.35    & 1.36  & 1.04 & 1.35\\
\hline
\hline
\end{tabular}
\caption{All the masses are in units of GeV and $\mu<0$.
All points satisfy the sparticle mass bounds, and B-physics constraints described in Section~\ref{sec:scan}. 
Points 1 and 2 represent the minimal value of $\Delta_{EW}$ not consistent and consistent with 
the WMAP9 5$\sigma$ bounds, 
while points 3-5  correspond respectively to the minimal value of $\Delta_{HS}$, best points with $t$-$b$-$\tau$ and
$b$-$\tau$ YU, an example of heavy gluino solution. Points 3 and 4 also satisfy the WMAP9 5$\sigma$ bounds. 
}
\label{table1}
\end{table}
\begin{table}[b]\hspace{-1.0cm}
\centering
\begin{tabular}{|c|cccc|}
\hline
\hline
                 & Point 1 & Point 2 & Point 3 & Point 4 \\

\hline
$m_{L}$        &2608.7  & 4050.1   & 3521.9  & 6589.5\\
$m_{R}$        &2270.8  & 2651.8   & 3022.8  & 6501\\
$M_{1} $       &-4372.3  & -2542.1 & -4090.1 & -2646.7\\
$M_{2}$        &-2924.2  & -3947.6 & -2720.5 & -6538.3\\
$M_{3}$        & -5426.5 & -1767.6 & -4397.9 & -487.58\\
$A_0$          & 2123.4  & 4103.5  & 4070.6  & 6423.1\\
$\tan\beta$    & 33.6    & 23.5    & 45.8    & 32.6\\
$m_{H_u}=m_{H_d}$ &2216.6& 3258.7  & 3615    & 4428.1\\
\hline
$\mu$            &-5691  & -2418 & -4801 & -3572\\
$\Delta_{EW}$    &7793   & 1489 &   5546 & 3070\\
$\Delta_{HS}$    & 8745  & 4049 &   8489 & 7761\\
\hline

\hline
$m_h$            &126 & 127 & 126 & 125 \\
$m_H$            & 5307 &4246  & 3827 & 5267\\
$m_A$            & 5272 &4218  & 3802 & 5233\\
$m_{H^{\pm}}$    & 5308 &4247 & 3828 & 5267\\

\hline
$m_{\tilde{\chi}^0_{1,2}}$
                 & 1958, 2437  &  1155, 2425  & 1863, 2285   & 1237, 3591\\

$m_{\tilde{\chi}^0_{3,4}}$
                 & 5718, 5720  & 2429, 3327  & 4815, 4817   & 3592, 5542\\

$m_{\tilde{\chi}^{\pm}_{1,2}}$
                 & 2444, 5719  & 2467, 3303  & 2289, 4817  & 3646, 5501\\
\hline
$m_{\tilde{g}}$  & 10673  & 3847 & 8807  & 1308 \\
\hline
$m_{ \tilde{u}_{L,R}}$
                 & 9526, 9362  & 5644, 4153    & 8369, 8082   & 7717, 6555\\
$m_{\tilde{t}_{1,2}}$
                 & 7694, 8601 & 1186, 4747   & 6173, 7097   & 3084, 6172\\
\hline
$m_{ \tilde{d}_{L,R}}$
                 & 9527, 9324  & 5645, 4120   & 8369, 8046   & 7717,6533 \\
$m_{\tilde{b}_{1,2}}$
                 & 8562, 8820  & 3751, 4817   & 7022, 7081   & 5649, 6210\\
\hline
$m_{\tilde{\nu}_{1,2}}$
                 &3242  & 4760 & 3958 & 7749\\
$m_{\tilde{\nu}_{3}}$
                 & 3081  & 4648 & 3505 & 7370\\
\hline
$m_{ \tilde{e}_{L,R}}$
                & 3256, 3089  & 4754, 2801  & 3964, 3367   & 7742, 6566  \\
$m_{\tilde{\tau}_{1,2}}$
                & 2356, 3089   & 2360, 4640  & 2097, 3504   &  5591, 7355\\
\hline

$\sigma_{SI}({\rm pb})$
                & $2.30\times 10^{-12}$ & $ 1.85\times 10^{-11} $ & $ 3.70\times 10^{-12} $ & $3.63\times 10^{-12}$\\

$\sigma_{SD}({\rm pb})$
                & $1.11\times 10^{-10}$ &$ 3.88\times 10^{-9} $ & $ 2.51\times 10^{-10} $ & $6.93\times 10^{-10}$ \\

$\Omega_{CDM}h^{2}$&  0.114 &0.114  & 0.121 & 0.055 \\
$R_{tb\tau}$       & 2.5    & 3.9   & 1.73  & 2.5 \\
$R_{b\tau}$        & 1.37   & 1.38  & 1.44  & 1.34 \\
\hline
\hline
\end{tabular}
\caption{All the masses are in units of GeV and $\mu<0$.
All points satisfy  the sparticle mass bounds, and B-physics constraints described in Section~\ref{sec:scan}.
Points 1, 2, 3 and 4 display neutralino-stau, neutralino-stop, $A$-resonance, and neutralino-gluino
coannihilation respectively. Point 4 is the case where relic density is below the WMAP9 5$\sigma$ bounds. 
}
\label{table2}
\end{table}
\begin{table}[b]\hspace{-1.0cm}
\centering
\begin{tabular}{|c|cccc|}
\hline
\hline
                 & Point 1 & Point 2 & Point 3 & Point 4\\

\hline
$m_{L}$        &4342.3  &2443.1  & 3047.1 &  2537.4\\
$m_{R}$        &3727.6  &1830.1  & 1485.7 &  2570.6\\
$M_{1} $       &-657.78 &-886.71  & 5041.4 & -4748.2\\
$M_{2}$        &4058.9  &-4758.7 &  1623.6 & -2511.2 \\
$M_{3}$        &2107.5  &-2435  &  5143.7 &  -5673.4\\
$A_0$          &-5395.7 & 3635.3 &  -2951.9 & 1813.3\\
$\tan\beta$    & 26     & 18.1 & 57.2 & 31.9\\
$m_{H_u}=m_{H_d}$  &5307.9 &3954.6  & 1689.5 & 1112.0\\
\hline
$\mu$            &-333 & -312 & -5696 & -6129\\
$\Delta_{EW}$    &79      & 38   &  7807 & 9039\\
$\Delta_{HS}$    &6869    & 3813 &  8100 & 9082\\
\hline

\hline
$m_h$            &126   & 126   &  125 & 125 \\
$m_H$            & 5140 & 4682 &  1834 & 5509\\
$m_A$            & 5106 & 4651 &  1822 & 5473\\
$m_{H^{\pm}}$    & 5141 & 4682 &  1836 & 5509\\

\hline
$m_{\tilde{\chi}^0_{1,2}}$
                 & 298, 345  &  307, 323  & 1347, 2287 & 2084, 2157\\

$m_{\tilde{\chi}^0_{3,4}}$
                 & 366, 3406 & 399, 3946  & 5596, 5596 & 6157, 6158\\
$m_{\tilde{\chi}^{\pm}_{1,2}}$
                 & 325, 3367  & 332, 3920   & 1350, 5548& 2091, 6158\\
\hline
$m_{\tilde{g}}$  & 4549  & 5075  & 10172& 11140\\
\hline
$m_{ \tilde{u}_{L,R}}$
                 & 6208, 5266   & 5705, 4643     & 9177, 8827 & 9826, 9820\\
$m_{\tilde{t}_{1,2}}$
                 & 1924, 4943  & 2038, 4824    & 7257, 7896 & 8226, 8947\\
\hline
$m_{ \tilde{d}_{L,R}}$
                 &6208, 5268   & 5706, 4644    & 9177, 8767 & 9827, 9784\\
$m_{\tilde{b}_{1,2}}$
                 &4775, 5014   & 4419, 4880    & 7413, 7856 & 8907, 9337\\
\hline
$m_{\tilde{\nu}_{1,2}}$
                 &5027  & 3850  & 3313 & 3063\\
$m_{\tilde{\nu}_{3}}$
                 & 4833  & 3783 & 3118 & 2936\\
\hline
$m_{ \tilde{e}_{L,R}}$
                & 5022, 3731   & 3849, 1849  & 3322, 2360 & 3077, 3093\\
$m_{\tilde{\tau}_{1,2}}$
                & 3161, 4828    & 1537, 3781   & 1637, 3112 & 2793, 2961\\
\hline

$\sigma_{SI}({\rm pb})$
                & $9.66\times 10^{-9}$ & $ 7.41\times 10^{-9} $ & $ 1.02\times 10^{-12} $ & $ 2.55\times 10^{-11} $\\

$\sigma_{SD}({\rm pb})$
                & $7.14\times 10^{-5}$ &$ 4.72\times 10^{-5} $ & $ 1.25\times 10^{-9} $ & $ 1.45\times 10^{-9} $\\

$\Omega_{CDM}h^{2}$&  0.119 &0.03  & 0.091 & $0.14\times 10^{-5}$\\
$R_{tb\tau}$       & 3.55    & 5.2   & 1.10& 2.0\\
$R_{b\tau}$        & 1.37   & 1.34  & 1.10 & 1.37\\
\hline
\hline
\end{tabular}
\caption{All the masses are in this table are in units of GeV and $\mu <0$.
All this points satisfy  the sparticle mass and B-physics constraints described in Section~\ref{sec:scan}.
Point 1 represents bino-like neutralino, point 2 displays higgsino like neutralino, point 3 and point 4 are
examples of wino-like neutralino. Point 2 and point 4 do not satisfy WMAP9 5$\sigma$ bounds.
}
\label{table3}
\end{table}
\begin{table}[b]\hspace{-1.0cm}
\centering
\begin{tabular}{|c|cccc|}
\hline
\hline
                 & Point 1 & Point 2 & Point 3 & Point 4\\

\hline
$m_{L}$        & 1992.4  &1541.5  &3525.1  & 3632.8  \\
$m_{R}$        & 1598.5  &1659.5  &3466.1   & 5259.2 \\
$M_{1} $       & -507.03 &-911.52  &-3825.5  & 6933.9  \\
$M_{2}$        & -3790.9 &-2613.2  &-2531.4  & 1916.5 \\
$M_{3}$        & -1951.9 &-922.33  &-2993.5  & 9495\\
$A_0$          &  2679.3 &2189.1   & 1957.5 & 525.13\\
$\tan\beta$    & 18      &22.4     & 31.7 & 40.2\\
$m_{H_u}=m_{H_d}$   & 3199.1  &1676.6  &662.98  & 4072.4\\
\hline
$\mu$            & 350 & 1083 & 4293& 8968\\
$\Delta_{EW}$    & 32  & 286 & 4435 & 19351\\
$\Delta_{HS}$    & 2502 &964 & 4438  & 22605\\
\hline

\hline
$m_h$            &125 & 123 & 124 & 126\\
$m_H$            &3789  &2344  & 3602  &8695 \\
$m_A$            &3765  &2328  & 3578 & 8638\\
$m_{H^{\pm}}$    &3790  &2345  & 3603 & 8696\\

\hline
$m_{\tilde{\chi}^0_{1,2}}$
                 & 209, 364  &  400, 1092  & 1737, 2124   & 1537, 3176 \\

$m_{\tilde{\chi}^0_{3,4}}$
                 & 366, 3133  & 1094, 2158  & 4303, 4304   & 8853, 8853 \\

$m_{\tilde{\chi}^{\pm}_{1,2}}$
                 & 342, 3101  & 1067, 2141 & 2130, 4304  & 1541, 8767 \\
\hline
$m_{\tilde{g}}$  & 4131 & 2097 & 6216 &18151\\
\hline $m_{ \tilde{u}_{L,R}}$
                 & 4631, 3830  & 2818, 2406    & 6451, 6320   & 15622, 16208 \\
$m_{\tilde{t}_{1,2}}$
                 & 1834, 3936  & 1063, 2321   & 5001, 5687   & 13969, 14327 \\
\hline 
$m_{ \tilde{d}_{L,R}}$
                 & 4632, 3833  & 2819, 2402   & 6451, 6272   & 15622, 16155 \\
$m_{\tilde{b}_{1,2}}$
                 & 3663, 3974  & 2230, 2340   & 5646, 5803   & 14293, 15613 \\
\hline
$m_{\tilde{\nu}_{1,2}}$
                 & 3104 & 2269  & 3915 & 3952\\
$m_{\tilde{\nu}_{3}}$
                 & 3053  & 2215 & 3794 & 3566\\
\hline
$m_{ \tilde{e}_{L,R}}$
                & 3104, 1603  & 2268, 1692  & 3918, 3734   & 3959, 5821 \\
$m_{\tilde{\tau}_{1,2}}$
                & 1391, 3052   & 1528, 2212  & 3467, 3799   & 3549, 5212 \\
\hline

$\sigma_{SI}({\rm pb})$
                & $6.00\times 10^{-10}$ & $ 1.07\times 10^{-11} $ & $ 2.36\times 10^{-12} $ & $5.58\times 10^{-14} $\\

$\sigma_{SD}({\rm pb})$
                & $1.75\times 10^{-5}$ &$ 9.92\times 10^{-8} $ & $ 3.35\times 10^{-10} $ & $1.52\times 10^{-10} $\\

$\Omega_{CDM}h^{2}$&  1.07 & 22.3 & 0.105 & 0.104\\
\hline
$R_{tb\tau}$     & 5.31   & 4.18 & 2.31  &2.86\\
$R_{b\tau}$      & 1.33   & 1.29 & 1.05 & 1.91\\
\hline
\hline
\end{tabular}
\caption{
All the masses  are in units of GeV and $\mu>0$.
All points satisfy  the sparticle mass bounds and B-physics constraints described in Section~\ref{sec:scan}.
Points 1-4 respectively correspond to the minimal value of $\Delta_{EW}$, minimal value of $\Delta_{HS}$, 
best point with $b$-$\tau$ YU, and
an example of heavy gluino solution. Points 3 and 4 also satisfy the WMAP9 5$\sigma$ bounds. 
}
\label{table4}
\end{table}

\begin{table}[b]\hspace{-1.0cm}
\centering
\begin{tabular}{|c|cccc|}
\hline
\hline
                 & Point 1 & Point 2 & Point 3 & Point 4\\

\hline
$m_{L}$        & 2517.8  &3734.8  &2660.3  & 5549.1   \\
$m_{R}$        & 2187.1  &1435    &2837.6  & 7667.2 \\
$M_{1} $       & 1843.3  &1871.4  &-4493   & -3325.8  \\
$M_{2}$        & 1796    & 1229.6 &-3017   & -4120.9 \\
$M_{3}$        & 3883    & 2215.4 &-5334.2 & -636.94\\
$A_0$          & -971.61 & -5825.6&1714.9  & 2617.6\\
$\tan\beta$    & 56.10    &  18.90  &36.80    & 51.23\\
$m_{H_u}=m_{H_d}$ &2985.9 &4082.3 &1226.1  & 1762.9\\
\hline
$\mu$            & 3773 & 2579 &5817 & 4968\\
$\Delta_{EW}$    & 3425  &1699 &8142 & 5938\\
$\Delta_{HS}$    & 5392  &5721 &8189 & 6542\\
\hline

\hline
$m_h$            &124 & 127 & 125 & 123\\
$m_H$            & 3296 & 4578 & 4047 & 2469 \\
$m_A$            & 3248 & 4548 & 4021 & 2453\\
$m_{H^{\pm}}$    & 3271 & 4578 & 4048 & 2470\\

\hline
$m_{\tilde{\chi}^0_{1,2}}$
                 & 814, 1488   &  833, 1033   & 2046, 2521    & 1542, 3501  \\

$m_{\tilde{\chi}^0_{3,4}}$
                 & 3793, 3793   & 2587, 2588   & 5831, 5832    & 4946, 4948  \\

$m_{\tilde{\chi}^{\pm}_{1,2}}$
                 & 1491, 3793   & 1032, 2612  & 2529, 5831   & 3505, 4939  \\
\hline
$m_{\tilde{g}}$  & 7817 & 4691  & 10511   &1662\\
\hline $m_{ \tilde{u}_{L,R}}$
                 & 7157, 6999   & 5476, 4206     & 9413, 9390   & 6139, 7788   \\
$m_{\tilde{t}_{1,2}}$
                 & 5672, 6310   & 896, 4523    & 7833, 8321    & 4343, 6129  \\
\hline 
$m_{ \tilde{d}_{L,R}}$
                 & 7158, 6994   & 5476, 4190    & 9413, 9349    & 6140, 7765  \\
$m_{\tilde{b}_{1,2}}$
                 & 6264, 6392   & 3906, 4596    & 8272, 8479    & 4417, 6749  \\
\hline
$m_{\tilde{\nu}_{1,2}}$
                 & 2753 & 3821  & 3327 & 6140\\
$m_{\tilde{\nu}_{3}}$
                 & 2332  & 3721  & 3195 & 5532\\
\hline
$m_{ \tilde{e}_{L,R}}$
                & 2761, 2284  & 3818, 1569   & 3334, 3272    & 6142, 7759  \\
$m_{\tilde{\tau}_{1,2}}$
                & 825, 2333   & 1009, 3724   & 2974, 3213    & 5533, 6765  \\
\hline

$\sigma_{SI}({\rm pb})$
                & $7.49\times 10^{-13}$ & $ 4.31\times 10^{-12} $ & $ 1.12\times 10^{-12} $ & $5.95\times 10^{-13} $\\

$\sigma_{SD}({\rm pb})$
                & $4.21\times 10^{-10}$ &$ 1.77\times 10^{-9} $ & $ 1.06\times 10^{-10} $ & $1.68\times 10^{-10} $\\

$\Omega_{CDM}h^{2}$&  0.137 & 0.296 & 0.126 & 0.178\\
\hline
$R_{tb\tau}$     & 2.13   & 4.96 & 1.83  &1.48\\
$R_{b\tau}$      & 2.12   & 1.40 & 1.08 & 1.21\\
\hline
\hline
\end{tabular}
\caption{All the masses  are in units of GeV and $\mu>0$.
All points satisfy  the sparticle mass bounds, and B-physics constraints described in Section~\ref{sec:scan}.
Points 1, 2, 3 and 4 display neutralino-stau, neutralino-stop, $A$-resonance and neutralino-gluino
coannihilation, respectively. Points 2 and 4 are the examples where relic density is little bit above the
WMAP9 5$\sigma$ bounds.
}
\label{table5}
\end{table}
\begin{table}[b]\hspace{-1.0cm}
\centering
\begin{tabular}{|c|cccc|}
\hline
\hline
                 & Point 1 & Point 2 & Point 3 & Point 4\\

\hline
$m_{L}$        &3620.5  &2136.8  & 2537.4  & 3216.5\\
$m_{R}$        &1475.8  &1248.8  & 2147.6  & 6057.0\\
$M_{1} $       &2080.6  &-713.72  & 5197.6 & -5042.8\\
$M_{2}$        &1109.9  &-3832.9 &  1992.9 & -2293.9\\
$M_{3}$        &2398.8  & -1991.3 & 6222.5 & -4933.4\\
$A_0$          &-5687.8 &3169.4  & -3322.2 & -1737.9\\
$\tan\beta$    & 16     & 19 &  16.57 & 11.12\\
$m_{H_u}=m_{H_d}$ & 4007.4& 3358.8 &  2721.1 & 3843.9\\
\hline
$\mu$            &2812 & 277 &  6507  & 5195\\
$\Delta_{EW}$    &1934 & 47   &  10189& 6493 \\
$\Delta_{HS}$    &5815 & 2770 &  11755& 9839\\
\hline

\hline
$m_h$            &126   & 126   &  125& 123\\
$m_H$            &4752 & 3884 &   7054& 6645\\
$m_A$            &4720 &3859  &  7008 & 6602\\
$m_{H^{\pm}}$    &4752 & 3885 &  7055 & 6646\\

\hline
$m_{\tilde{\chi}^0_{1,2}}$
                 & 926, 930   &  267, 289   & 1636, 2359 & 1902,2295 \\

$m_{\tilde{\chi}^0_{3,4}}$
                 & 2819, 2820  & 327, 3178  & 6430, 6430 & 5237, 5238\\
$m_{\tilde{\chi}^{\pm}_{1,2}}$
                 & 927, 2843   & 272, 3149    & 1644, 6373 & 1906, 5233\\
\hline
$m_{\tilde{g}}$  & 5044  & 4199  & 12141 & 9899\\
\hline
$m_{ \tilde{u}_{L,R}}$
                 & 5620, 4515   & 4756, 3755     &10601, 10561 & 8936, 10358  \\
$m_{\tilde{t}_{1,2}}$
                 & 1805, 4723   & 1394, 4003    & 8684, 9770 & 7989, 8885\\
\hline
$m_{ \tilde{d}_{L,R}}$
                 & 5621, 4495   & 4756, 3756    & 10602, 10511 & 8937, 10314  \\
$m_{\tilde{b}_{1,2}}$
                 & 4304, 4780   & 3543, 4055    & 9728, 10370 & 8061, 10228 \\
\hline
$m_{\tilde{\nu}_{1,2}}$
                 & 3697 & 3218  & 2925 & 3610\\
$m_{\tilde{\nu}_{3}}$
                 & 3626  & 3159 & 2864 & 3573\\
\hline
$m_{ \tilde{e}_{L,R}}$
                & 3696, 1642   & 3217, 1262   & 2950, 2854 & 3599, 6325\\
$m_{\tilde{\tau}_{1,2}}$
                & 1313, 3628    & 897, 3156   & 2729, 2886 & 3580, 6286\\
\hline

$\sigma_{SI}({\rm pb})$
                & $2.36\times 10^{-11}$ & $ 9.03\times 10^{-9} $ & $ 2.03\times 10^{-12} $ & $ 8.48\times 10^{-12} $\\

$\sigma_{SD}({\rm pb})$
                & $1.16\times 10^{-8}$ &$ 1.32\times 10^{-4} $ & $ 6.81\times 10^{-10} $ & $ 3.13\times 10^{-9} $\\

$\Omega_{CDM}h^{2}$&  0.101 & 0.028 & 0.109 & $0.16\times 10^{-5}$\\
\hline
$R_{tb\tau}$     & 6.05   & 4.98 & 6.25 & 8.2\\
$R_{b\tau}$      & 1.41   & 1.33 & 1.51 & 1.25\\
\hline
\hline
\end{tabular}
\caption{
All the masses are in this table are in units of GeV and $\mu>0$.
All this points satisfy  the sparticle mass and B-physics constraints described in Section~\ref{sec:scan}.
Point 1 represents bino-like neutralino, point 2 displays higgsino like neutralino, point 3 and point 4
are examples of wino-like neutralino. Point 2  and point 4 do not satisfy WMAP9 5$\sigma$ bounds.
}
\label{table6}
\end{table}
In Fig.~\ref{input_params1}, we present graphs for various parameter given in Eq.~(\ref{input_param_range}). 
The left and the right panels show solutions for $\mu <0$ and $\mu>0$ scenarios, respectively. 
Color coding is given as, grey points satisfy
REWSB and neutralino as an LSP conditions. Aqua points satisfy the mass bounds and B-physics bounds. Magenta
points are subset of aqua points and also represent $123~{\rm GeV}\leqslant m_h\leqslant 127~{\rm GeV}$.
Red points are subset of magenta points and also satisfy WMAP9 5$\sigma$ bounds. 

We see that in our scans, in $\Theta_1-\Theta_2$ plane for both cases, the range of red points for 
$\Theta_1$ is $-0.6\lesssim \Theta_1 \lesssim 0.6$, but most of the points are concentrated 
in the range -0.4 to 0.4, while
for $\Theta_2$ most of the points are in the range of large values 0.4-0.8. But we also have 
some red points -0.6 to -0.4. 
On the other hand, magenta points can be more or less anywhere in the plot. We see that for $\Theta_1$, we have solutions 
for its entire range in contrast to $\Theta_2$ where points mostly have relatively large absolute values.
In $\Theta_1-\Theta_3$ plane we see that red points favor positive values of $\Theta_1$ and $\Theta_3$ as we have also
seen in $\Theta_1-\Theta_2$ plane.
We also see some red points for small negative values of $\Theta_1$ and but large negative values of $\Theta_3$.
Magenta points are every where but in contrast to $\Theta_1-\Theta_2$ plane, here large density of points are around 
the centre of the plot. In the last panel we have plot in $\Theta_3-\Theta_2$ plane. Here too, we see that the red points 
lie mostly in large positive ranges of $\Theta_2$ and $\Theta_3$. In case of magenta points, as compared 
to other panels, here we
have some kind of polarisation and we do not have magenta points in the center.

We calculate (SSB) parameters using Eqs.~(\ref{input_param_range}) and (\ref{ssb}). We present 
our results in Fig.~\ref{funda_params2}. Color coding and panel description are same 
as in  Fig.~\ref{input_params1}.
In the top left and right panels we present plots in $M_1-M_2$ plane. We note that there are some minor 
differences. In left panel we see that there is a patch of red points around $M_1 \sim$ -1 TeV and $M_2 \sim$ 4 TeV
 as compare to right panel where we have some red points around $M_1 \sim$ -5 TeV and $M_2 \sim$ 2 TeV. In case of magenta points, 
there are points up to $M_2 \sim$ -12 TeV in the left panel as compared to right panel. Similarly, we also note
minor differences in other panels of the figure.
In Fig.~\ref{funda_params3}, plots in $m_L-m_R$, $\tan\beta-m_{H_{u,d}}$ and $\tan\beta-A_0$ planes are displayed. Color coding is same as
in Fig.~\ref{input_params1}. The left panels represent $\mu <0$ and the right panels represent $\mu >0$ cases.
In $m_L-m_R$ plane we see that the left and right panels have almost similar data spread with some minor differences.
For example, in the left panel we have more points around $m_R \sim 12$ TeV, while in the right panel the maximum value of $m_L\approx$ 10 TeV.
In the $\tan\beta-m_{H_{u,d}}$ plane, right panel seems to be more populated in red points as compared to the left panel. 
We can see that in the right panel red points are $10 \lesssim \tan\beta \lesssim 60$ with 
$0 \lesssim m_{H_{u,d}} \lesssim 7$ TeV. This apparent difference is due to lack of data in the case of $\mu <0$. By
generating more data, we can reduce the apparent differences. In $\tan\beta-A_0$ plane too, we see the same situation.
But one thing is clear from both panels which is that data favours $A_{0} <0$. 

Plots in $\mu-\Delta_{EW}$ and $\Delta_{HS}-\Delta_{EW}$ planes are shown in Fig.~\ref{delew}. Color coding is 
same as in Fig.~\ref{input_params1}.
The top left and right panels depict plots with large ranges of parameters as compared 
to the bottom left and right panels. 
Moreover, the left and right panels represent $\mu<0$ and $\mu>0$ scenario, respectively. With large parameter ranges,
the top two panels almost look like the mirror images of each other. But from the left panel we see that it is relatively
easy to have WMAP9 compatible red points with $\mu<0$ as compared right panel with $\mu>0$ where 
the minimal value of $\Delta_{EW}$ for red points is about 2800. In order to investigate further we redraw the same plot with small ranges of parameters. 
We immediately note that there are some red points below $\Delta_{EW} \lesssim 200$. We also note that the minimal
values of $\Delta_{EW}$ with and without WMAP9 bounds are 56(1.78$\%$ FT) and 24(4.1$\%$ FT) respectively. 
On the other hand in the
right panel we see that the minimum value of $\Delta_{EW}$ for magenta points is 31(3.2$\%$ FT). 
We have also checked that 
in the right panel, points with relatively small values of $\Delta_{EW}$ have relic density of about 1. This shows that
if we try more harder we can get some solutions with small $\Delta_{EW}$ and compatible with the WMAP9 bounds. In the bottom 
left and right panels we show plots in $\Delta_{HS}-\Delta_{EW}$ plane. Here we see that for the entire data 
$\Delta_{HS} \gtrsim \Delta_{EW}$. We note that for $\mu<0$ case the minimal value of $\Delta_{HS}$ is
1125 (0.08$\%$ FT) with $\Delta_{EW}$ value of 297(0.33$\%$ FT), while we have 963(0.1$\%$ FT) and 285(0.35$\%$ FT) for $\Delta_{HS}$ and $\Delta_{EW}$ respectively for $\mu>0$. It was shown in dedicated studies of natural supersymmetry 
\cite{Gogoladze:2012yf,Gogoladze:2013wva} that with 
the above definitions of $\Delta_{EW}$ and $\Delta_{HS}$ it is possible to have values for both the measures $\lesssim$ 50
simultaneously.

In Fig.~\ref{m32} we show graphs in $m_h-\Delta_{EW}$, $m_h-\mu$ and $m_h-m_{3/2}$
planes. Color coding is same as in  Fig.~\ref{input_params1} with only one exception that there are no 
magenta color points. In these plots vertical solid black lines
represent Higgs mass bounds of 123 GeV and 127 GeV. Here we want to show $\Delta_{EW}$, $\mu$, $m_{3/2}$ and 
corresponding $m_h$ values. We note that most of our 
solutions in both the left and right panels are around $m_h \approx$125 GeV. Plots in
$m_h-m_{3/2}$ show that the gravitino mass $m_{3/2}$ has to be more than 2 TeV in both cases to have solutions
consistent with bounds on Higgs mass 123 to 127 GeV.

We know that the LHC is a color particle producing machine. Among the color particles, gluinos are the smoking guns
for the SUSY signals. Recent analysis have put limits of gluino mass $m_{\tilde g} \gtrsim$ 1.7 TeV 
(for $m_{\tilde g} \sim m_{\tilde q}$) and $m_{\tilde g}\gtrsim$ 1.3 TeV (for $m_{\tilde g} \gtrsim m_{\tilde q})$ \cite{Chatrchyan:2013wxa, Aad:2014wea}. 
In Fig.~\ref{glumu} we present plots in $m_{\tilde g}-\Delta_{EW}$ and $m_{\tilde g}-\mu$ planes. Color coding is 
same as in Fig.~\ref{input_params1} except we do not apply gluino mass bounds mentioned in Section~\ref{sec:scan}. The top left and right panels depict plots with large ranges of parameters as 
compared to the bottom left and right panels. Moreover, the left and right panels represent $\mu<0$ and $\mu>0$ scenario 
receptively. Here we show that in both scenarios we have heavy gluinos as $M_3$ is a free parameter in our model. Such
solutions can easily evade the above mentioned LHC bounds on gluino and squarks. In top left frame,
we see that we have $m_{\tilde g} \gtrsim$ 3 TeV for small values of $\Delta_{EW}$ in case of red points. 
Interestingly, there exists a region of parameter space with $|\mu|\lesssim$ 500 GeV and $\Delta_{EW}\lesssim$ 300, where gluino masses are from 3 to 7 TeV, and
the first two-generation squarks and sleptons are respectively 
in the mass ranges $[4,~7]$ TeV 
and $[2,~4]$ TeV. Because such parameter space is natural from low-energy fine-tuning definition while the gluino and first two-generation squarks/sleptons can not be
probed at the 14 TeV LHC, this will provide a strong motivation for
33 TeV and 100 TeV proton-proton colliders. In the 
top right frame, we have red points around $m_{\tilde g}\sim$ 5 TeV with $\Delta_{EW}\sim$ 2000. Even if we consider magenta points, 
we see that we lose very tiny amount of data because of LHC bounds  on gluino mass and most of our data remains intact. We also note that
in our model $\Delta_{EW}$ can be small over the gluino mass range of 2 to 10 TeV (magenta points). It is shown in \cite{cern_note1} that the squarks/gluinos of 2.5 TeV, 3 TeV and 6 TeV may be probed by the LHC14, High Luminosity (HL)LHC14 and High Energy (HE) LHC33, respectively. This clearly shows that our models have testable predictions. 
Moreover, in future if we have collider
facility with even higher energy, we will be able to probe over even larger values of sparticle masses.

We present results with neutralino mass verses $\tilde \tau_1$, $A$ and $\tilde \chi_{1}^{\pm}$ masses in Fig.~\ref{spectrum1}. 
Color coding is same as in  Fig.~\ref{input_params1} and same panel description. 
Solid black lines are just to guide the eyes, where we can expect to have coannihilation and resonance solutions.
 In the top left and right panels we have plots in $m_{\tilde \chi_{1}^{0}}-m_{\tilde \tau_1}$ plane. 
We here note that in the left panel, if do not care about the stringent WMAP9 5$\sigma$ bounds we have 
$\tilde \tau_1$ nearly degenerate with $\tilde \chi_{1}^{0}$ from 0.3 TeV to 2.8 TeV. But the solutions satisfy WMAP9 5$\sigma$ bounds have stau mass in the range of 1.2 TeV to 2.8 TeV. In the right panel, stau mass range is 0.8 TeV to 2.6 TeV while
we see the solutions without WMAP9 bonds have the have same stau mass range as we have in the left panel. 
Here we also note that the next to NLSP (NNLSP) $m_{\tilde \chi_{1}^{\pm}}$ is close to NLSP $m_{\tilde \tau}$ in mass. Their masses also lie within the 20$\%$ of LSP $\tilde \chi_{1}^{0}$ mass.
In the bottom left and right panels of the figure we present plots in $m_{\tilde \chi_{1}^{0}}-m_A$ plane.  We see that,
in both panels we have A-resonance solutions for more than 1 TeV $m_A$ without WMAP9 bounds. But if WMAP9 5$\sigma$ consistent points
have $m_A \gtrsim$ 2 TeV.

Graphs in $m_{\tilde \chi_{1}^{0}}-m_{\tilde t_1}$ and $m_{\tilde \chi_{1}^{0}}-m_{\tilde g}$ 
planes are shown in Fig.~\ref{spectrum2} with the same color coding and panel description given in Fig.~\ref{input_params1}, except in middle and bottom panels we do not apply gluino bounds mentioned in Section~\ref{sec:scan}. From top left panel
we see that we have two red points compatible with the 
WMAP9 bounds and representing neutralino-stop coannihilation scenario
with mass around 570 GeV and 1.2 TeV respectively. On the other hand in the right panel we do not have red points along the line but we know that it is just because of lack of statistics. In the middle left and right panels we show graphs in 
$m_{\tilde \chi_{1}^{0}}-m_{\tilde g}$. In both cases we see that there are no WMAP9 compatible red points. But we do 
note that we have some magenta solutions where gluino and neutralino masses are almost degenerate and $\Omega h^2 < 1$. In the right panel we 
see only one magenta point near the black line but we can always generate more data around this point. Graphs in
the bottom panels show that in our model, we can accommodate gluinos as heavy as 18 TeV consistent with WMAP9 5$\sigma$ bounds.
Such a scenario suggest that there should be very high energy collider in order to probe such model points.

We quantify $t$-$b$-$\tau$ and $b$-$\tau$ the Yukawa coupling unification (YU) via the R-parameter
\begin{equation}
R_{tb\tau}\equiv \frac{ {\rm max}(y_t,y_b,y_{\tau})} { {\rm min} (y_t,y_b,y_{\tau})},
R_{b\tau} \equiv \frac{ {\rm max}(y_b,y_{\tau})} { {\rm min} (y_b,y_{\tau})},
\label{eq:R}
\end{equation}
where $y_{t}$, $y_{b}$ and $y_{\tau}$ are Yukawa couplings at the scale
of the Grand Unified Theory (GUT). $R_{tb\tau}=1$ ($R_{b\tau}=1$) 
means $y_{t}=y_{b}=y_{\tau}$ ($y_{b}=y_{\tau}$) that is a solution with perfect $t$-$b$-$\tau$ ($b$-$\tau$) YU.

In Fig.~\ref{YU} we present graphs in $\tan\beta-R_{tb\tau}$ and $\tan\beta-R_{b\tau}$ planes. Color coding is
same as in Fig~\ref{input_params1}. The left panels represent graphs in $\tan\beta-R_{tb\tau}$ and $\tan\beta-R_{b\tau}$ 
planes in case of $\mu<0$ and the right panels represent  plots in $\tan\beta-R_{b\tau}$ and $\tan\beta-R_{tb\tau}$ for
$\mu >0$. The horizontal black dashed line represents 10$\%$ or better $t$-$b$-$\tau$ ($b$-$\tau$) YU. We see in
the top left panel that in our scans we have 10$\%$ or better YU solutions for $\tan\beta \sim 50-60$. The minimal value
for $R_{tb\tau}$ we have is 1.05 (5$\%$ YU). We note that 
red points below the dashed line have $m_h \sim$ 125 GeV, gluino in the mass range of 8 TeV to 10 TeV, the first 
two generation squarks and sleptons are in the mass ranges of 8 TeV to 9.5 TeV and 3 TeV 3.5 TeV respectively. 
The third generation squarks and sleptons lie in the mass ranges of 5 TeV to 7.5 TeV and 1.3 TeV to 3 TeV.  
They also have large values for $|\mu|$ ($\sim$ -5 TeV to -4 TeV) 
and $\Delta_{EW} \sim $ 4000 to 8500. More or less magenta points also have the similar mass spectrum. It was shown 
in more exhausted studies (see e.g \cite{Gogoladze:2010fu} and references there in) with non-universal gaugino masses that one can have 100$\%$ YU with the LHC testable predictions.

In the bottom left panel we have $b$-$\tau$ YU solutions. Since this is a less constraint situation, we have 10$\%$ or better 
YU solutions for a wider range of $\tan\beta$, {\it i.e.}, $30 \lesssim \tan\beta \lesssim 60$. Here, 
the minimal value of 
$R_{b\tau}$ is about $1.04$ (4$\%$ YU). 
Moreover, the particle mass spectra also have slightly wider ranges as compared to 
$t$-$b$-$\tau$ YU case. We also note that those magenta points,
which do not satisfy WMAP9 bounds,  have more or less the same mass ranges as given above.

In the top right panel, we see that we do not have even magenta solution with 10$\%$ 
or better $t$-$b$-$\tau$ YU with $\mu >0$. It was noticed  that 
in a SUSY $SO(10)$ GUT with non-universal
SSB gaugino masses at $M_{\rm GUT}$ and $\mu>0$, $t$-$b$-$\tau$ Yukawa unification \cite{big-422} can lead one to predict 
the lightest CP even Higgs boson mass to be 125 GeV \cite{Gogoladze:2011aa}. Even if we consider 
gaugino-universality 10$\%$ or better $t$-$b$-$\tau$ YU can be achieved consistent with the LHC bounds \cite{Baer:2012cp}
but very hard to satisfy relic density bounds.
In our scans, we have solutions with 12$\%$ $t$-$b$-$\tau$ YU, if we do dedicated searches for better YU solutions, 
we can get them. Since here, we are not so keen to have 100$\%$ YU 
solutions but to give a flavor that our model can admit such solutions. In the bottom right panel we see that we 
have only three red points. On the other hand we have plenty of magenta points with 10$\%$ or better $b$-$\tau$ YU.
In fact, we have $R_{b\tau}=$1, {\it i.e.}, 100$\%$ $b$-$\tau$ YU solutions for $\tan\beta \sim$ 30-55. 
In order to save such solutions we have to add some extra physics to the MSSM.
In such scenario we can consider SUSY models augmented with Peccei-Quinn(PQ) solution to the 
strong CP problem \cite{pqww} (PQMSSM). In SUSY context the axino field is just
one element of an axion supermultiplet. The axion supermutiplet contains a complex scalar field, whose real part is 
the $R$-parity even saxion field $s(x)$ and whose imaginary part is the axion field $a(x)$. The supermutliplet also contains
an $R$-parity odd spin half Majorana field, the axino $\tilde a(x)$\cite{axino_rev}. In case where $\Omega h^2 \gtrsim 1$, one way to have relic density within the observed range
if we assume the $\tilde \chi_{1}^{0}$ may not be the LSP, but instead decays to much lighter state, such as
$\tilde \chi_{1}^{0}\rightarrow \gamma \tilde a$, where $\tilde a$ is axino. 
In such a scenario we have mixed axion/axino ($a\tilde a$) 
dark matter \cite{bs}.
In this way the neutralino abundance is converted into an axino abundance with \cite{ckkr}
\begin{equation}
\Omega_{\tilde a} h^{2}=\frac{m_{\tilde a}}{m_{\tilde \chi_{1}^{0}}}\Omega_{\tilde \chi_{1}^{0}}^{2}.
\label{oh2}
\end{equation}  
It is important to know the life time ($\tau$) of decaying neutralino. If it is more than 1 second, it can disturb
Big Bang Nucleosynthesis (BBN) (see \cite{Choi:2013lwa} and references there in).
We first calculate $m_{\tilde a}$ for a given ${m_{\tilde \chi_{1}^{0}}}$ and its relic density $\Omega_{\tilde \chi_{1}^{0}}^{2}$ by assuming relic density of axino $\Omega_{\tilde a} h^{2}=$0.11 by using Eq.~\ref{oh2}. We then follow \cite{Baer:2010kw} 
to calculate the lifetime for the decaying NLSP neutralino.
We use $C_{aYY}= 8/3$ in the DFSZ model \cite{Dine:1981rt}, $N=$6 (the color anomaly model dependent factor). We present 
our calculations in Fig.~\ref{LT}, where we display the NLSP bino-like 
neutralino mass ($m_{\tilde \chi_{1}^{0}}$) versus its lifetime ($\tau$). Panel description is same as in 
Fig.~\ref{input_params1}. Here points in various colors correspond to various choices of 
the axion decay constant $f_a$ values. The
light-green, light-blue, orange, black, dark-green and blue represent $f_a=10^{10}-10^{15}\,{\rm GeV}$, respectively.
The black solid line show $\tau=$1 second. In the left panel, the plot appears somewhat flat as compare to the right
plot in right panel because of small mass range.
From both the frames, it is clearly visible that for $f_a=10^{15}\,{\rm GeV}$, life time of NLSP bino is more than 1 second.
For $f_a=10^{14} \,{\rm GeV}$, in the right panel, points with bino mass greater than 1 TeV are allowed. 

 In another approach to reduce relic density is to assume the additional late decaying scalar fields are present in the model. 
These fields may get produced at large rates via coherent oscillations. If they temporarily dominate the energy density of the
Universe, and then decay to mainly SM particles, they may inject considerable entropy into
the cosmic soup, thus diluting all relics which are present at the time of decay. Entropy
injection can occur at large rates for instance from saxion production in the PQMSSM \cite{shafi,bls}, or from moduli production and decay, as is expected in string theory \cite{gg}.
However, it was shown in \cite{Bae:2013qr} that the efforts to dilute the relic density of neutralino below the observed dark
matter relic density through entropy injection from saxion decays such as saxion decays to gluon violate the CMB bound
on $\Delta N_{eff}$, where $\Delta N_{eff}$ is the apparent number of additional effective neutrinos.

On the other hand, the
 solutions with good YU may also have small relic density $\Omega h^{2} \sim ~ 10^{-5}-10^{-2}$. In such cases 
the neutralino abundance can be augmented in the PQMSSM case where $m_{\tilde a} > m_{\tilde \chi_{1}^{0}}$
and additional neutralinos are produced via thermal axino
production and decay $m_{\tilde a} \rightarrow m_{\tilde \chi_{1}^{0}} \gamma$ \cite{bls}. In these cases, the CDM tends
to be neutralino dominated with a small component of axions.

In Fig.~\ref{charNeu} we show graphs in $m_{\tilde \chi_{1}^{0}}-m_{\tilde \chi_{1}^{\pm}}$ plane with the same
panel description as in Fig.~\ref{input_params1}. The top left and right frames have same color coding as in
Fig.~\ref{input_params1}.
From these frames, it is apparent  that we have solutions from 0.1 TeV to 2.8 TeV. In bottom frames we further analyse these points on the basis of
neutralino composition. Here orange, green and brown points represent neutralino with more than $90\%$ wino,
more than $80\%$ bino and more than $50\%$ higgsino composition, respectively. It is to be noted that orange and the green
points satisfy all constraints given Section~\ref{sec:scan} but brown point do not satisfy relic density bounds.
Here, we want to show that in our scans where the neutralino and chargino masses are almost degenerate, and
neutralino LSP can be of bino, wino and higgsino like. We immediately see that in both cases ($\mu<0$ and $\mu>0$), 
wino-type neutralino have masses more than 1 TeV. On the other hand bino-like solutions have masses less than 1 TeV 
while higgsino-type solutions have mass range of 150 to 600 GeV. It is shown in \cite{Fan:2013faa,Cohen:2013ama} that 
for NFW and Einasto distribution, the entire mass range of thermal wino dark matter from 0.1 to 3 TeV may be excluded.
In a recent study \cite{Hryczuk:2014hpa}, wino as dark matter candidate is excluded in the mass range bellow
800 GeV from antiproton and between 1.8 TeV to 3.5 TeV from the absence of a $\gamma$-ray line feature toward
the galactic center. Since our bino-like points have some admixture of higgsinos and that is why they have large 
nucleon-neutralino scattering cross section. Such solutions are also under stress because of 
the current upper bound set by XENON100 \cite{Xenon100}. 
Here, we argue that such wino-like (bino-like) neutralino solutions may avoid the above mentioned bounds. For example, 
the wino-like neutralino density is smaller than the observed density. Otherwise,
instead of treating them as the LSPs we assume that they are the NLSP and may decay to axino and $\gamma$ as we have 
discussed above. Similarly, we can also assume the mechanism of late decaying fields 
via coherent oscillations or production 
of moduli and their decay as we argued previously. In addition to it, we can also invoke $R$-parity violation scenario,
where the bino LSP and similarly wino-like neutralino can decay to the SM fermions via
sfermion exchange. In order to address the issue of underabundance of higgsino-like solutions we argue 
that mainly higgsino-like neutralino
 by itself does not make a good cold dark matter candidate 
and we need additional dark matter candidates to match the
observed dark matter relic density. For this purpose we assume  that the higgsino could make up 
only a fraction of the relic dark matter and the remaining abundance is comprised of 
axions produces through the vacuum misalignment mechanism \cite{vacmis}.
This is why we could expect the higgsino relic density somewhat suppressed between $1-15$
in the present universe. This not only provides us with the opportunity to look for higgsinos, despite the fact that they
would only constitute a fraction of the measured relic dark matter abundance 
but also the possibility to detect axions.
We would also like to mention that 
our higgsino-like solutions especially for $\Delta_{EW}\lesssim$50 more or less look like
the solutions form radiative natural SUSY \cite{Baer:2012cf}. Since such solutions tend to have large direct and 
indirect neutralino detection rates, let us check the 
 status of our higgsino-like solutions. We will follow \cite{Baer:2013vpa}.
In the left panel of Fig.~\ref{SISD} we plot rescaled higgsino-like neutralino spin-independent cross section $\xi \sigma^{SI}(\tilde Z_{1}p)$ versus $m({\rm higgsino})$ (in this figure for both panels we have combined solutions with $\mu<0$ and $\mu>0$). 
The orange solid line represents the current
 upper bound set by the CDMS experiment and black solid line depicts upper bound set by XENON100 \cite{Xenon100},
 while the orange (black) dashed line represents future reach
 of SuperCDMS \cite{Brink:2005ej} (XENON1T \cite{Aprile:2012zx}). We rescale our result by a factor $\xi=\Omega_{\tilde Z_1}h^{2}/0.11$ in order to account for the 
fact that the local relic density might be much less than the usually assumed value $\rho_{local}\simeq 0.3 \,{\rm GeV}/cm^{3}$
as pointed out in \cite{Bottino:2000jx}. Here, we see that all the points lie below the current upper bounds set by CDMS XENON100 experiments. It is very clear that the future experiments like XENON1T will be able to probe almost all of our model points. 
This shows our results are in agreement with \cite{Baer:2013vpa} where it was shown that all higgsino points could be tested 
by the XENON1T and one could discover neutralino (WIMPs) or exclude the concept of electroweak naturalness in $R$-parity conserving natural SUSY models.  
In right panel of Fig.~\ref{SISD}, we have a plot of (non-rescaled) higgsino-like neutralino spin-dependent 
cross section $\sigma^{SD}(\tilde Z_{1}p)$ versus $m({\rm higgsino})$. 
The IceCube DeepCore and future IceCube DeepCore bounds are
shown in black solid line and black dashed line \cite{IceCube}.
Color coding is same as in left panel. Here we do
not rescale our results because the IceCube detection depends on whether the Sun has equilibrated its core abundance
between capture rate and annihilate rate \cite{Jungman:1995df}. It was shown in \cite{Niro:2009mw} that for the Sun, equilibrium is reached for almost all of SUSY parameter space. In this plot we see that the future
IceCube DeepCore searches will be able to probe our entire set of solutions in our present scans.

In Tables~\ref{table1}-\ref{table3}, we list benchmark points for $\mu<0$ case. All of these points satisfy the sparticle mass, B-physics and Higgs mass constraints described in Section~\ref{sec:scan}. In Table~\ref{table1}, point 1(2) represents the minimal value 
of $\Delta_{EW}$ not consistent and consistent with WMAP9 5$\sigma$ bounds, while points 3-5 respectively correspond 
to the minimal value of $\Delta_{HS}$, best point with $t$-$b$-$\tau$ and
$b$-$\tau$ YU, an example of heavy gluino solution. Points 3 and 4 also satisfy WMAP9 5$\sigma$ bounds. In 
Table~\ref{table2}, points 1, 2, 3 and 4 display neutralino-stau, neutralino-stop, $m_A$-resonance and neutralino-gluino
coannihilation, respectively. Point 4 is the case where relic density is below WMAP9 5$\sigma$ bounds. In Table~\ref{table3},
point 1 represents bino-like neutralino, point 2 displays higgsino like neutralino, point 3 and point 4
are examples of wino-like neutralino. Point 2 and point 4 do not satisfy WMAP9 5$\sigma$ bounds.  

In Tables~\ref{table4}-\ref{table6}, we display benchmark points for $\mu>0$ case consistent with the sparticle 
mass, B-physics and Higgs mass constraints described in Section~\ref{sec:scan}. In Table~\ref{table4}, points 1-4 respectively correspond to the minimal value of $\Delta_{EW}$, minimal value of $\Delta_{HS}$, best point with $b$-$\tau$ YU,
an example of heavy gluino solution. Points 3 and 4 also satisfy WMAP9 5$\sigma$ bounds. Table~\ref{table5} and 
Table~\ref{table6} have similar description as Table~\ref{table2} and Table~\ref{table3}.
\section{Discussions and Conclusion}

The three-family Pati-Salam models have been constructed systematically in 
Type IIA string theory on the $\mathbf{T^6/(\Z_2\times \Z_2)}$
orientifold with intersecting D6-branes~\cite{Cvetic:2004ui}.
It was found that one model has a realistic phenomenology~\cite{Chen:2007px, Chen:2007zu}. 
Considering the Higgs boson mass around 125 GeV and the LHC supersymmetry search constraints,
we have revisited this three-family Pati-Salam model in details. We systematically scanned the viable
parameter space for $\mu<0$ and $\mu>0$, and found that in general the gravitino mass is heavier 
than about 2 TeV for both cases because of the Higgs boson mass low bound 123 GeV.
In particular, we identified a natural region of parameter space where 
the electroweak fine-tuning 
can be as small as $\Delta_{EW}\sim$ 24-32 (3-4$\%$). 
Also, we found another interesting region of 
parameter space with $|\mu|\lesssim$ 500 GeV and $\Delta_{EW}\lesssim$ 300, where the mass ranges for the gluino, 
and first two-generation squarks and sleptons are $[3,~7]$ TeV, $[4,~7]$ TeV, 
and $[2,~4]$ TeV, respectively. This will provide a strong motivation for
33 TeV and 100 TeV proton-proton colliders since it is natural from 
low-energy fine-tuning definition while the gluino
and first two generation squarks/sleptons are heavy.
In the whole viable parameter space which
is consistent with all the current experimental constraints including the dark matter relic density bounds, 
the gluino mass range is $[3, ~18]$ TeV, the first 
two-family squarks have masses from 3 to 16 TeV, and the first two-family sleptons have masses
from 2 to 7 TeV. Thus, the viable parameter space with heavy gluino and squarks is even out of 
reach of the 100 TeV proton collider~\cite{Cohen:2013xda}. 
On the other hand, for the third-family sfermions, 
the NLSP light stop satisfying $5\sigma$ WMAP bounds is in the mass range 
$[0.5,~1.2]$ TeV, and
the light stau can be as light as 800 GeV. We also showed various coannihilation and resonance scenarios 
through which the observed dark matter relic density can be achieved. Interestingly,
the certain portions of parameter space have excellent $t$-$b$-$\tau$ and $b$-$\tau$ Yukawa coupling 
unification. Also, we highlighted the regions of parameter space
where the LSP neutralino can be a bino, wino, or higgsino. We discussed various scenarios in which
such solutions may avoid recent astrophysical bounds in case if they satisfy 
or above the correct dark matter relic density bounds. 
Prospects of finding higgsino-like neutralino in direct and indirect searches 
were shown and discussed as well. To be concrete, we displayed
six benchmark tables depicting various interesting features of our model.
Furthermore, because the LSP neutralino can be heavier than 1 TeV and up to about 2.8 TeV, 
how to test such scenario at the 14 TeV LHC 
is still a big question. Therefore, the 33 TeV and 100 TeV proton-proton colliders are indeed needed 
to probe our D-brane model.

\label{conclusions}
\section*{Acknowledgements}
We would like to thank Howard Baer, Eung-Jin Chun, Bin He, and Azar Mustafayev very much
for helpful discussions. The work of TL, SR and XW is supported in part by by the Natural Science
Foundation of China under grant numbers 10821504, 11075194, 11135003, and 11275246, and by the National
Basic Research Program of China (973 Program) under grant number 2010CB833000.
And the work of DVN was supported in part by the DOE grant DE-FG03-95-ER-40917.


\end{document}